\pdfoutput=1
\documentclass[11pt, oneside]{mnthesis}
\usepackage{epsfig,epic,eepic,units}
\usepackage{hyperref}
\usepackage{url}
\usepackage{longtable}
\usepackage{mathrsfs}
\usepackage{multirow}
\usepackage{bigstrut}
\usepackage{amssymb}
\usepackage{graphicx}

\newcommand{\be}{\begin{equation}}
\newcommand{\ee}{\end{equation}}
\newcommand{\ba}{\begin{eqnarray}}
\newcommand{\ea}{\end{eqnarray}}
\newcommand{\bd}{\begin{displaymath}}
\newcommand{\ed}{\end{displaymath}}

\def\rootg{\sqrt{-g}}

\def\DL{\mathcal{D}_L}

\def\thalf{{\textstyle{\frac{1}{2}}}}

\begin{document}
\bibliographystyle{utphys} 

\phd

\title{\bf Hydrodynamics of Strongly Coupled Non-conformal Fluids
from Gauge/Gravity Duality}
\author{Greggory Todd Springer}
\campus{University of Minnesota} 
\program{Physics} 

\submissionmonth{August} 
\submissionyear{2009} 

\abstract{
The subject of relativistic hydrodynamics is explored
using the tools of gauge/gravity duality.  A brief literature review
of AdS/CFT and gauge/gravity duality is presented first.  This is
followed by a pedagogical introduction to the use of these methods in
determining hydrodynamic dispersion relations, $w(q)$, of perturbations
in a strongly coupled fluid.

Shear and sound mode perturbations are examined in a special class of
gravity duals: those where the matter supporting the metric is scalar
in nature.  Analytical solutions (to order $q^4$ and $q^3$
respectively) for the shear and sound mode dispersion relations are
presented for a subset of these backgrounds.

The work presented here is  based on previous publications 
by the same author \cite{Kapusta:2008ng}, \cite{Springer:2008js}, and
\cite{Springer:2009wj}, though some previously unpublished
results are also included.  In particular, the subleading term in the
shear mode dispersion relation is analyzed using the AdS/CFT
correspondence without any reference to the black hole membrane
paradigm.

}
\words{331}    
\copyrightpage 
\acknowledgements{
I would like to thank R. Anthony, E. Aver, A. Buchel, T. Cohen, A. Cherman,
T. Kelley, M.  Natsuume, D. T. Son, and A. Starinets for helpful
comments, suggestions and discussions regarding the work presented
here.
\newline
\newline
\noindent I thank the members of my examination committee:
 D. Cronin-Hennessey, L.L.R. Williams, and M. Peloso for their time
 and consideration
\newline
\newline
\noindent I would also like to thank T. Gherghetta, R. Fries and S. Barthel for
their help in transitioning to life after graduate school.
\newline
\newline
\noindent Finally, I gratefully acknowledge the support and guidance of my
adviser J. I. Kapusta who has provided invaluable insight and feedback
on all aspects of my graduate school career.
\newline
\newline
\noindent This work was was supported by the US Department of Energy (DOE) under
Grant. NO. DE-FG02-87ER40328, and by the Graduate School at the
University of Minnesota under the Doctoral Dissertation Fellowship.

}
\dedication{To my parents, who have provided me with unconditional support over the years.
}


\beforepreface 

\figurespage

\afterpreface            


\chapter{Introduction}
\label{intro_chapter}
Each of the fundamental forces in nature is currently understood in
the context of a particular theory.  Gravity is understood in terms of
Einstein's theory of general relativity (the quantum theory of gravity
is currently unknown, and highly sought after); while the
electromagnetic and weak forces are both understood in the context of
the standard model of electroweak interactions.  The subject of this
research is the theory of the strong nuclear force, Quantum
Chromodynamics (QCD).

QCD is the theory of the interaction that binds protons and neutrons
together to form atomic nuclei.  Quarks (fermions), and gluons (gauge
bosons of the underlying SU(3) gauge group) are the fundamental
constituents of the theory. It is believed that if one heats normal
matter to high enough temperatures, or compresses it to high enough
densities, a new state of matter is formed wherein the boundaries
between particular nucleons are no longer well defined.  Instead of a
sea of nucleons, one finds a ``soup'' of quarks and gluons. We will
refer to this state of matter as \emph{quark-gluon plasma} (QGP) It is
believed to have dominated the universe in the first few microseconds
after the Big Bang.
	
Experimentalists are attempting to re-create this environment
using particle accelerators which smash nuclei into each other
at very high energies.  In the last few years, there is
evidence that the Relativistic Heavy Ion Collider (RHIC) at
Brookhaven National Laboratory has created such a
state of matter \cite{Adams:2005dq,Adcox:2004mh,Arsene:2004fa,Back:2004je}.
	
The plasma created at RHIC exhibits some surprising properties.  In
particular, observation of strong collective behavior (elliptic flow),
and the large energy loss of high energy particles traversing the
medium (jet quenching) indicate that the plasma interacts very
strongly with itself, and is thus referred to as \emph{strongly
coupled}.  This presents a problem for theoretical physicists, because
the equations of QCD cannot be solved analytically in this regime;
traditional approximation techniques involve a perturbative expansion
in the coupling constant which measures the strength of the
interaction.  If the coupling constant is large
(i.e. $\mathcal{O}(1)$), such a perturbative expansion is unreliable,
as each successive term in the series becomes larger than the previous
one, and the approximation scheme breaks down.  This difficulty with
QCD has plagued theorists for over 30 years since the inception of the
theory.
	
However, in the late 1990's, new techniques were developed which
attempt to address these non-perturbative problems
\cite{Maldacena:1997re,Witten:1998qj,Gubser:1998bc}.  In particular,
it has been conjectured that there is a duality between certain
strongly coupled gauge theories and weakly coupled string theories.
By this, it is meant that both theories may describe the same physics,
but calculations may be easier in one theory than the other.
As mentioned above, it has been argued that the duality is a weak/strong
type, such that when one theory is strongly coupled, the other is
weakly coupled and vice-versa.  One can immediately see the possible
application to QCD; if a dual theory to QCD is found, one can do
computations in the dual theory (at weak coupling, where calculations
are possible), instead of in QCD itself (at strong coupling where
calculations are difficult).

Much work needs to be done before we can apply these ideas to
real-world QCD.  At the present time, the duality (referred to in
the title as the \emph{gauge/gravity} correspondence) is well developed for
a theory which shares some properties with QCD, but lacks some of the
essential features that we observe in the real world.  Attempts are
currently being made to both search for a dual theory of QCD, and to
modify the idealized theory to make it more physically relevant.

The ultimate goal of this line of research is the calculation of
observable properties of the quark-gluon plasma.  This is still a
difficult task, because the string dual to QCD is not currently known.
In light of this fact, there are two avenues which one may pursue.
First, one can attempt to compute observables in phenomenologically
motivated extra-dimensional models of QCD.  The drawback here is that at the
present time, such models are lacking from a theoretical point of view.
Secondly, one can attempt to look for universal features of strongly
coupled theories, which might then also be applicable to QCD.  

Here we employ the latter approach; we search for universal behavior
in hydrodynamic transport coefficients which describe a
fluid's response to small perturbations.  Hydrodynamics is an
effective theory of fluids in thermal equilibrium.  It is applicable
when macroscopic length and time scales of interest are much longer
than any microscopic scale.  In this regime, the effective
stress-energy tensor can be constructed as a derivative expansion in
the fluid velocity.  Naturally, such an effective construction
introduces unknown coefficients.  These are the transport coefficients
mentioned above, and which will be discussed in much greater detail in
the coming chapters.

These coefficients are necessary input for hydrodynamical simulations
of the quark-gluon plasma.  Currently, we understand that a heavy ion
collision occurs in roughly three distinct phases.  Before the
collision, the two colliding nuclei are highly Lorentz contracted, but
have not yet entered into thermal equilibrium; in this phase, the
nuclei can be described within the framework of, for example, the
\emph{Color Glass Condensate} \cite{McLerran:2008uj}.  After the
collision, some of the energy of the colliding nuclei is transformed
into a bath of new particles, which rapidly thermalizes.  Once the new
system has reached thermal equilibrium, a hydrodynamic description is
appropriate.  The newly created region of quark-gluon plasma then
expands and cools; once it has cooled enough, it drops out of thermal
equilibrium and the matter hadronizes into particles which are then
detected.  Describing the initial (pre-equilibrium) and final (freeze
out) stages of the heavy ion collision is a difficult and very active
area of research, it is also beyond the scope of this thesis - for
reviews, see \cite{Weigert:2005us, Shuryak:2008eq,
Banerjee:2008nt}. Luckily, it appears that there is a stage of the
heavy ion collision when the matter is thermally equilibrated, and
thus admits a hydrodynamic description.  It is in this phase of the
collision that one needs to specify the relevant transport
coefficients.  It is thus desirable to calculate these coefficients,
but as mentioned above, the strong coupling of the plasma renders
traditional perturbative calculations unreliable.  Thus, one is led to
try using gauge/gravity duality methods to compute such quantities.

The thesis is organized as follows

\begin{itemize}

\item In Chapter \ref{GGDReview_chapter}, we give a brief review of the
motivation, and some of the basic tools of gauge/gravity duality.
While hydrodynamics and transport coefficients are the main focus of
this work, in this section we also discuss some other applications
of gauge/gravity duality to strongly coupled plasma.  Particular
attention is paid to thermodynamic applications, some of which are
useful in later chapters.

\item In Chapter \ref{TCReview_chapter}, we explain the relevant
background material regarding hydrodynamics.  We also discuss some of the
standard methods of computing transport coefficients from
gauge/gravity duality, and explain in detail the method which will be used
in deriving the central results of the thesis.  

\item In Chapters \ref{ShearMode_chapter} and \ref{SoundMode_chapter},
the techniques described above are applied to a particular class of
dual theories.  These sections contain the central results of this
thesis.  We derive equations which are applicable to a wide variety of
gravitational dual theories, and could in principle be used within the
context of phenomenological models of QCD.  We also apply these
equations to a few analytically solvable special cases, and examine
the results to determine if any universal features of strongly coupled
plasmas are manifest.  The methodology employed in each of these
chapters is the same, but each one concerns a different type of
hydrodynamic perturbation.  In Chapter \ref{ShearMode_chapter} we
study shear perturbations while sound (or compressional) perturbations
are studied in Chapter \ref{SoundMode_chapter}.

\item Finally, in Chapter \ref{conclusion_chapter}, we discuss the main
conclusions of the analysis and also mention open questions and
prospects for future investigation.

\end{itemize}


\chapter{Gauge/Gravity Duality: An Overview}
\label{GGDReview_chapter}

In this chapter, we first explain the basics of gauge/gravity duality,
and the AdS/CFT (Anti-de Sitter/Conformal Field Theory)
correspondence.  The focus of this thesis is phenomenology, and
particular applications of the correspondence, so many of the details
regarding the foundations of the duality will not be presented here.
Relevant references are provided for the interested reader.  Some
excellent reviews of the subject which parallel the present discussion
can be found in
\cite{Aharony:1999ti,Gubser:2009md,Myers:2008fv,Edelstein:2009iv,
Klebanov:2005mh}.

Once the basics of the duality have been established, we present a
review of the relevant literature regarding applications of the
duality (mass spectra, form factors, jet quenching, etc.).  The main
application discussed in this thesis (hydrodynamic dispersion
relations and transport coefficients) will be explored in more detail
in the next chapter.

Finally, we discuss in some detail applications of the duality to
thermodynamic aspects of the relevant gauge theory.  Many of these
results will be useful in subsequent chapters.
\section{What is Gauge/Gravity Duality?}
The original motivation for gauge/gravity duality came from
considerations of type IIB string theory, which exists in ten
dimensions.  It is not necessary to understand the details of string
theory for this derivation, but one should appreciate that it is a
theory which contains strings, but also extended membrane like
objects, \emph{branes}.  D-branes are such objects which can serve as
endpoints for the strings (`D' stands for Dirichlet).  The theory has
two types of excitations, closed strings which have no endpoints, and
open strings which are allowed to have endpoints on a D-brane.
Strings are characterized by their length $l_s$, and the coupling
constant which controls the strength of their interaction is denoted
 $g_s$.  The string coupling can be related to the $D$ dimensional
Newton's gravitational constant $G_D$ by 
\be 
	G_D \propto g_s^2 l_s^{D-2}. 
\ee

To proceed, let us consider type IIB string theory in the background
of a system of $N$ D3-branes stacked on top of one another.  (A
D3-brane is a D-brane which extends in three spatial dimensions).  We
will now consider the low energy dynamics of this system from two
different viewpoints.

\begin{figure}[h]
\centering
\includegraphics[width=1.0\textwidth]{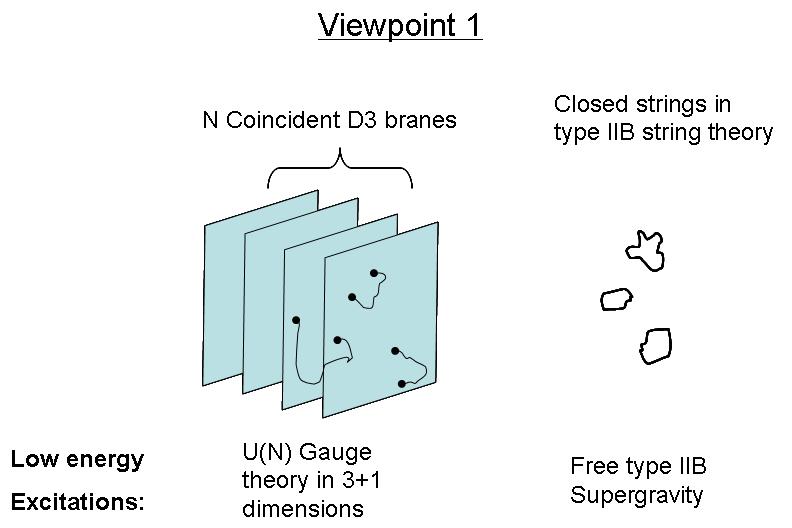}
\caption{Type IIB string theory in the background of a set of N coincident D3 branes.  Low energy excitations of this system
decouple into the two pieces shown above; the low energy excitations of the
brane system are those of a gauge theory. }
\label{viewpoint1_figure}
\end{figure}

First, the low energy excitations of open strings on the system of $N$
D3-branes can be described by a $SU(N)$ gauge theory.  Specifically, the
relevant low energy effective theory is $\mathcal{N}=4$ supersymmetric
Yang-Mills (hereafter SYM) theory.\footnote{One should take care to
distinguish $\mathcal{N}$ (the number of supercharges in the
supersymmetric gauge theory) from $N$ (the number of D-branes and the number 
of colors in the gauge theory).}The
gauge coupling $g_{YM}$ is related to the string coupling $g_s$ as
\cite{Witten:1995im},
\begin{equation}
	g^2_{YM} = 4 \pi g_s.  
\label{couplingrelation}
\end{equation}
If we consider the string length to be small, then we can neglect any
interactions between the branes and the closed strings which exist
outside of the branes (in the region called the \emph{bulk}).  We can
also neglect any interactions among the closed strings themselves.
This is because all interaction terms will contain positive powers
of the gravitational coupling $G_D$.  If we are interested in
processes which have typical energy $E$ much smaller than the Planck
mass $M_p$, these interaction terms will be suppressed by powers of
$E/M_p$.  Thus, from this point of view, the low energy limit of this
entire system consists of two decoupled pieces: free low energy closed
string theory (type IIB supergravity) in the bulk, and $\mathcal{N} =
4$ SYM theory on the branes.  This point of view is presented in
graphical form in Fig. \ref{viewpoint1_figure}.

Now, consider this system from an alternative point of view.  If $N$
is very large, the stack of branes will have a considerable amount of
energy, and thus will curve space-time in accordance with general
relativity.  In \cite{Horowitz:1991cd}, it was shown
that one can find black hole type solutions to these supergravity
equations.
The classical metric which solves the supergravity equations can be written
\begin{equation}
	ds^2 = \frac{1}{\sqrt{1+L^4/r^4}} \left(-dt^2 + d\vec{x}^2\right) 
	+ \sqrt{1+\frac{L^4}{r^4}}\left(dr^2 + r^2 d\Omega_5^2 \right),  
\end{equation}
where $L$ is the only scale, and is the so called \emph{curvature
radius}.  The vector $\vec{x}$ runs over the three spatial
coordinates, and $\Omega_5$ is the five dimensional angular element.
In addition to this metric, there is a scalar field (dilaton), and a
five-form, but these additional fields are unimportant for the
argument here.  From this point of view, the system now has two types
of low energy excitations: closed string excitations which are far
away from the brane which is located at $r=0$, or any sort of
excitation near the location $r=0$.  From the point of view of an
observer at $r=\infty$, these excitations appear to be low energy due
to the gravitational redshift from the metric.  In the near horizon
regime $r<<L$, the metric becomes
\begin{equation}
	ds^2 = \frac{r^2}{L^2} \left(-dt^2 + d\vec{x}^2\right) 
	+ \frac{L^2}{r^2} dr^2 + L^2 d\Omega_5^2, 
\label{10Dconformalmetric}
\end{equation}
which is the product of five dimensional anti-de Sitter space, with a
5-sphere ($AdS_5 \times S_5$).  (Anti-de Sitter space is a maximally
symmetric metric which is a solution to Einstein's equations with a
negative cosmological constant).

\begin{figure}[h]
\centering
\includegraphics[width=1.0\textwidth]{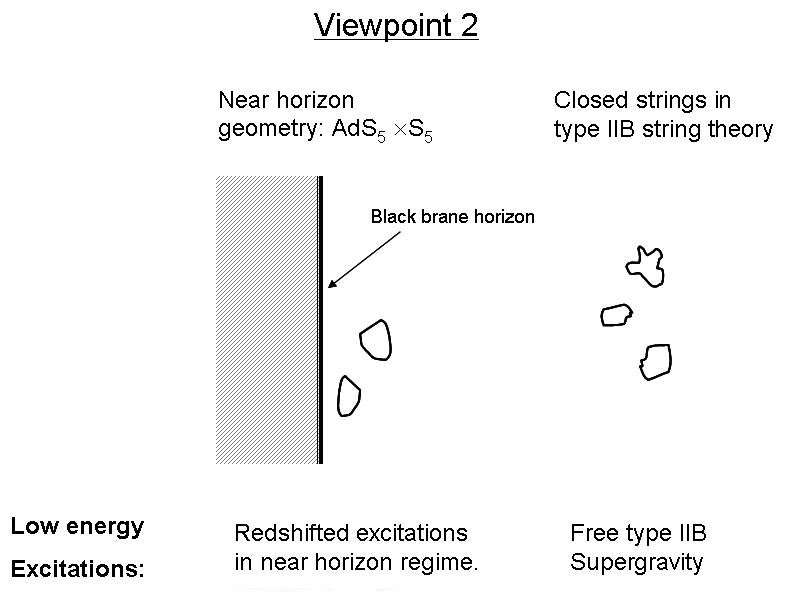}
\caption{The same system as in Fig. \ref{viewpoint1_figure}, viewed
in a different way.  If the stack of branes is large, they curve
space-time and can be described by a classical metric.  The gravitational
redshift causes excitations near the horizon to appear very low in energy.}
\label{viewpoint2_figure}
\end{figure}

Furthermore, both of the types of excitations listed above decouple if
the energy is low enough, because the wavelength of the closed string
excitations will be larger than the gravitational size of the branes
($\sim L$).  This point of view is presented graphically in
Fig. \ref{viewpoint2_figure}.

Now let us compare the two viewpoints of this system.  In each
description there are two decoupled pieces.  In the first system, the
decoupled pieces are the $\mathcal{N}=4$ SYM gauge theory, and low
energy closed string excitations.  In the second description, the
decoupled pieces are the excitations in the near horizon regime of the
black hole, and low energy closed string excitations far away from the
horizon.  Maldacena's conjecture is that these two systems should
describe the same physics \cite{Maldacena:1997re}.  Each of the two
descriptions has low energy closed string excitations as one element,
thus we should equate the remaining two elements.  The result is the
conjecture that \textbf{$\mathcal{N}=4$ SYM theory is dual to type IIB
string theory on $\mathbf{AdS_5 \times S_5}$}.

To see why this is useful, consider describing the type IIB string
theory by a classical supergravity metric (\ref{10Dconformalmetric})
and associated supergravity fields.  In other words, we neglect all stringy
effects and treat the theory classically.  When is such an approximation valid?
Clearly, one must have the string length much less than the curvature radius:
\begin{equation}
	l_s \ll L.
\label{validregime}
\end{equation}
We can relate this to the parameters of the gauge theory by examining
the total energy/volume of the stack of branes within these two
different viewpoints.

On one hand, the total energy/volume of the brane system is just $N$
times the tension of one brane.  The branes extend in three spatial
dimensions, thus by dimensional analysis,
\begin{equation}
	M/V \sim N/l_s^4.
\label{branemass}
\end{equation}
Clearly, the tension must scale linearly with $N$, and the string
length is the only relevant scale available to get the dimensions
correct.

On the supergravity side, one can compute
the total mass of the D-brane system by using the ADM mass formula
\begin{equation}
	M \sim \int \, d^D x\, T^{00},
\end{equation}
where $D$ is the number of dimensions, and $T^{00}$ is a particular
component of an energy-momentum tensor.  (We are not being rigorous here, 
we are doing dimensional analysis only).  Now, by Einstein's equations, 
we have 
\begin{equation}
	T_{\mu \nu} \sim G_{\mu \nu} / G_{D} 
\end{equation}
where $G_{\mu \nu}$ is the Einstein tensor. The components of the
Einstein tensor in question will in general be functions of the
space-time coordinates, but the ADM mass formula integrates over these
coordinates.  The only scale that can enter is the curvature scale
$L$.  Recalling that in $D$ dimensions, Newton's constant has mass
dimension $D-2$, by dimensional analysis
\begin{eqnarray}
	M &\sim& L^{D-3} / G_{D},  \\  
	M/V &\sim& L^{D-6}/G_D.  
\end{eqnarray}
Furthermore, Newton's constant can be related to the string length,
since the supergravity description is only a low energy limit of
string theory.  By dimension,
\begin{equation}
	G_D \sim l_s^{D-2},
\end{equation}
finally, 
\begin{equation}
	M/V \sim L^{D-6} / l_s^{D-2}.
\end{equation}
Specifying to $D=10$, and equating this result with (\ref{branemass}),
we have
\begin{equation}
	N \sim (L/l_s)^4.  
\end{equation}

The above analysis was done on dimensional grounds only, which makes
it impossible to track dimensionless quantities like the string
coupling $g_s$.  After a more careful analysis
\cite{Polchinski:1998,Lu:1993vt,Peet:2000hn}, one finds that the brane
tension goes as $g_s^{-1}$, and that the gravitational constant $G_D$
goes as $g_s^2$ as mentioned previously.  Including these factors, we have the useful
relation
\begin{equation}
	g_s N \sim g^2_{YM} N \sim L^4/l_s^4.
\end{equation}
Here we have used (\ref{couplingrelation}) in the first step.  Going
back now to (\ref{validregime}), one finds that the supergravity
description is valid for $l_s \ll L$ which in turn implies
\begin{equation}
	g^2_{YM} N \gg 1.
\end{equation}
The quantity $g^2_{YM} N$ is referred to as the t'Hooft coupling.  We
have just shown that our classical approximation is valid when dual
gauge theory is strongly coupled!

 We also need to suppress quantum gravitational corrections
in order to use the classical metric.  To do so, we require that
\begin{equation}
	L \gg l_p
\end{equation}
where $l_p$ is the Planck length.  It is related to Newton's constant 
as
\begin{equation}
	G_D \sim l_p^{D-2}.  
\end{equation}
Thus we have the relation
\begin{equation}
	L^4/l_p^4 \sim L^4/g_s l_s^4 \sim N. 
\end{equation}
Hence, we can neglect quantum corrections provided that
\begin{equation}
	N \gg 1.  
\end{equation}
This is referred to the t'Hooft limit.  It is the limit where the rank
of the gauge group is large $N \gg 1$, and the t'Hooft coupling is
fixed and large $g^2_{YM} N \gg 1$.  This is the limit where the
supergravity approximation is valid.

To summarize, Maldacena's conjecture is that a certain kind of string
theory on a curved $AdS$ background is equivalent to  
$\mathcal{N} =4$ SYM theory.  If we neglect stringy effects, we can
describe the string theory by a classical metric (and other classical
fields).  It turns out that the regime when this works is the regime
when the gauge theory is strongly coupled.  Hence, we see the
usefulness of this approach; the strong coupling regime of the gauge
theory is exactly the regime where it is difficult to use traditional
perturbative techniques.  However, this is precisely the regime in the
dual string theory where calculations are easy, because the theory can
be described classically!

We will discuss how one can use the dual gravity theory to do
calculations in future sections.  In the next section, we will discuss
how such a duality may be relevant to QCD.
\section{Quantum Chromodynamics and AdS/CFT}
A phenomenologist would like to apply these techniques to QCD, a real
theory of nature, as opposed to the supersymmetric Yang-Mills theory
mentioned in the previous chapter.  Are these theories so different?  

For one thing, $\mathcal{N}=4$ SYM theory is a conformal field theory.
Such a theory has no intrinsic energy scale (unlike QCD,
where the running of the coupling introduces the scale
$\Lambda_{QCD}$).  Thus the SYM theory has a gauge coupling
which does not run, and the theory is not confining.  It is also supersymmetric,
and has $N \gg 1$ whereas QCD has $N=3$.  Furthermore, QCD contains
quarks whereas the only fermions in the SYM theory are the
super-partners of the gauge bosons.  These fermionic partners transform
in the same way as the gauge bosons, but this is not how quarks
transform.  (In more pedestrian language, the gauge bosons have two
``color indices'', as opposed to quarks which only have one).

At first sight, these two theories are thus very different, and one
may be skeptical about the application of this duality to QCD.  In
general, there are three philosophies regarding the application of
gauge/gravity duality to QCD.

\subsection{Top Down Approach}
The first approach, usually referred to as the \emph{top down} approach,
focuses on attempting to discover other dualities where the gauge
theory more closely resembles QCD.  Here, one must modify the existing
string theory in some way.  For example, one can address the lack of
quarks by introducing `flavor branes' in the bulk
\cite{Aharony:1998xz,Karch:2002sh}.  One must break conformal symmetry
in order to introduce confinement; some examples of string duals which
do so are those of Witten \cite{Witten:1998zw}, Polchinski and
Strassler \cite{Polchinski:2000uf}, Maldacena and Nunez
\cite{Maldacena:2000yy}, and Klebanov and Strassler
\cite{Klebanov:2000hb}.  The addition of flavor into Witten's model by
Kruczenski \emph{et al}. \cite{Kruczenski:2003uq} and by Sakai and Sugimoto
\cite{Sakai:2004cn} allows for chiral symmetry breaking.  These latter
constructions share many features with QCD and are sometimes called
\emph{Holographic QCD}.

\subsection{Bottom Up Approach}
The second approach (sometimes called the \emph{bottom up} approach, or
\emph{AdS/QCD}) examines the duality from a phenomenological perspective.
The idea is to modify by hand the existing classical $AdS_5$ background in order
to introduce some essential features of QCD.  The resulting modified
backgrounds may or may not be embeddable into string theory.  The
benefits of such an approach are twofold.  First, one may view the
resulting background as a \emph{model} of QCD, which may be
systematically improved and can be used to make predictions, even
though this model was not derived from a fundamental theory.  Second,
by examining the properties of the backgrounds which lead to QCD like
behavior, one might gain information that may be useful in constructing top down
string duals.  Most often, these phenomenological
modifications are \emph{ad hoc} constructions introduced to break the
conformal symmetry and allow the introduction of the QCD scale.  This
was first investigated by Polchinski and Strassler
\cite{Polchinski:2001tt}, who truncated the $AdS_5$ metric at
some finite radius $r_{\rm min}$.  This approach is called the \emph{hard-wall model}, and was further investigated by Erlich \emph{et al}. in
\cite{Erlich:2005qh}.  Rather than chop off a piece of the $AdS$
space, Karch \emph{et al}. \cite{Karch:2006pv} showed that the introduction
of a non-trivial dilaton field $\phi(r) \propto 1/r^2$ also breaks the
conformal symmetry.  This latter approach is called the \emph{soft wall
model}, and is an improvement because the resulting meson spectrum
displays the desired phenomenological behavior.  One can also examine
warped $AdS$ metrics, which often have phenomenologically interesting
thermodynamics.  Some notable examples are the works of Gubser
\cite{Gubser:2008ny,Gubser:2008yx}, Andreev \cite{Andreev:2007zv} and
Kajantie \cite{Kajantie:2006hv}.  Finally, the work of Gursoy \emph{et al}.
\cite{Gursoy:2007cb,Gursoy:2007er,Gursoy:2008bu,Gursoy:2008za,
Gursoy:2009kk} attempts to incorporate many of the features of QCD
(e.g. running coupling) and are probably the most sophisticated
examples of models built using the bottom up approach.

\subsection{Universality Approach}
Finally, one may hope that many or all strongly coupled theories share some common
features.  If in the course of working with other strongly coupled theories
one discovers universal behavior which is model
independent, one could then conjecture that the behavior would also 
be applicable to strongly coupled QCD. 

This approach has been especially fruitful in the case of the shear
viscosity, which is a transport coefficient necessary for the
hydrodynamic description of the quark-gluon plasma (more details
will be given in Chapter \ref{TCReview_chapter}).  Indeed, Kovtun, Son
and Starinets showed that for a wide variety of gravity dual theories,
the ratio of the shear viscosity $\eta$ to the entropy density $s$
has the universal value $1/4\pi$.  It was later shown that this
relation holds for all theories of Einstein gravity
\cite{Iqbal:2008by, Buchel:2003tz} (assuming the dual gauge theory is
infinitely strongly coupled).  It is quite remarkable that this result
holds for both conformal and non-conformal theories, and is
independent of the number of dimensions.

One outgrowth of this universal behavior is the famous
\emph{KSS bound conjecture} \cite{Kovtun:2003wp} which states that for
all physical substances,
\begin{equation}
  \eta/s \geq 1/4\pi.  
\end{equation}
This has observable consequences for the plasma created at RHIC, even
though the gravitational dual to QCD is not
currently known.  Hydrodynamic simulations of the quark-gluon plasma
which attempt to fit the data from RHIC indicate that the QGP nearly
saturates this bound
\cite{Romatschke:2007mq,Song:2007fn,Romatschke:2007eb,Song:2007ux}.
It is thus desirable to find other such universal behavior from
gauge/gravity duality in hopes that there may be implications for heavy
ion collisions at RHIC and at the Large Hadron Collider (LHC).

The central results of this thesis employ this universality approach.
In Chapters \ref{ShearMode_chapter} and \ref{SoundMode_chapter}, we
present the results for second order hydrodynamics for a wide variety
of gravity duals.  While these theories are not especially similar to
QCD, one can examine the results to see if any other universal
behavior is manifest.  Some of the results presented in the
aforementioned chapters could be applied to certain phenomenological
bottom up models.
\section{Applications of the duality}
The idea that a strongly coupled gauge theory has a dual description
in terms of an extra-dimensional gravitational theory has many
practical applications.  We will now review some of these
applications.  The application to hydrodynamic descriptions of the
quark gluon plasma is the central focus of this thesis, and thus it
will be presented in more detail in the next chapter.  We will not
dwell unnecessarily on the details of the other applications
considered in the remainder of this chapter.  Instead, a basic
conceptual understanding of the relevant approaches is given, along
with relevant references.

\subsection{Thermodynamics}
One can examine a strongly coupled gauge theory at finite temperature using
gauge/gravity duality.  In order to do so, one has to introduce a
temperature into the gravitational background.  The obvious way to do
this is to include a black brane (an extra-dimensional generalization of
a black hole) into the background.  Black holes have well defined
thermodynamic properties, such as entropy and temperature; these
thermodynamic properties are dual to the thermodynamics of the gauge
theory.  

One gravitational background which is often used is the usual $AdS_5 \times S_5$ with
the addition of a black brane horizon.  The metric (in the near horizon limit $r \ll L$) takes the form
\begin{eqnarray}
	ds^2 &=& \frac{r^2}{L^2} \left(-f(r) dt^2 + d\vec{x}^2\right) 
	+ \frac{L^2}{f(r) r^2} dr^2 + L^2 d\Omega_5^2, 
\label{10DconformalmetricfiniteT} \\
	f(r) &=& 1-\frac{r_0^4}{r^4}.
\end{eqnarray}
Notice that this reduces to (\ref{10Dconformalmetric}) if $f(r) = 1$.
The position $r_0$ is the horizon; the time component of the metric
vanishes at the horizon, while the $r$ component diverges.  We will
often refer to this metric as the \emph{Schwarzschild $AdS$ black hole}, 
or simply as the \emph{$SAdS$} metric.    

Rather than restrict ourselves to this metric, we will begin to use
a more general black brane metric as originally considered by \cite{Kovtun:2003wp}, 
\begin{equation}
  ds^2 = g_{tt}(r) dt^2 + g_{xx}(r) dx_i dx^i + g_{rr}(r) dr^2. 
  \label{blackbranemetric}
\end{equation}
Here, $r$ is the coordinate denoting the extra dimension, and we
assume $r \, \epsilon \, (r_0, \infty)$.  The index $i$ runs over all
spatial coordinates; let us denote the number of spatial coordinates
by $p$.  So that $i = 1,2...p$.  The metric components are assumed
only to be functions of $r$.  There is $p$ dimensional
isometry along the spatial directions.

The position of a horizon at $r=r_0$ is assumed.  We define this
position as the place where $g_{tt}$ vanishes.  If there is more than
one such position, we take $r_0$ to be the maximum of these.  Furthermore, 
we will need to specify the behavior of the metric components near the horizon:  
\begin{eqnarray}
	g_{tt}(r \to r_0) &\approx& -\gamma_0 (r-r_0) + \mathcal{O}(r-r_0)^2,
	\label{ttNH}\\
	g_{rr}(r \to r_0) &\approx& \frac{\gamma_r}{r-r_0} + \mathcal{O}(1),
	\label{rrNH}\\
	g_{xx}(r \to r_0) &\approx& g_{xx}(r_0) + \mathcal{O}(r-r_0)
	\label{xxNH}.
\end{eqnarray}
The quantities $\gamma_0, \gamma_r$ and $g_{xx}(r_0)$ are independent of $r$, but
can in principle depend on $r_0$.  

The subject of black hole thermodynamics was pioneered by Bekenstein
and Hawking \cite{Bekenstein:1973ur,Bekenstein:1974ax,Hawking:1974sw}.  The Bekenstein-Hawking
entropy of a black hole for any theory of Einstein gravity is given by
\begin{equation}
  S = \frac{A_h}{4 G_d},
\end{equation}
where $A_h$ is the area of the horizon.  This quantity can be calculated from
\begin{equation}
  A_h \equiv \int \sqrt{-\gamma}\, dV
\end{equation}
where $\gamma_{ij}$ is the induced metric on the horizon, and $dV$ is the induced
volume element.  These can be straightforwardly found from our metric by setting
$r=r_0$ and $dr=0$.  Then, for our metric, 
\begin{equation}
  \frac{S}{V} \equiv s  = \frac{g_{xx}(r_0)^{p/2}}{4 G_D}.
\label{bhentropy}
\end{equation}

There are many ways to derive the Hawking temperature from the metric.
The approach listed below is based on the presentation of Zee
\cite{Zee:2003}.  Let us examine our metric in the near horizon
regime.  We will only need the $t$ and $r$ components for this
argument:
\begin{equation}
  ds^2_{NH} = -\gamma_0(r-r_0) dt^2 + \frac{\gamma_r}{r-r_0} dr^2 + ...
\end{equation}
Now, define a coordinate $\rho$ so that
\begin{eqnarray}
  \rho &=& 2\sqrt{\gamma_r (r-r_0)},\\
  d\rho^2 &=& \frac{\gamma_r dr^2}{(r-r_0)}.  
\end{eqnarray}
Then the metric is
\begin{equation}
  ds^2_{NH} = -\frac{\gamma_0}{4\gamma_r} \rho^2 dt^2 + d\rho^2 + ...
\end{equation}
We now switch to an imaginary time variable $\tau \equiv i t$:  
\begin{equation}
    ds^2_{NH} = \frac{\gamma_0}{4\gamma_r} \rho^2 d\tau^2 + d\rho^2 + ...
\end{equation}
In these coordinates, this piece appears to be written in polar
coordinates, where $\rho$ is the radial variable, and $\tau$ is an
angular variable.  Examine the arc length $\alpha$ of a circle at
constant $\rho = \rho_0$, and denote the period of the angular variable by
$\beta$.  The arc length is
\begin{equation}
  \alpha^2 = \frac{\gamma_0}{4\gamma_r} \rho_0^2 \beta^2
\end{equation}
The distance $\alpha$ should be equal to the circumference of the
circle, unless there is a conical singularity at the origin.  We
desire regularity at the origin, so
\begin{eqnarray}
  (2 \pi \rho_0)^2 &=& \frac{\gamma_0}{4\gamma_r} \rho_0^2 \beta^2,\\
  \frac{1}{\beta} &=& \frac{1}{4 \pi} \sqrt{\frac{\gamma_0}{\gamma_r}}.
\end{eqnarray}
Finally, as is customary in the case of finite temperature field theory, 
one equates the period of Euclidean time with the inverse temperature   
\begin{equation}
  \beta^{-1} = T,
\end{equation}
which gives us an expression for the Hawking temperature
\begin{equation}
  T = \frac{1}{4 \pi} \sqrt{\frac{\gamma_0}{\gamma_r}}.
\label{hawkingtemp}
\end{equation}

Once we know the temperature and the entropy, it is easy to determine
other quantities using thermodynamic identities.  Throughout this
thesis, we focus only on systems with zero chemical potential.  In 
this case, the relations
\begin{eqnarray}
  s(T) &=& \frac{dP(T)}{dT},\\
  T s(T) &=& \epsilon(T) + P(T)
\end{eqnarray}
allow us to determine the pressure $P$ and the energy density $\epsilon$.  Another
quantity of interest is the speed of sound $v_s$ in a thermally equilibrated medium:
\begin{equation}
  v_s^2 \equiv \frac{\partial P}{\partial \epsilon}.
\end{equation}
Again, in the case of zero chemical potential, we have
\begin{equation}
  v_s^2 = \frac{dP/dT}{d\epsilon/dT} = \frac{s}{T ds/dT}.
\end{equation}
Generically, the temperature is a function of $r_0$, and thus
we can instead write
\begin{equation}
  v_s^2 = \frac{s(r_0)(dT(r_0)/dr_0)}{T(r_0)(ds(r_0)/dr_0)}.
\end{equation}
Substituting in the previous relations for the temperature 
and the entropy (\ref{bhentropy}), (\ref{hawkingtemp}), we have
\begin{equation}
  v_s^2 = \frac{1}{p} \frac{\frac{d}{dr_0} Log[\frac{\gamma_0(r_0)}{\gamma_r(r_0)}]}{\frac{d}{dr_0} Log[g_{xx}(r_0)]}.
\label{vsBHthermo}
\end{equation}
This is a simple way to determine the speed of sound from the metric.
We shall see in the next chapter that one can also compute the speed
of sound by determining the hydrodynamic dispersion relation.  This
latter approach, which is much more complicated than the formula
above, is of greater utility since it allows one to determine other
transport coefficients of interest as well.

These thermodynamic tools are indispensable when attempting to study a
strongly coupled gauge theory using gauge/gravity duality.  As
mentioned previously, these formulas for the entropy, temperature and
speed of sound are used to study the thermodynamic properties of
phenomenological models in \cite{Gubser:2008ny,Gubser:2008yx,
Andreev:2007zv, Kajantie:2006hv, Gursoy:2008bu, Gursoy:2008za}

In addition to the thermodynamic aspects above, methods exist for
examining phase transitions within the context of gauge/gravity
duality.  Specifically, the gravity dual of a phase transition is the
\emph{Hawking-Page transition} which occurs if more than one
gravitational background satisfies the equations of motion
derived from the action.  In this case, one must examine the
conditions under which each background is energetically favorable over
the other(s).  If at a critical temperature, one background is suddenly favored over the
existing background, a transition occurs.  This transition can be
equated to the phase transition in the dual field theory.  This
methodology has led to holographic predictions of the deconfinement
temperature \cite{Herzog:2006ra,BallonBayona:2007vp}.

\subsection{Hadronic physics}
Much work has been done in describing bound hadronic states within the
context of gauge/gravity duality.  In the top down approach, as
mentioned above, the introduction of flavor branes is necessary in
order in include fundamental matter.  Strings which begin and end on
the D7 branes can be thought of as a quark/anti-quark pair.  (Since
the strings have no endpoints on the $N$ D3 branes, they have no color
indices, and are thus color singlets).  Thus, excitations of this
system correspond to the different hadronic states.  There is a large
literature on mesons in AdS/CFT, and an excellent review has recently
been written \cite{Erdmenger:2007cm}.

One can also describe hadronic physics using the bottom up approach.
This is done by introducing appropriate fields into the existing
$AdS_5$ background.  The bulk fields are necessarily dual to operators
in the field theory.  The normalizable solutions to the equations of
motion for these fields correspond to the hadronic states.  For
example, consider the left-handed current operator in QCD,
$\bar{q}_{L}\gamma^{\mu} t^a q_L$, where $q_L$ is a quark operator,
and $\gamma^\mu$, and $t^a$ are the usual Dirac matrices and group
generators respectively.  A suitable five dimensional field dual to
this operator must be a gauge field $A^{\mu \,(a)}_L$ due to the
Lorentz structure of the operator.\footnote{The mass of the dual field
is set by the rules of the AdS/CFT correspondence.  Here, $m=0$ \cite{Witten:1998qj,Gubser:1998bc}.}Similarly, a right-handed gauge field which is dual to
the right-handed quark current operator is introduced.  These two
gauge fields can be combined into vector and axial fields $V^{\mu \,
(a)} \propto A^{\mu \, (a)}_L + A^{\mu \, (a)}_R$ and $A^{\mu \, (a)}
\propto A^{\mu \, (a)}_L - A^{\mu \, (a)}_R$.  One then constructs a
Lagrangian involving these fields, and solves the equations of motion
for the gauge field excitations.  The normalizable solutions for the
excitations of the vector field $V^{\mu \, (a)}$ are interpreted as
the vector mesons in QCD.  Similarly, the solutions for the axial
field $A^{\mu \,(a)}$ correspond to axial mesons.  Other techniques
exist which allow for the calculation of other observables such as
form factors and decay constants.

These bottom up models are \emph{ad hoc}.  The backgrounds are often
imposed by hand and not dynamically generated as the solution to any
equations of motion; any back reaction which the fields have on the metric
is usually neglected.  The backgrounds are also not embedded within
a high energy theory such as string theory.  Despite these
shortcomings, the phenomenological accuracy of such models is often
close to 10\%.  The interested reader is referred to the review
already mentioned \cite{Erdmenger:2007cm}, as well as the primary
works \cite{Erlich:2005qh, Karch:2006pv, Gursoy:2007er, DaRold:2005zs, DaRold:2005vr, 
deTeramond:2005su, Gherghetta:2009ac}.

The behavior of mesons at finite temperature has also been
investigated using gauge/gravity duality.  By examining different
configurations of the D7 flavor branes, one finds an interesting
phenomenon related to the disassociation of mesons in a plasma at
finite temperature.  The idea is that the flavor branes may or may not end
on the black hole.  To determine which setup is
energetically favorable, one needs to examine the free energy of each
configuration.  It has been shown that there exists a critical
temperature above which the flavor branes end on the horizon (\emph{black
hole embedding}), and below which the flavor branes do not end on the
horizon (\emph{Minkowski embedding}).  The spectrum of the mesons is very
different in each of these two cases.  In the low temperature phase,
the spectrum is discrete and the mesons are stable; in the high temperature phase the
spectrum is continuous and there is no mass gap.  Thus, one observes a
phase transition for the fundamental matter, which gives a holographic
prediction for the temperature at which mesons disassociate in the
plasma.  This phenomenon was noticed in
\cite{Mateos:2006nu,Hoyos:2006gb}, which built upon the earlier works
\cite{Babington:2003vm,Kirsch:2004km}.

\subsection{Jet quenching and energy loss}
A jet is a large collection of particles which arise from the
hadronization of a freely moving quark.  Usually two jets are produced from
a quark/anti-quark pair, and these jets are both
seen in the detector.  At RHIC, often only one jet is seen.  The
physical interpretation is that the missing jet lost energy while
traversing the strongly coupled medium in order to reach the detector
on the other side.  This phenomenon is not observed in proton-proton
collisions because a large region of quark gluon plasma is not
created.

We can learn a lot about a medium by examining how energetic particles
traverse it.  For this reason, it is desirable to examine how
energetic quarks lose energy when traversing a strongly coupled
plasma.  This subject also helps us examine the phenomenon of jet
quenching in heavy ion collisions at RHIC.  From the point of view of
gauge/gravity duality, this problem is usually approached in top
down models.  A quark, as mentioned previously, is the endpoint of a
string on one of the flavor D7 branes.  One can study a single quark
by having one endpoint on the flavor brane, and the other on the color
D3 brane (at the horizon).  Alternatively, one can study a
quark/anti-quark pair by having both ends of the string on the D7
brane.  In either of these cases, the string connecting the two
endpoints is described classically by the Nambu-Goto action.  One can
then examine the dynamics of this classical string in various
situations (e.g. freely moving in one spatial dimension, falling
toward the horizon, or under the influence of a constant force).  The
dynamics of the string and its corresponding energy tell us about the
dynamics of the quark(s) in the dual field theory.

The pioneering papers in this regard are those of
\cite{Herzog:2006gh,Gubser:2006bz}, as well as \cite{Liu:2006ug} which
uses another approach; this latter work allows the calculation of a
transport coefficient $\hat{q}$, called the \emph{jet-quenching
parameter} using Wilson loops.  These original papers have been
extended in numerous ways, and this is still an active area of
research.  We refer the reader to the previously mentioned review
\cite{Gubser:2009md} for a more complete set of references.

\subsection{Out of equilibrium and real time dynamics}
So far, we have only discussed the thermally equilibrated medium (QGP)
that is supposedly created in heavy-ion collision. Of course, this
grossly simplifies the problem, and in order to have a full
description of the collision, other effects must be taken into
account.  For example, we have considered an infinitely extended
thermal medium whereas in a heavy ion collision the region has finite
extent and expands after it is created.  The introduction of flow into
the gravity picture originated with \cite{Janik:2005zt,Janik:2006gp}.

Before and immediately after the collision, the system is not in
thermal equilibrium, and the method by which it quickly equilibrates
is not presently understood.  From the point of view of the gravity dual,
thermalization can be examined by perturbing a black hole background,
and examining how long it takes to return to equilibrium.  This
approach has led to estimates of the thermalization time for a heavy
ion collision.  Typical estimates of this time are in the range of 0.3
- 0.5 fm/c, (see for example \cite{Friess:2006kw}).  In addition to the
thermalization time, one may also be interested in the isotropization
time scale, and it can be accessed using similar methods
\cite{Kovchegov:2007pq, Chesler:2008hg}.

\section{Summary}
In this chapter, we have attempted to present the basics of the
foundations of gauge/gravity duality and AdS/CFT.  We have also
explained three main philosophical approaches (top down, bottom up and
universality) which attempt to adapt the existing AdS/CFT
correspondence to the real world theory of QCD.  Finally, we presented
some example applications (and relevant references) of the duality to QCD
and quark-gluon plasma phenomenology.

The literature on this subject is vast and grows every day.  This
chapter is not meant to be comprehensive, but can serve as a starting
point for the interested reader.  This chapter also serves to present
a context for the central results of this thesis.  We have not yet
discussed the application of gauge/gravity duality to hydrodynamics;
this topic will be the focus of the remaining chapters.  


\chapter{Hydrodynamics from Gauge/Gravity Duality}
\label{TCReview_chapter}
This chapter is an introduction to the methods used in the central
calculations of this thesis.  We first review the theory of viscous
hydrodynamics, and thus explain the origin of the transport
coefficients; one of the main goals of this thesis is to calculate
such coefficients.

Secondly, we explain the methodology for calculating hydrodynamic
dispersion relations within the context of gauge/gravity duality.
These dispersion relations give access to the transport coefficients.
The necessary linearized Einstein equations are presented in full
generality here.  These equations will be solved in specific cases in
the remaining chapters.

\section{Hydrodynamics}
In this section, we review the theory of hydrodynamics.  To this
end, we discuss how to construct the effective energy momentum tensor and the
relevant normal modes which result from perturbations of this tensor.
The discussion here follows excellent reviews
presented in
\cite{Landau1987,Weinberg:1972,Kapusta2006,Son:2007vk,Kovtun:2003vj}.

\subsection{Hydrodynamic energy momentum tensor}
Hydrodynamics is best thought of as an effective theory which
describes a thermal fluid when the length and time scales of interest
are much longer than any relevant microscopic scale.  In that case, one is able
to smooth over the microscopic physics and instead use a
perfect fluid description with viscous corrections.  To lowest order, the energy
momentum tensor takes the form of a perfect fluid
\begin{equation}
  T^{\mu \nu}_{PF} = -P \eta^{\mu \nu} + u^{\mu} u^{\nu} (\epsilon + P).  
\end{equation}
The subscript ``PF'' stands for ``perfect fluid''.
Here, $P$ is the pressure, $\epsilon$ is the energy density, and 
$u^{\mu}$ is the fluid 4 velocity 
\begin{equation}
  u^{\mu} = (\gamma c, \gamma \vec{v}),
\end{equation}
which takes the form $(1,0,0,0)$ in the fluid rest frame.  The flat
space Minkowski metric is denoted by $\eta_{\mu \nu}$; the fluids we
consider will always be in flat space, while the extra dimensional
gravity dual will have some non-trivial curvature.  In general our
conventions are those of Weinberg \cite{Weinberg:1972}.  Our metric
signature is $\eta_{\mu \nu} =$ diag(1,-1,-1,-1) in four dimensions,
with an obvious generalization in higher dimensions.  Greek indices
indicate both space and time coordinates, while Latin indices
$(i,j...)$ indicate spatial coordinates only.

One can proceed to add extra terms proportional to derivatives of
the fluid velocity.  In order to be in the hydrodynamic regime, the
length and time scales of interest must be much larger than the
microscopic scale, which is the inverse temperature $T^{-1}$.  Because
the fluid variations are assumed to happen on long length and time
scales, any these derivative terms
are small corrections to the perfect fluid case.  First
order hydrodynamics comes from adding terms which contain at most one
derivative.  Let us now write these corrections in the local fluid
rest frame where $u^{i} = 0$, though in general derivatives of these
quantities may not vanish.  We define such a rest frame as the one
where there is no flow of energy or momentum in the fluid; this is the
so called Landau formulation.  In such a frame, the fluid is at rest,
so $T^{00}$ is still the energy density, and $T^{0i}$ must vanish
because there is no flow of momentum energy and momentum.  Thus, the
derivative terms can only show up in the spatial components of the
energy momentum tensor:
\begin{equation}
  T^{ij} = T^{ij}_{PF} + \Delta T^{ij}.  
\end{equation}
The viscous correction $\Delta T^{ij}$ is symmetric in $i$ and $j$,
and must be composed out of derivatives of the fluid velocity.
Clearly, the two relevant terms which one can write are $\partial^i
u^j + \partial^j u^i$ and $\eta^{ij}\partial_k u^k$.  Thus, the most
general form of the first order derivative corrections is
\begin{equation}
  \Delta T^{ij} = \eta \left( \partial^i u^j + \partial^j u^i - \frac{2}{p} \eta^{ij} \partial_k u^k \right) + \zeta \eta^{ij} \partial_k u^k.  
\label{constitutiverelation}
\end{equation}
This equation is sometimes referred to as the \emph{constitutive
relation}.  Here, we have introduced the variable $p$ which denotes
the number of spatial dimensions.  We have written $T^{ij}$ as above
so that the first term which is proportional to $\eta$ vanishes under
the trace operation.  The coefficients $\eta$ (shear viscosity) and
$\zeta$ (bulk viscosity) are the low energy constants of the effective
theory, and are referred to as transport coefficients.  One should
take care not to confuse the shear viscosity with the Minkowski
metric.  

Physically, these coefficients describe a fluid's resistance to flow
under stress.  Shear viscosity is relevant for applied shear stress.
As an example of shear stress, consider the following setup: lower a
small, solid cylinder into a fluid and apply an external torque to the
cylinder so that it rotates with a constant angular velocity.
Naturally, the fluid very near to the cylinder will move with a
greater velocity than the fluid further away.  This velocity gradient
will vary with the shear viscosity of the fluid.  In fact, a similar
technique is used to measure the shear viscosity of laboratory fluids,
by measuring the amount of torque required to reach a certain angular
velocity.

Bulk viscosity is relevant for applied volume stress.
To apply such stress, one should take a fluid element and compress it
to a smaller volume (or expand it to a larger volume) without changing
the shape of the fluid element.  Bulk viscosity is thus only relevant
for fluids which are compressible.

Clearly, we are not in a position to directly measure these
coefficients for the QGP, since it exists for an exremely short time.
Instead, one must infer transport properties of the fluid from the
multitude of hadrons which reach the detector after a heavy ion
collision.  Theoretically, these transport coefficients can in
principle be calculated from the microscopic physics, but if the fluid
is strongly coupled, perturbative methods used to compute these
quantities fail.  Fortunately, gauge/gravity duality allows one to
gain information about these coefficients.  We will explain this how
this is done in Sec. \ref{prescription_section}.

If the fluid under consideration has a conserved charge, one can also
construct a similar relation for the current $j^{\mu}$ as an expansion
in derivatives.  Such an expansion will necessarily come with its own
transport coefficients (e.g. the thermal conductivity $\kappa$).
Throughout this thesis we avoid any such complications; all theories
under consideration here will have no chemical potential, and thus no
conserved charge.  In this case, the energy momentum tensor
(\ref{constitutiverelation}) is all that is necessary.

\subsection{Hydrodynamic modes}
\label{hydromodes_section}
Let us consider a fluid's response to hydrodynamic perturbations.  To
this end, we introduce fluctuations of the energy momentum tensor
$\delta T^{\mu \nu}$ and examine the equations of motion for such
perturbations.  We assume that all perturbations are of the plane wave
type, $e^{i(q z - w t)}$ where $q$ is the momentum and $w$ is the
energy of the perturbation.  Without loss of generality, we have
chosen our coordinate system so that $z \equiv x_p$ points in the
direction of the momentum.  By assumption, we require the energy and 
momentum of these perturbations to be small compared with the temperature $w,q \ll T$ (otherwise
we are no longer in the hydrodynamic regime).  

Expanding $T^{\mu \nu}$ to linear order in these perturbations gives
\begin{eqnarray}
  \delta T^{0 \mu} &=& - \delta P \eta^{0 \mu} + (\delta \epsilon + \delta P) u^0 u^\mu + (\epsilon + P) \left(\delta u^0 u^\mu + u^0 \delta u^\mu\right)\\
  &=& -\delta P \eta^{0 \mu} + (\delta \epsilon + \delta P) u^\mu + (\epsilon + P) \left(\delta u^0 u^\mu + \delta u^\mu \right) 
\label{deltaT0mu}
\end{eqnarray}
and
\begin{eqnarray}
  \delta T^{i j} = -v_s^2 \eta^{ij} \delta \epsilon + \eta \left( \partial^i \delta u^j + \partial^j \delta u^i - \frac{2}{p} \eta^{ij} \partial_k \delta u^k \right) + \zeta \eta^{ij} \partial_k \delta u^k.  
\label{deltaTij}
\end{eqnarray}
Here, we have introduced the speed of sound $v_s$ defined as
\begin{equation}
  v_s^2 \equiv \frac{\delta P}{\delta \epsilon}.  
\end{equation}

The equations of motion for such fluctuations can be found from the conservation of the energy momentum tensor
\begin{equation}
  \partial_\mu \left( \delta T^{\mu \nu} \right) = 0.
  \label{conservation}
\end{equation}
First, let us consider the case of $\nu = 1$, recalling the assumed plane wave style dependence of the perturbations.  We have
\begin{eqnarray}
  \partial_t \delta T^{0 1} + \partial_z \delta T^{1p} &=& 0 \\
  (-i w) \delta T^{0 1} + (i q)  \delta T^{1p} &=& 0.
\end{eqnarray}
Substituting results from (\ref{deltaT0mu}) and (\ref{deltaTij}),
\begin{eqnarray}
  \delta T^{01} &=& (\epsilon + P) \delta u^1  \\
  \delta T^{1p} &=& \eta \left( \partial^z \delta(u^1) \right) = -(i q) \eta \delta u^1,  
  \label{shearconservationequations}
\end{eqnarray}
we have
\begin{equation}
  (-i w) (\epsilon + P) \delta u^1 + q^2 \eta \delta u^1 = 0,
\end{equation}
which requires a dispersion relation of the form
\begin{equation}
  w(q)_{\rm shear} = - i \frac{\eta}{\epsilon + P}q^2 + \mathcal{O}(q^4). 
  \label{firstorderhydroshear}
\end{equation}
The corrections of order $q^4$ to the dispersion relation would
come from higher order dissipative corrections.  Since we have only written the
energy momentum tensor to first order in derivatives, our dispersion
relation is only valid to the leading order in $q$.  This is the
\emph{shear mode} perturbation; its dispersion relation depends on the
shear viscosity.  We derived it by assuming $\nu = 1$ in the
conservation equation, but we would get the same relation if we used
$\nu = 2, 3...p-1$.

Next, consider the conservation equation (\ref{conservation}) with
$\nu = 0$ and $\nu = p$.  We then get two coupled equations
\begin{eqnarray}
  -i w \delta T^{00} + iq \delta T^{0p} &=& 0,\\
  -i w \delta T^{0p} + iq \delta T^{pp} &=& 0.
  \label{soundconservationeqns}
\end{eqnarray}
One can now go back to the constitutive relations for $\delta T^{\mu \nu}$ (\ref{constitutiverelation})
and substitute them here.  Before doing, so, however, it is useful to
note that when introducing these perturbations, we desire preservation of
the normalization of the velocity vector.  This means
\begin{eqnarray}
  \delta \left( \eta_{\mu \nu} u^\mu u^\nu \right) &=& 0\\
  \eta_{\mu \mu} \left(2 u^\mu \delta u^\mu \right) &=& 0\,\,\, \rm{(no\, summation)}\\
  \delta u^0 &=& 0 \,\,\,\rm{(fluid \,  rest \, frame)}.  
\end{eqnarray}
In the last line, we have specified to the local rest from where only
$u^0$ is non-zero.  Using this fact along with (\ref{deltaT0mu}), (\ref{deltaTij}), we can now write
\begin{eqnarray}
  \delta T^{00} &=& \delta \epsilon, \\
  \delta T^{0p} &=& \left(\epsilon + P \right) \delta u^p, \\
  \delta T^{pp} &=& v_s^2 \delta \epsilon - i q \delta u^p \left[\zeta + 2\left(\frac{p-1}{p} \right)\eta \right].
\end{eqnarray}
Plugging these back into (\ref{soundconservationeqns}) gives
\begin{eqnarray}
  -w \delta \epsilon + q \left(\epsilon + P\right) \delta u^p &=&0,\\
  -w \left(\epsilon + P\right) \delta u^p + q \left\{ v_s^2 \delta \epsilon - i q \delta u^p \left[\zeta + 2\left(\frac{p-1}{p} \right)\eta \right]\right\} &=& 0,
\end{eqnarray}
which combine to give
\begin{equation}
  \delta u^p \left( \epsilon + P \right) \left\{ -w^2 + q^2 v_s^2 - \frac{i q^2 w}{\epsilon + P} \left[\zeta + 2\left(\frac{p-1}{p} \right)\eta \right] \right\} =0.
\end{equation}
After solving for $w(q)$ and expanding in powers of $q$ we find
\begin{equation}
  w(q)_{\rm sound} = \pm v_s q - \frac{i q^2}{2(\epsilon + P)} \left[\zeta + 2\left(\frac{p-1}{p} \right)\eta \right] + \mathcal{O}(q^3).
  \label{firstorderhydrosound}
\end{equation}
Our hydrodynamic dispersion relation is only accurate to first order
in the dissipative terms.  In order to go to higher order in $q$, we
would need to add terms with more than one derivative to the effective
energy momentum tensor.  

This latter dispersion relation is the \emph{sound mode} (or compressional
mode).  It corresponds to a perturbation moving with speed $v_s$ relative to the fluid; the
dissipation of the wave is controlled by a combination of the shear
and bulk viscosities.

\subsection{Second order (causal) hydrodynamics}
\label{secondorderhydro_section}

In the preceding section, we discussed first order hydrodynamics; in
this section we discuss the next hydrodynamic order.  This subject was
first approached by M\"{u}ller \cite{Muller:1967} and later Israel and
Stewart \cite{Israel:1976tn,Israel:1979wp}.  It is desirable to
extend the theory of hydrodynamics to the next order, because the
first order theory has problems with causality \cite{Hiscock:1983zz}.
Formally, the issues with causality have been shown to exist only for
modes which are outside the hydrodynamic regime
\cite{Geroch:1995bx,Geroch:2001xs,Kostadt:2000ty,Kostadt:2001rr}, but
from a practical standpoint, such issues are unacceptable when
attempting to do numerical simulations
\cite{Kostadt:2000ty,Kostadt:2001rr,Hiscock:1985zz}.  These are the
issues which the Israel-Stewart formulation attempts to resolve, and
for this reason the approach is sometimes called \emph{causal hydrodynamics}.

The examination of the next hydrodynamic order is also attractive from a
theoretical standpoint.  Gauge/gravity duality has led to the important
observation that in all Einstein gravity duals, the ratio
of the shear viscosity to entropy density $\eta / s$ takes on the
universal value $1/4\pi$ (this will be discussed in more detail in the
subsequent chapters).  Given this success, it is interesting to
inquire whether other such universal relations exist at the next
hydrodynamic order.

Israel introduced five new transport coefficients that appear
in the hydrodynamic expansion of the energy momentum tensor.  In what
follows, we use the same notations and conventions as
\cite{Natsuume:2007ty}.  Three of these five transport coefficients
are relaxation times associated with the diffusive, shear, and sound
mode, and are denoted by $(\tau_J, \tau_{\pi}, \tau_{\Pi})$
respectively.  There are two other transport coefficients which are
related to coupling between the different modes, $\alpha_0, \alpha_1$.

We will not write down the full expression for the second order energy
momentum tensor in Israel-Stewart theory, but refer the interested
reader to the original works mentioned above.  For our purposes, we
need the expression for the shear and sound mode dispersion relations
in terms of the transport coefficients.  These relations were worked
out by Natsuume \emph{et al}. \cite{Natsuume:2007ty} for the case of a background
without a conserved charge.  (As mentioned previously, this is the case which we focus on
here.  A more complete list of
assumptions regarding this dispersion relation can be found in
\cite{Natsuume:2007ty}).
The results for the sound mode are
\begin{eqnarray}
  w(q)_{\rm sound} &=& w_1 q + w_2 q^2 + w_3 q^3 + \mathcal{O}(q^4), \\
  w_3 &=&  \pm \frac{\eta }{2 v_s T s}\left\{\frac{p-1}{p}\left[2 v_s^2 \tau_{\pi} - \left(1-\frac{1}{p}\right)\frac{\eta }{T s}\right] \right.\\
   &+& \left. \frac{\zeta }{\eta }\left[v_s^2 \tau_{\Pi}- \left(1-\frac{1}{p}\right)\frac{\eta }{T s}-\frac{\zeta }{4 T s}\right]\right\} \nonumber
  \label{w3transport}
\end{eqnarray}
with $w_1$ and $w_2$ unchanged from
(\ref{firstorderhydrosound}).\footnote{Note that there are actually two
solutions here, corresponding to the plus/minus sign in the formulas
for $w_1$ and $w_3$.  These two solutions only differ by the relative
direction of the momentum $q$.  Throughout this thesis, we will always
assume that our coordinate system is chosen so that $v_s > 0$.}Furthermore,
\begin{equation}
  w(q)_{\rm shear} = - i \frac{\eta}{\epsilon + P} q^2 - i \left(\frac{\eta}{\epsilon + P} \right)^2 \left(\tau_\pi + \Delta \right)q^4 + \mathcal{O}(q^6).
  \label{sheardispersionsecondorder}
\end{equation}

A few words are in order about the parameter which we have denoted as $\Delta$ above.  
Suppose that we are after the next order correction to the shear mode dispersion 
relation.  Let us return to the conservation equation for the shear mode
(\ref{shearconservationequations})
\begin{equation}
   -w \delta T^{0 1} + q  \delta T^{1p} = 0.
\end{equation}
In the rest frame which we consider, $\delta T^{01} \sim \mathcal{O}(1)$, it does
not depend on $q$.  If we are working within the context of second order
hydrodynamics, $\delta T^{1p}$ can have terms proportional to one or two derivatives.  In general, 
it will have the form
\begin{equation}
  \delta T^{1p} = A q + B w + C w^2 + D w q + E q^2,
\end{equation}
where $A$, $B$, $C$... are functions of the fluid four velocity, but are independent of $q$. The conservation equation
becomes
\begin{equation}
    -w \delta T^{01} + A q^2 + B w q + C w^2 q + D w q^2 + E q^3 = 0.
\end{equation}
Expanding $w(q)$ in even powers of $q$, one finds that the term
which is relevant for determining the next correction to the
dispersion relation is the term containing $D$, since this is the only
piece which has a term proportional to $q^4$.  However, it is clear
that there is another way to produce a term which is proportional to
$q^4$, and that is if $T^{1p} \sim q^3$.  Clearly, such a term could
occur from three spatial derivatives, and is thus only possible within
the context of \emph{third order hydrodynamics}.  Hence, we have shown
that the correction to the shear mode dispersion relation is dependent
on the third order hydrodynamic expansion of the energy momentum
tensor.  This fact was first realized in \cite{Baier:2007ix}.  One
might be surprised by the fact that in order to consistently compute
the subleading term in the hydrodynamic dispersion relation, one needs
to use third order hydrodynamics.  This is simply a consequence of the
fact that the to lowest order $w(q)_{\rm shear} \propto q^2$.

Currently, there is no formulation for third order hydrodynamics.  We
have introduced the parameter $\Delta$ which appears in the shear mode
dispersion relation (\ref{sheardispersionsecondorder}) to denote the
sum of all such contributions from third order hydrodynamics.

\section{Dispersion relations from gravity}
\label{prescription_section}
Observations of elliptic flow of the plasma created at RHIC 
indicate that the system exhibits collective motion characteristic of a
thermally equilibrated system, and is thus
describable by hydrodynamics
\cite{Adams:2005dq,Adcox:2004mh,Arsene:2004fa,Back:2004je,
Molnar:2001ux, Huovinen:2001cy,Huovinen:2003fa}.  The transport
coefficients mentioned in the previous section are necessary input for
hydrodynamical simulations of this  matter.  As outlined in Chapter
\ref{GGDReview_chapter}, gauge/gravity duality allows one to map a
strongly coupled gauge theory to a classical gravity theory in an
extra dimension.  Thus, there exist methods for determining transport
coefficients, and the shear and sound mode dispersion relations within
this framework.

There are many ways to calculate the transport coefficients and
dispersion relations of a strongly coupled gauge theory using an
extra-dimensional gravity dual.  One can compute correlation functions
of the stress-energy tensor and use Kubo formulas or examine the poles
of such correlators \cite{Policastro:2001yc,
Son:2002sd,Policastro:2002se, Policastro:2002tn, Herzog:2002fn,
Herzog:2003ke}.  Alternatively, one can examine the behavior of the
gravitational background under perturbations and determine the
dispersion relation for such perturbations by applying appropriate
boundary conditions.  Comparison with the expected dispersion
relations from
(\ref{firstorderhydroshear}),(\ref{firstorderhydrosound}) yields
formulas for the transport coefficients \cite{Springer:2008js,
Kovtun:2005ev, Mas:2007ng}.  In addition, the black hole membrane
paradigm has been employed to calculate the hydrodynamic properties of
the stretched horizon of a black hole.\footnote{This is the idea that
a black hole's influence on the outside world can be encoded in an
effective membrane which lies just outside the horizon.  This idea is
similar to replacing a spherical mass distribution with a point mass;
provided one remains outside the event horizon, the membrane appears
physically equivalent to the actual black hole.  This effective
membrane is sometimes called the \emph{stretched horizon} and can be
endowed with thermodynamic and hydrodynamic properties such as
electrical conductivity and viscosity \cite{Thorne:1986}.}In many
cases, the transport coefficients calculated on the stretched horizon
coincide with the transport coefficients in the dual gauge theory
\cite{Kapusta:2008ng, Iqbal:2008by, Kovtun:2003wp, Saremi:2007dn,
Fujita:2007fg, Natsuume:2008gy}.\footnote{For an example of a
situation where the stretched horizon transport coefficients differ
from those in the dual field theory, one can consider the bulk
viscosity.  The bulk viscosity on the stretched horizon is negative,
whereas the bulk viscosity of the dual field theory is non-negative \cite{Iqbal:2008by}.}Recently, the
work of \cite{Bhattacharyya:2008jc} provides yet another way to
compute hydrodynamic transport coefficients (sometimes referred to as
\emph{fluid/gravity correspondence}), by deriving the equations of
fluid dynamics directly from gravity.  This work has proved quite
influential, and has led to much subsequent research
\cite{VanRaamsdonk:2008fp, Haack:2008cp, Bhattacharyya:2008mz,
Erdmenger:2008rm, Banerjee:2008th, Bhattacharyya:2008ji}.

In the past few years, much work has been done to extend previous
analyses to second order hydrodynamics \cite{Kapusta:2008ng,
Natsuume:2007ty, Baier:2007ix, Natsuume:2008gy, VanRaamsdonk:2008fp,
Haack:2008cp, Bhattacharyya:2008mz, Erdmenger:2008rm, Banerjee:2008th,
Bhattacharyya:2008ji}.  Most of the work on second order hydrodynamics
has focused on conformal theories.  It is notable that a
universal relation between second order hydrodynamic transport
coefficients of a conformal theory was presented in
\cite{Haack:2008xx}, though it is not known whether this relation
still holds for non-conformal theories.

In this thesis, we use the gravitational perturbation approach similar
to \cite{Kovtun:2005ev, Mas:2007ng} to compute the shear and sound
mode dispersion relations for a special class of gravity duals.  In
this section we explain this methodology.  We will always work with a
black brane type metric (\ref{blackbranemetric}) as discussed in
Chapter \ref{GGDReview_chapter}.

The idea behind this method is conceptually simple.  Fluctuations in
the strongly coupled plasma should be dual to fluctuations of the
gravitational background.  The latter are introduced by adding perturbations
in the metric 
\begin{equation}
  g_{\mu \nu} \rightarrow g_{\mu \nu} + h_{\mu \nu}
\end{equation}
and any matter fields present.  For example, if a scalar field is present, 
one must introduce the fluctuation
\begin{equation}
  \phi \rightarrow \phi + \delta \phi.
\end{equation}
In later chapters, we will specify the matter supporting the metric as
scalar in nature.  For now, though, we will work with a general energy
momentum tensor.  It is important to note that the energy momentum
tensor discussed here describes the matter which supports the gravity
dual.  It is \emph{not} the same as the hydrodynamic energy momentum
tensor discussed in the previous section.  This latter quantity is
defined on the (p+1) dimensional boundary of the bulk space-time.  In
short, one should take care to distinguish the energy momentum tensor
which supports the extra-dimensional metric, and the energy momentum
tensor of the dual field theory.

Returning now to our perturbations of the gravity background, one must 
take the background Einstein equations
\begin{equation}
  G^{(0)}_{\mu \nu} = -8 \pi G_{p+2} T^{(0)}_{\mu \nu}, 
\end{equation}
and expand them to linear order in these perturbations.  Symbolically,
we write
\begin{equation}
   G^{(1)}_{\mu \nu} = -8 \pi G_{p+2} T^{(1)}_{\mu \nu}. 
\end{equation}
Here, $G_{\mu \nu}$ is the Einstein tensor and $G_{p+2}$ is the $p+2$
dimensional Newton's constant.  The superscripts $(0)$ and $(1)$
denote the order of the perturbation.  Explicit expressions for the
background and perturbed Einstein tensor are given in Appendix
\ref{app_EinsteinEqns}.  (In addition, the background equations for
the matter fields must also be expanded to first order in the
perturbation).

The resulting set of linearized equations can then be solved for the
perturbations.  The perturbations of the hydrodynamic energy momentum
tensor were considered to be plane waves in the previous section.
Hence, we should make the same assumption here, but there is an extra
complication owing to the fact that the metric lives in one extra
dimension.  Thus, we make the ansatz
\begin{equation}
  h_{\mu \nu} = h_{\mu \nu}(r) e^{i(q z - w t)},
\end{equation}
and similarly for the matter perturbations.  One must then solve the
linearized Einstein equations for the functions $h_{\mu \nu}$
perturbatively in $q$ (since we are still working in the hydrodynamic
regime $w,q \ll T$).  Applying appropriate boundary conditions to
these solutions will result in a dispersion relation $w(q)$.  One can
then read off relations for the transport coefficients by comparing
the resulting dispersion relations to those expected from the
hydrodynamic fluid
(\ref{firstorderhydroshear}),(\ref{firstorderhydrosound}).  This
allows one to write expressions for the quantities in the dispersion
relation ($w_1, w_2,...$) in terms of the metric components ($g_{tt},
g_{rr}, g_{xx}$) and any background fields.  Let us now discuss
the individual steps of this method in more detail.

\subsection{Classification of hydrodynamic modes}
In Sec. \ref{hydromodes_section}, we described how perturbations of
the hydrodynamic energy-momentum tensor can be decomposed into two
normal modes: shear and sound.  Since we assume that our system
admits a dual gravitational description, we need to do the same thing
for the gravitational perturbations.  Which components of the metric
fluctuation $h_{\mu \nu}$ correspond to shear/sound perturbations in
the dual field theory?

These components are determined by classifying the perturbations under
the rotation group $SO(p-1)$ \cite{Kovtun:2003wp, Policastro:2002se, 
Mas:2007ng}.  We have singled out the $z$ direction as the direction
of our momentum, but there is still symmetry in the remaining $p-1$ spatial 
dimensions, and thus such a classification is possible.  A
general transformation matrix for a spatial rotation about the z-axis
is given by
\begin{equation}
  U(\theta)^\mu_{\, \, \nu} = 
       	\left(
  	\begin{array}{ccccccc}
    		1 & 0 & ... & 0 & 0 & 0 \\
    		0 & \lambda_{1,1}(\theta) & ... & \lambda_{1,p-1}(\theta) & 0 &0  \\
		: & : & ... & : & : & : \\
		: & : & ... & : & : & : \\
		0 & \lambda_{p-1,1}(\theta) & ... & \lambda_{p-1,p-1}(\theta) &0 &0 \\
		0 & 0 & ... &0 &1 &0\\
		0 & 0 & ... &0 &0 &1\\
   	\end{array}
   	\right).
\end{equation}
Here $\mu$ denotes the row, and $\nu$ denotes the column. The indices are ordered such that
$\mu = 0,1,...p-1,p,p+1$ corresponds to $t,x_1,x_2...,x_{p-1},z,r$.  The components $\lambda(\theta)$ describe
how the various spatial components are rotated into one another.  (As a simple illustration, in the 
case of three spatial dimensions, $\lambda_{1,1} = \lambda_{2,2} = \cos(\theta)$, and $ \lambda_{1,2} = - \lambda_{2,1} = \sin(\theta)$).   
Let us now examine
how the components of $h_{\mu \nu}$ transform under this transformation
\begin{equation}
  \tilde{h}_{\mu \nu} = U^{\alpha}_{\,\, \mu} U^{\beta}_{\,\, \nu} h_{\alpha \beta}. 
\end{equation}
For example, consider 
\begin{equation}
  \tilde{h}_{\mu r} =  U^{\alpha}_{\,\, \mu} U^{\beta}_{\,\, r} h_{\alpha \beta} 
  = U^{\alpha}_{\,\, \mu} h_{\alpha r}. 
\end{equation}
In what follows, we use the Latin indices ($a,b,c...$) to denote the spatial coordinates $x_1 ... x_{p-1}$.  Then, 
it is clear that 
\begin{eqnarray}
  \tilde{h}_{tr} &=& h_{tr}\\
  \tilde{h}_{zr} &=& h_{zr}\\
  \tilde{h}_{rr} &=& h_{rr}\\
  \tilde{h}_{ar} &=& U^{b}_{\,\,a} h_{br}.
\end{eqnarray}
Notice that the components $h_{tr}, h_{rr}, h_{zr}$ transform into themselves, and are
thus referred to as scalars under this transformation.  The components $h_{br}$ transform
with \emph{one} factor of $U$, which is how a vector transforms.  Continuing with other components, 
one finds that $h_{zz}, h_{tz}, h_{tt}$ belong to the scalar mode, while $h_{ta}$, and $h_{za}$ belong
to the vector sector.  The only unaccounted for elements are the $h_{ab}$ components.  

The components of $h_{ab}$ can be split into a trace and a trace-free part.  How does
the trace piece transform under the spatial rotation?  Clearly, our rotation is a unitary 
transformation.  Recalling that the trace of a matrix is unchanged under a unitary transformation, 
it is required that
\begin{equation}
  \tilde{h}_{tt} + \sum_a \tilde{h}_{aa} + \tilde{h}_{zz} + \tilde{h}_{rr} =  
  h_{tt} + \sum_a h_{aa} + h_{zz} + h_{rr}.    
\end{equation}
Since we've already shown that $h_{tt}, h_{zz}$, and $h_{rr}$ are
scalars under the transformation, we conclude that the trace of
$h_{ab}$ is also a scalar.

The remaining (trace-free) part of $h_{ab}$ transforms as a rank two
tensor.  These components constitute a separate mode (the \emph{tensor
mode}), which we will not be concerned with in this thesis because the
the resulting dispersion relation does not have a well defined hydrodynamic limit (see for example
\cite{Natsuume:2007ty}).  In full, we have the following tensor decomposition
for the metric perturbations
\begin{equation}
  \begin{array}{ll}
  h_{tt}, \sum_a h_{aa}, h_{zz}, h_{rr}, h_{tz}, h_{tr}, h_{rz} & \mbox{ scalar mode (sound mode) }\\
  h_{ra}, h_{ta}, h_{za} & \mbox{ vector mode (shear mode) } \\
  h_{ab} - \delta_{ab} \frac{1}{p-1} \sum_c h_{cc} & \mbox{ tensor mode }.
  \end{array}
  \label{modeclassification}
\end{equation}

\subsection{Shear mode equations}
In the previous subsection, we found that to describe the shear mode, 
we should make only the following components of the metric perturbation
non-zero: $h_{ta}$, $h_{za}$, and $h_{ra}$.  We now make the ansatz mentioned
in the Sec. \ref{prescription_section} and define the functions $M(r)$ and $N(r)$ as
\begin{eqnarray}
  h_{ta} &=& h_{ta}(r) e^{i(q z - w t)} \equiv  g_{xx}M(r) e^{i(q z - w t)}, \label{Mdef}\\
  h_{za} &=& h_{za}(r) e^{i(q z - w t)} \equiv  g_{xx}N(r) e^{i(q z - w t)}. \label{Ndef}
\end{eqnarray}
There is always relativistic gauge freedom when one deals with
gravitational perturbations.  We are free to choose our coordinate system so that $h_{\mu r}$
vanishes.  This is referred to as the radial gauge.  It is important to note that this choice
only partially fixes the gauge; later we will define a set of gauge invariant variables to 
deal with the remaining gauge freedom.   

In Appendix \ref{app_EinsteinEqns}, the relevant formulas are given
which allow one to compute the Einstein equations to linear order in
the metric perturbation.  Using these formulas with only the shear
perturbations non-vanishing results in three non-trivial Einstein
equations.\footnote{One may wonder why there are only three equations,
when at first sight there are $3(p-1)$ degrees of freedom in this
channel.  The equations for $a=x_1$, for example, are identical to
those with $a=x_2$ up to a change of variable.  This is why we have
defined $M$ and $N$ as above, and not taken care to distinguish $M_1,
M_2$, etc.}These equations come from the ($ta$), ($za$) and
($ra$) components of the linearized Einstein equations 
(\ref{masterlinearizedeqn}).  They can be written:
\be
	\frac{1}{\rootg} \partial_r \left[ \rootg g^{rr}g^{tt}g_{xx} M' \right] - q g^{tt} \left(q M + w N\right)
	= -16 \pi G g_{xx} \left[T^{ta}_{(1)} + T^{tt}_{(0)} M \right], \label{sheararray1}
\ee
\be
	 \frac{1}{\rootg} \partial_r \left[ \rootg g^{rr} N' \right] -wg^{tt} \left(q M + w N \right) 
	 = -16 \pi G g_{xx} \left[ T^{za}_{(1)} + T^{xx}_{(0)} N \right], \label{sheararray2}
\ee
\be
	q g^{xx} N' - w g^{tt}M' = -i 16 \pi G T^{ra}_{(1)} \label{sheararray3}.
\ee
Here, the prime denotes derivatives with respect to $r$.  When not
indicated otherwise, $G$ is short for $G_{p+2}$.  Of course,
components of the background energy momentum tensor $T^{\mu
\nu}_{(0)}$ can be written in terms of the metric components using the
relations (\ref{Ftdef} - \ref{Frdef}).

\subsection{Sound mode equations}
\label{soundmodeeqns_subsection}
In contrast to the shear mode, the relevant perturbations which must
be considered for the sound mode are $h_{tt}, h_{tz}, h_{zz}$, and $h_{aa}$.  Again we
have partially fixed the gauge by setting $h_{\mu r} = 0$.  
We proceed as before by defining
\begin{eqnarray}
  	h_{tt}(r) & \equiv & g_{tt}(r)A(r), \label{Adef}\\
	\frac{1}{p-1}\sum_{a=1}^{p-1} h_{aa}(r) & \equiv & g_{xx}(r) B(r), \label{Bdef}\\
	h_{zz}(r) & \equiv & g_{xx}(r)C(r),\label{Cdef} \\
	h_{tz}(r) & \equiv & g_{tt}(r)D(r),\label{Ddefinition}.
\end{eqnarray}
Ostensibly, there are $5+p$ Einstein equations.  These come from the
$(tt)$, $(zz)$, $(rr)$, $(tz)$, $(rz)$, $(tz)$ components of the linearized Einstein
equation (\ref{masterlinearizedeqn}), plus the $(p-1)$ equations for the $(aa)$ components of this
equation.  However, each of these latter equations is identical
up to a change of variable.  For example, label two of the $a$
coordinates as $x$ and $y$.  Then the equation involving the $(xx)$
component of (\ref{masterlinearizedeqn}) is identical to the equation equation
involving the $(yy)$ component up to a replacement $h_{xx} \rightarrow
h_{yy}$.  We can add all of these $p-1$ equations together and
get a single equation which involves only the variable $B$ defined
above.  Thus, there are seven Einstein equations in total.  To simplify
the presentation of these equations, we make the following definitions
\begin{equation}
  f(r) \equiv -g_{tt}g^{xx}.
  \label{fdef}
\end{equation}
\begin{equation}
	\DL[X(r)] \equiv \frac{X'(r)}{X(r)}.  
	\label{DLdef}
\end{equation}
Then, the relevant seven Einstein equations (in the radial gauge) can be written
\begin{eqnarray}
  	\frac{g_{rr}}{\rootg} \sqrt{f} \partial_r \left[\frac{\rootg g^{rr}}{\sqrt{f}} ((p-1)B'+C') \right]
	&-&(p-1) q^2 g_{rr}g^{xx} B \label{g00eqn}\\
	&=& 16 \pi G g_{rr}g_{tt} \left(T^{tt}_{(1)} + T^{tt}_{(0)} A \right), \nonumber
\end{eqnarray}
\begin{eqnarray}
  	\label{gxxeqn}
	\frac{g_{rr}}{\sqrt{-g f}}\partial_r \left[\sqrt{-g f} g^{rr} A' \right] 
	&+& \frac{g_{rr}}{\rootg}\partial_r \left[\rootg g^{rr}\left((p-2)B'+C'\right) \right] \\ 
	&-& g_{rr}g^{xx}\biggl\{q^2 \left[A+ (p-2)B+\frac{2w}{q} D\right] -  \frac{w^2}{f} \left[(p-2)B+C\right] \biggr\} \nonumber \\
	&=& 16 \pi G g_{rr}g_{xx} \left( \frac{1}{p-1}\sum_{b=1}^{p-1} T^{bb}_{(1)} + T^{zz}_{(0)} B \right) \nonumber ,
\end{eqnarray}
\begin{eqnarray}
  	\label{gzzeqn}
	\frac{g_{rr}}{\sqrt{-g f}}\partial_r \left[\sqrt{-g f} g^{rr} A' \right] 
	&+& \frac{g_{rr}}{\rootg}\partial_r \left[\rootg g^{rr}(p-1)B' \right] \\
	&-& (p-1) w^2g_{rr}g^{tt}B = 16 \pi G g_{rr}g_{xx} \left( T^{zz}_{(1)} + T^{zz}_{(0)} C \right),
	\nonumber 
\end{eqnarray}
\begin{equation}
  	\label{g0zeqn}
	\frac{g_{rr}}{f \rootg} \partial_r \left [ \rootg g^{rr} f D' \right] + (p-1) q w g^{tt}g_{rr} B = 
	-16 \pi G g_{xx}g_{rr} \left(T^{tz}_{(1)} + T^{zz}_{(0)} D \right),
\end{equation}
\begin{eqnarray}
  	\label{grreqn}
	\DL \left[(g_{xx})^p\right]A' &+& \DL \left[g_{tt}(g_{xx})^{p-1}\right]((p-1)B'+C') \\ 
	&-& 2g_{rr}\left\{w^2g^{tt}\left[(p-1)B+C \right] + q^2 g^{xx}\left[A+(p-1)B + \frac{2w}{q}D\right] \right\} \nonumber \\
	&=& 32 \pi G (g_{rr})^2 T^{rr}_{(1)}, \nonumber  
\end{eqnarray}
\begin{equation}
  \label{g0reqn}
  w \sqrt{f} \partial_r\left[ \frac{1}{\sqrt{f}} ((p-1)B+C) \right] -q f D' 
  = - i \left( 16 \pi G g_{rr}g_{tt}T^{tr}_{(1)} \right),
\end{equation}
\begin{equation}
  \label{grzeqn}
  \frac{q}{\sqrt{f}}\partial_r \left[ \sqrt{f} A \right] + \frac{w}{f} \partial_r \left[f D\right]+ q (p-1)B'
  = i\left( 16 \pi G g_{rr}g_{xx}T^{rz}_{(1)} \right).  
\end{equation}
These equations are the Einstein equations which involve
the $(tt)$, $(aa)$, $(zz)$, $(tz)$, $(rr)$, $(tr)$ and $(rz)$ components of the
(\ref{masterlinearizedeqn}) respectively.  We emphasize that these
equations are valid for \emph{any} matter distribution which is
minimally coupled to the metric (in other words, they are valid for Einstein gravity, and any
matter distribution).  When a particular form of matter is
chosen, the components of the energy momentum tensor on the right side of the above equations will have to be
explicitly evaluated.  

Whatever matter fields are present will obey separate background
equations.  Introducing perturbations into these equations will result
in equations for the matter fluctuations (e.g. $\delta \phi$ for a
scalar field, $\delta A_\mu$ for a gauge field, etc...).  We will
determine the form of these equations for scalar matter in Chapters 
\ref{ShearMode_chapter} and \ref{SoundMode_chapter}.

\subsection{Gauge invariance}
When examining perturbations of a general relativistic background, it becomes
difficult to tell the difference between physical perturbations, and
those which could be interpreted as a coordinate transformation.  The
ability to change ones coordinates at will, is referred to as relativistic gauge
freedom.  In particular, metric perturbations which are related to each
other under the transformation
\begin{equation}
	h_{\mu \nu} \rightarrow h_{\mu \nu} - \nabla^{(0)}_\mu \xi_\nu - \nabla^{(0)}_\nu \xi_\mu  
\label{gaugetransformation}
\end{equation}
describe the same physics.  Here $\xi_\mu$ is any vector, and
$\nabla^{(0)}$ denotes the covariant derivative with respect to the
background metric (see Appendix \ref{app_EinsteinEqns}).  In order to
preserve the assumed space-time dependence of our perturbations, the
vector $\xi_\mu$ should also contain the same dependence
\begin{equation}
  \xi_\mu = \xi_{\mu}(r)e^{i(q z - w t)}.  
  \label{xiansatz}
\end{equation}

In order to address the issue of gauge freedom, one can take two
approaches: either fix the gauge completely, or do the analysis in
terms of gauge invariant variables.  In this thesis, we use the latter
approach.  Hence, before we determine our dispersion relation, we will
need to create gauge invariant combinations of the perturbations
which do not transform under (\ref{gaugetransformation}).  The
particular choice of gauge invariant variables will be given
explicitly in the later chapters, as they depend on the type of matter
fields that are present.  Gauge invariant variables have been used
extensively within the context of cosmology
(cf. \cite{Mukhanov:2005}); in the context of gauge/gravity duality,
they were first introduced in \cite{Kovtun:2005ev}.  In this thesis,
we use $Z$ to denote gauge invariant variables; sometimes we will also
add an additional subscript (e.g. $Z_0, Z_{\phi}$) when it is
necessary to make a distinction between several such variables.

When constructing such gauge invariant combinations, one has two
possible choices.  The first option is to first solve for the
perturbations $h_{\mu \nu}(r)$, and then construct gauge invariant
combinations from these solutions.  The second option is to first take
combinations of the linearized Einstein equations and reduce them to a
set of equations which depend only on the gauge invariant variables,
and then solve these equations.  In our analysis of the shear and sound
modes, we will employ the latter approach.  

\subsection{Boundary conditions}
\label{boundarycondition_section}
After deriving the relevant equations for the perturbations, one must
solve these equations perturbatively (order by order) in the variable
$q$ by first expanding the perturbations in powers of $q$, and also
inserting an ansatz for the dispersion relation $w(q)$.  We are
ultimately interested in the dispersion relation (not the solutions
for $h_{\mu \nu}(r)$ or $Z(r)$).  One must apply appropriate
boundary conditions on the solutions in order to determine $w(q)$.

Two obvious places to apply boundary conditions are at the horizon
and at the boundary $r\rightarrow \infty$.  The correct boundary
conditions that need to be applied on the gauge invariant variables
were worked out in \cite{Kovtun:2005ev}.  In this work, the authors constructed
gauge invariant equations for the $AdS_5$ background.  By examining
the behavior of the equations in the near horizon regime, one finds that the 
gauge invariant variable $Z$ must behave as
\begin{eqnarray}
  Z(r \rightarrow r_0) &\sim& (r-r_0)^\alpha + \mathcal{O}(r-r_0)^{\alpha + 1}\\  
  \alpha &=& \pm \frac{i w}{4 \pi T}.  
\end{eqnarray}
One interprets the positive/negative exponent by stating that the perturbation
corresponds to either an outgoing/incoming wave.  Because classically nothing can be
emitted from the horizon of a black brane, we should choose the minus sign.  This is 
referred to as the incoming wave boundary condition.  It can be implemented by making the
ansatz 
\begin{equation}
  Z(r) = f(r)^{-i w / 4 \pi T} Y(r),
  \label{incomingwaveansatz1}
\end{equation}
where the function $Y(r)$ is regular at the horizon - it either
vanishes there or approaches a constant.  Another way to implement
this condition is to take the logarithmic
derivative of the above equation (using the notation defined in
(\ref{DLdef})),
\begin{equation}
  \DL[Z(r)] = -\frac{i w}{4 \pi T} \DL[f(r)] + \DL[Y(r)].
\end{equation}  
Let us examine the leading behavior of this equation near the horizon,
using the near horizon behavior of the black brane (\ref{ttNH} -
\ref{xxNH}) and the fact that $Y(r)$ is regular at the horizon.  This
leads to
\begin{equation}
  Z'(r \to r_0)(r-r_0) = -\frac{i w}{4 \pi T} Z(r \to r_0).  
  \label{incomingwaveansatz2}
\end{equation}

A condition at the boundary ($r \rightarrow \infty$) is also needed.  The authors of
\cite{Kovtun:2005ev} show that the correct boundary condition to apply is 
\begin{equation}
  Z(r \rightarrow \infty) = 0.  
\end{equation}
The authors of the above mentioned paper argue this is the correct
boundary condition because if one applies this condition, one
correctly reproduces the dispersion relation that appears as a pole
in correlation functions of the energy-momentum tensor.  In other words, 
if one applies this boundary condition, one finds agreement between this 
method and other methods of computing the dispersion relation.  

\subsection{Summary}
To summarize, in order to compute hydrodynamic dispersion relations using gauge/gravity
duality, one must follow the following prescription.
\begin{enumerate}
  \item
    Depending on which hydrodynamic mode is being studied, one must
    introduce the relevant metric perturbations
    (\ref{modeclassification}), as well as perturbations of any and
    all matter fields present.
  \item
    With the relevant perturbations turned on, one must linearize the
    background Einstein equations, and equations for the matter
    fields.  The linearized Einstein equations for our particular
    black brane metric are given explicitly in (\ref{sheararray1} -
    \ref{sheararray3}) for the shear mode, and (\ref{g00eqn} -
    \ref{grzeqn}) for the sound mode.
  \item
    Either:
    \begin{itemize}
      \item
	Reduce this set of equations down to equations which depend
	only on gauge invariant variables and then solve for the
	gauge invariant variables order by order in the momentum of
	the perturbation, $q$. 
    \end{itemize}
 Or:
 \begin{itemize}
      \item
	Solve for the perturbations $h_{\mu \nu}(r)$ order by order in
	$q$, and construct gauge invariant combinations from these
	solutions.
    \end{itemize}
  \item
  For the resulting solutions, apply the incoming wave boundary
  condition in the form of (\ref{incomingwaveansatz1}) or
  (\ref{incomingwaveansatz2}).  Also apply a Dirichlet boundary
  condition at the boundary ($r \rightarrow \infty$).
\end{enumerate}
In the next two chapters, we will illustrate this procedure for the shear
and sound modes assuming the matter fields present are scalar fields.  

\chapter{Shear mode}
\label{ShearMode_chapter}

\section{Introduction}
In this chapter, we will apply the prescription outlined in the
previous chapter to the shear mode.  We assume that the matter which
supports the black brane metric is scalar in nature.  To this end, we
must first explicitly compute the components of the energy momentum
tensor to first order in the perturbations.  Once this is done, we
proceed to construct gauge invariant equations, solve them, and apply
the appropriate boundary conditions to determine the shear mode
dispersion relation.  Shear mode perturbations are strongly
over-damped, and the shear mode dispersion relation is an expansion in
even powers of the momentum $q$.  Let us parametrize the dispersion
relation by introducing two real constants $D_{\eta}$ and
$\tau_{\rm shear}$.
\begin{equation}
	w(q)_{\rm shear} = -i D_{\eta} q^2 \left( 1 + \tau_{\rm shear} D_{\eta} q^2 + ... \right).
	\label{sheardispersion}
\end{equation}
In what follows, we will determine $D_{\eta}$ and $\tau_{\rm shear}$
in terms of the metric components $g_{tt}, g_{xx}$, and $g_{rr}$.
Through comparison with the hydrodynamic expectations for the shear
mode (\ref{sheardispersionsecondorder}), the computation of $D_{\eta}$
allows one to derive a formula for the shear viscosity $\eta$, and the
result for $\tau_{\rm shear}$ provides information the relating second
order transport coefficient $\tau_\pi$ to the correction from third
order hydrodynamics $\Delta$ which was introduced in section
\ref{secondorderhydro_section}.

The formula presented here for $D_{\eta}$ was first derived by Kovtun, Son 
and Starinets in \cite{Kovtun:2003wp} using the membrane paradigm, and later in
\cite{Starinets:2008fb} using techniques similar to those we employ
here.  The formula for $\tau_{\rm shear}$ was first derived by Kapusta
and Springer in \cite{Kapusta:2008ng} again using the membrane
paradigm.  This chapter is based on this previous work, though here we
do not make any reference to the membrane paradigm.  As far as we are aware,
this is the first instance in the literature where the formula for
$\tau_{\rm shear}$ is derived without any reference to the black hole
membrane paradigm.

Finally, once we have the formulas for $D_{\eta}$ and $\tau_{\rm shear}$ in
hand, we will consider a few applications of these formulae to
specific gravitational backgrounds.  

\section{Scalar matter}
\subsection{Background equations}
Let us assume that the matter supporting the metric is a set of $n$ minimally coupled
scalar fields.  
In other words, we assume the action is of the form
\be
	\mathcal{S} = \frac{1}{16 \pi G_{p+2}} \int \, d^{p+2}x \rootg 
	\left( R  - \thalf \partial_\mu \phi_k \partial^\mu \phi_k - U(\phi_1,\phi_2...\phi_n) \right). 
	\label{scalaraction}
\ee
We assume a summation over the repeated index $k$, and we use superscripts $(0)$ and $(1)$ to 
denote background quantities and quantities that are first order in the perturbation respectively.  
Then, the background equations are
\ba
	G^{(0)}_{\mu \nu} &=& -8 \pi G_{p+2} T^{(0)}_{\mu \nu} \\
	\Box^{(0)} \phi_k &=& \frac{\partial U}{\partial \phi_k}.
	\label{scalarbackgroundequations}
\ea 
Here $\Box^{(0)}$ is defined as in the appendix (cf. \ref{boxdef}).  
We ignore any subtleties regarding boundary terms that come from
integration by parts.  Such terms can be taken care of by adding
additional boundary terms to the action.

The energy-momentum tensor derived from the action is
\ba
	8 \pi G_{p+2} T^{(0)}_{\mu \nu} 
	&=& \frac{1}{2} \left(\partial_\mu \phi_k \partial_\nu \phi_k - g_{\mu \nu} \mathcal{L}_{\phi} \right), 
	\label{scalarTuv}
	\\
	\mathcal{L}_{\phi} &\equiv& \frac{1}{2} \partial_\lambda \phi_k \partial^\lambda \phi_k + U(\phi_1,\phi_2...\phi_n).
\ea
Because the background metric only depends on the extra dimensional
coordinate $r$, if $n=1$ it is clear that the single field $\phi$ must also only a
function of $r$.  In the case of multiple scalar fields ($n>1$), it might be
possible to have fields which depend on the other coordinates,
provided that all such dependence cancels out in the combination that
appears in $T^{(0)}_{\mu \nu}$.  We will not consider such special cases,
and will always assume that the scalar fields only depend on $r$.

One can write the fields $\phi_{k}(r)$ in terms of the metric components by noting that
\be
	g^{tt}G^{(0)}_{tt} - g^{rr}G^{(0)}_{rr} = F_{t}(r) - F_{r}(r),
\ee
where the $F$ functions are defined in the Appendix \ref{app_EinsteinEqns}.  Furthermore,
\ba
	8 \pi G_{p+2} \left(g^{tt} T^{(0)}_{tt} -g^{rr} T^{(0)}_{rr} \right) &=&
	\frac{1}{2} \left[
	- \mathcal{L}_{\phi}  - \left( \sum_{k=1}^n g^{rr} \phi_k'(r)^2 -
	\mathcal{L}_\phi (r) \right) \right] 
	\\ &=& - \frac{1}{2} g^{rr} \sum_{k=1}^n \phi_k'(r)^2.
\ea
Thus we have the relation
\be
	\label{phikeqn} 
	\sum_{k=1}^n \phi_k'(r)^2 = 2g_{rr}\left[F_t(r) - F_r(r)\right].
\ee
It is also noteworthy that this background has the special property $F_t = F_x$ as can be seen by considering
\ba
	g^{tt}G^{(0)}_{tt} - g^{xx}G^{(0)}_{xx} &=& -8 \pi G_{p+2} \left(g^{tt}T^{(0)}_{tt} - g^{xx}T^{(0)}_{xx}\right)\\
	F_{t}(r) - F_x(r) &=& \mathcal{L}_{\phi}(r) - \mathcal{L}_{\phi} (r) = 0.  
	\label{F0Fx}
\ea
Because of this fact, and the general theorem given in
\cite{Buchel:2003tz}, \textbf{all backgrounds we consider saturate the
conjectured shear viscosity bound: $\eta/s = 1/4 \pi$.}

\subsection{A note on phenomenological model building}
\label{SoftWall_section}
Let us make a small digression and examine implications of these
background equations for phenomenological model building.  As
mentioned in Chapter \ref{GGDReview_chapter}, the bottom up
approach (or AdS/QCD) often employs metrics and scalar fields which
are chosen to reproduce some essential features of QCD.  Often, these
metrics are imposed by hand, and are not dynamically generated from any matter distribution.

One such popular background is the soft-wall model
of \cite{Karch:2006pv}.  At zero temperature, this five dimensional
model has a (string frame) metric and background scalar field
(dilaton) of the form
\ba 
	ds^2_{\rm string} &=& \frac{r^2}{L^2} \left( -dt^2 + dx_i dx^i \right) + \frac{L^2}{r^2} dr^2\, ,
	\label{softwallmetric} \\
	\phi(r) &=& \frac{c L}{r^2}.
\ea
Here, $L$ is the usual $AdS$ curvature radius and $c$ is a dimensionful constant
which allows for conformal symmetry breaking and the introduction of the QCD scale.  
In \cite{Batell:2008zm}, it was shown that such a background can be
realized dynamically with the addition of a second scalar field
$\chi(r)$, and an appropriate scalar potential $V(\phi, \chi)$.

These results have not yet been generalized to finite
temperature.  Here we will show that an obvious generalization of this
metric to finite temperature is not possible.  Because the string
frame metric is exactly $AdS_5$, a naive expectation is that at finite
temperature, the metric should take the form of the Schwarzschild
$AdS$ black hole.  Thus, we might expect a finite temperature generalization
of the metric to be
\ba 
	ds^2_{\rm string} &=& \frac{r^2}{L^2} \left( -f(r) dt^2 + dx_i dx^i \right) + \frac{L^2}{r^2 f(r)} dr^2\, ,
	\label{finiteTsoftwallmetric} \\
	f(r) &=& 1- \frac{r_0^4}{r^4}.
\ea

The dilaton is a scalar field which is coupled
to gravity by a term $e^{b \phi} R$ in the (string frame) action where $b$ is a
constant, and $R$ is the Ricci scalar.  Thus, in the string frame, the
action is not of the form (\ref{scalaraction}).  However, it is easy
to make a conformal transformation of the metric which brings the
action into the form (\ref{scalaraction}) (cf. Appendix G of \cite{Carroll:2004}).
This is called the Einstein frame; in this frame, the metric takes the
form
\be
	g^{\mu \nu}_{\rm Einstein} = e^{a \phi(r)} g^{\mu \nu}_{\rm string}.
\label{softwallEinstein}
\ee
$a$ is a constant which depends on $b$ and the number of dimensions of
the theory; its value is unimportant for our purposes.

With the Einstein frame metric in hand, we can use the results of the previous
section.  For example, if this metric is generated by scalar fields, it must
obey the constraint (\ref{F0Fx}).  An explicit evaluation of this equation
yields
\be
	-\frac{3 a e^{a \phi (r)}{r_0^4} \phi'(r)}{L^2 r^3} = 0.
\label{aeqn}
\ee   
Clearly, this is inconsistent as long as $a$ is non-zero and the
dilaton $\phi$ has some non-trivial profile.  The conclusion is that
the metric (\ref{finiteTsoftwallmetric}) cannot be generated
by scalar fields alone.  One can phrase this another way by stating that
one must modify the metric away from the $SAdS$ form in order to
generate a finite temperature soft wall model using scalar fields alone.  
The reason why there is no problem at zero temperature is that one can arrive
at the zero temperature $AdS_5$ metric by taking the limit $r_0 \to 0$.  In this
limit, the equation (\ref{aeqn}) is satisfied and there is no inconsistency.  

\section{Shear mode equations for scalar matter}
Let us now move beyond the background equations and examine the relevant
shear mode perturbations.  
In order to make use of the general Einstein equations given in the
previous chapter (\ref{sheararray1} - \ref{sheararray3}), one must
determine the components of the perturbed energy momentum tensor
$T^{\mu \nu}_{(1)}$.  For the background, we have
\be
	8 \pi G_{p+2} T^{\mu \nu}_{(0)} 
	= \frac{1}{2} \left(g^{\mu \alpha} g^{\nu \beta} \partial_\alpha \phi_k \partial_\beta \phi_k - g^{\mu \nu} \mathcal{L}_{\phi} \right).	
\ee   
Including the perturbations $g^{\mu \nu} \rightarrow g^{\mu \nu} - h^{\mu \nu}$, and $\phi_k \rightarrow \phi_k + \delta \phi_k$, to linear order we have,
\ba
	8 \pi G_{p+2} T^{\mu \nu}_{(1)} 
	&=& \frac{1}{2} \Bigl\{
	g^{\mu \alpha} g^{\nu \beta} \left[ \partial_\alpha (\delta \phi_k) \partial_\beta \phi_k 
	+ \partial_\alpha \phi_k \partial_\beta (\delta \phi_k) \right] \Bigr. \nonumber \\
	&-& h^{\mu \alpha} g^{\nu \beta} \partial_\alpha \phi_k \partial_\beta \phi_k
	- g^{\mu \alpha} h^{\nu \beta} \partial_\alpha \phi_k \partial_\beta \phi_k \nonumber \\
	&-& \Bigl. g^{\mu \nu} \left( \delta \mathcal{L}_{\phi}\right) + h^{\mu \nu} \mathcal{L}_{\phi} \Bigr\}.  
\label{perturbedTuv} 	
\ea   
With the choice of the radial gauge $h^{\mu r} = 0$, the second line
vanishes due to the fact that the background scalar fields only depend
on $r$.  For the shear mode, we need to evaluate $T^{ta}, T^{za}$, and $T^{ra}$.
For linear response, it is natural to assume that the perturbed scalar
fields have the same space-time dependence as the metric perturbations
\be
	\delta \phi_k = \delta \phi_k(r) e^{i(q z - w t)}.  
\ee  
Then, these three components of the energy momentum tensor are straightforwardly given as
\ba
	8 \pi G_{p+2} T^{t a}_{(1)} &=& \thalf  h^{t a} \mathcal{L}_{\phi}, \\
	8 \pi G_{p+2} T^{z a}_{(1)} &=& \thalf  h^{z a} \mathcal{L}_{\phi}, \\
	8 \pi G_{p+2} T^{r a}_{(1)} &=& 0.
\ea
Noting that for the background, 
\ba
	8 \pi G_{p+2} T^{xx}_{(0)} &=& -\thalf  g^{xx} \mathcal{L}_{\phi},  \\
	8 \pi G_{p+2} T^{tt}_{(0)} &=& -\thalf  g^{tt} \mathcal{L}_{\phi},  
\ea
we have the relations
\ba
	T^{t a}_{(1)} &=& -h^{t a} g_{tt}T^{tt}_{(0)} = - M(r) T^{tt}_{(0)} \\
	T^{z a}_{(1)} &=& -h^{z a} g_{xx}T^{xx}_{(0)} = - N(r) T^{xx}_{(0)}.  
\ea
Here we have used the definitions of $M,N$ in (\ref{Mdef}),(\ref{Ndef}).  
Thus, the equations that we need to solve are simply (\ref{sheararray1} - \ref{sheararray3}) with the 
right hand side set to zero:
\begin{eqnarray}
  \frac{1}{\rootg} \partial_r \left[ \rootg g^{rr}g^{tt}g_{xx} M' \right] - q g^{tt} \left(q M + w N\right)
  &=& 0 \label{scalarsheararray1}\\
  \frac{1}{\rootg} \partial_r \left[ \rootg g^{rr} N' \right] -wg^{tt} \left(q M + w N \right) 
  &=& 0 \label{scalarsheararray2} \\
  q g^{xx} N' - w g^{tt}M' &=& 0 \label{scalarsheararray3}.
\end{eqnarray}  

\section{Gauge invariant variables}
As previously mentioned, it is desirable to work only with physical
degrees of freedom, and this is accomplished by introducing gauge
invariant variables which do not transform under
(\ref{gaugetransformation}).  Within the context of the shear mode,
one such gauge invariant combination is
\be
	Z(r) \equiv q M(r) + w N(r).
\label{sheargaugevariable}
\ee
This can be seen by using the definition of the covariant derivative and
the Christoffel symbols in Appendix \ref{app_EinsteinEqns}:
\be
	 \nabla^{(0)}_\mu \xi_\nu + \nabla^{(0)}_\nu \xi_\mu = \partial_\mu \xi_\nu + \partial_\nu \xi_\mu 
	 - 2\Gamma^{\lambda}_{\mu \nu} \xi_\lambda.  
\ee
Using the presumed form of the vector $\xi$ (\ref{xiansatz}), we have
\ba
 	\nabla^{(0)}_t \xi_a + \nabla^{(0)}_a \xi_t &=& \partial_t \xi_a + \partial_a \xi_t 
	- 2\Gamma^{\lambda}_{t a} \xi_\lambda \nonumber \\
	&=& -i w \xi_a,
\ea
and similarly
\be 
	\nabla^{(0)}_z \xi_a + \nabla^{(0)}_a \xi_z = i q \xi_a.
\ee
Thus, under the gauge transformation (\ref{gaugetransformation}) the 
shear mode metric perturbations behave as
\ba
	h_{ta} &\rightarrow& h_{ta} + iw \xi_a \\
	h_{za} &\rightarrow& h_{za} - iq \xi_a,  
\ea
or equivalently
\ba
	g_{xx}M(r) &\rightarrow& g_{xx} M(r) + i w \xi_a(r), \\
	g_{xx}N(r) &\rightarrow& g_{xx} N(r) - i q \xi_a(r).
\ea
It is clear that the variable $Z(r)$ from (\ref{sheargaugevariable}) does not
transform and is thus gauge invariant.  

It is quite straightforward to take combinations of
the equations (\ref{scalarsheararray1} - \ref{scalarsheararray3})
and reduce them to a single equation for $Z(r)$.  Eliminating $M$ from
(\ref{scalarsheararray3}) results in 
\be
	N'(r) = \frac{ w g_{xx} Z'(r)}{q^2 g_{tt} + w^2 g_{xx}}
\ee
substituting this into (\ref{scalarsheararray2}) gives an equation which only involves $Z$.  
It can be written
\be
	\frac{g_{rr} \left(q^2 g_{tt} + w^2 g_{xx} \right)}{\rootg} 
	\partial_r \left[ \frac{\rootg g_{xx}}{g_{rr} \left(q^2 g_{tt} + w^2 g_{xx} \right)} Z' \right]
	- g_{rr} g^{tt}\left( q^2 g_{tt} + w^2 g_{xx} \right) Z = 0.
\label{sheargaugeequation}
\ee
We now insert the incoming wave ansatz,
\ba
	Z(r) &=& f(r)^{-i w / 4 \pi T}\left(Y_0(r) + q^2 Y_2(r) + q^4 Y_4(r) + \mathcal{O}(q^6)  \right), \\
	w(q) &=& w_2 q^2 + w_4 q^4 + \mathcal{O} \left( q^6 \right),
\ea
and expand (\ref{sheargaugeequation}) in powers of $q$.  The $Y$ functions must be regular
at the horizon, and must also satisfy a Dirichlet boundary condition at $r \rightarrow \infty$
as explained in Sec. \ref{boundarycondition_section}.  

\section{Solution for $w_2$ and shear viscosity} 
The lowest order equation for $Y_0$ is 
\be
	\partial_r \left[ \frac{\rootg g^{rr}}{f} Y_0'\right] = 0,
\ee
which is easily solved as
\be
	Y_0(r) = a_1 \int_\infty^r \frac{f(r')}{\sqrt{-g(r')} g^{rr}(r')} dr'  
\label{Y0shearsoln}
\ee
where $a_1$ is a constant, and the Dirichlet boundary condition has
been applied at the boundary $r \rightarrow \infty$.  Notice that the
function $Y_0$ already meets the requirement of regularity at the
horizon.  Near the horizon, the integrand approaches a finite value
due to (\ref{ttNH}), (\ref{rrNH}), and (\ref{fdef}) and thus there are
no divergences which need to be removed.
For convenience, we define the quantity
\be
	\xi \equiv \frac{i w_2}{4 \pi T}.  
\label{xidef}
\ee
Then, expanding (\ref{sheargaugeequation}) to the next order in $q$ gives an 
equation for $Y_2$,
\be
	\partial_r \left[\frac{\rootg g^{rr}}{f}
	  \left(Y_2' + \frac{w_2^2}{f} Y_0' - \xi \partial_r \left(Y_0 \log[f] \right)\right) \right] 
	+ Y_0 \rootg g^{tt}=0.
\ee
Performing a single integration gives
\be 	
	Y_2' + \frac{w_2^2}{f} Y_0' - \xi \partial_r \left(Y_0 \log[f] \right) = 
	\frac{f}{\rootg g^{rr}} \left[ a_2 - \int_\infty^r Y_0 \rootg g^{tt} dr'\right]=0 .
\label{Y2sheareqn}
\ee

Perhaps the simplest way to implement the requirement of regularity of $Y_2$ is to require
that, $Y_2'(r \rightarrow r_0)$ should be at most logarithmically divergent.  To this end, 
let us examine (\ref{Y2sheareqn}) near the horizon.  To leading order we have 
\be
	Y_2'(r_0) + \frac{1}{r-r_0} \left[-\frac{w_2^2}{\gamma_0 g^{xx}(r_0)}Y_0'(r_0) - \xi Y_0(r_0) \right] 
	+ \mathcal{O}(\log(r-r_0)) = 0.  
\label{Y2shearNHeqn}
\ee
In deriving the above expression, one can estimate the divergence of the right hand side of
(\ref{Y2sheareqn}) by simply inserting the near horizon behavior of each function.  For example, 
the integrand 
\be
	Y_0(r') \sqrt{-g(r')} g^{tt}(r') \sim \mathcal{O} \left(\frac{1}{r'-r_0}\right), 
\ee
so that after the integration over $r'$, this term is at most logarithmically
divergent, and does not contribute to the leading order in (\ref{Y2shearNHeqn}).  
By assumption, $Y_2$ is regular at the horizon, thus the coefficient of the
leading divergence in (\ref{Y2shearNHeqn}) must vanish.  Recalling the definition of $\xi$ from (\ref{xidef})
we obtain the solution for $w_2$:
\be
	w_2 = \frac{-i \gamma_0 g^{xx}(r_0)}{4 \pi T} \frac{ Y_0(r_0)}{Y_0'(r_0)}.
\label{w2shearsoln}
\ee
Using the definition of the temperature (\ref{hawkingtemp}), and the
solution for $Y_0$ (\ref{Y0shearsoln}), we find
\be
	w_2 = -i \frac{\sqrt{-g(r_0)}}{\sqrt{\gamma_0 \gamma_r}} \int_{r_0}^\infty \frac{f(r')}{\sqrt{-g(r')}g^{rr}(r')}dr'.
\ee
Comparison with (\ref{sheardispersion}) and
(\ref{firstorderhydroshear}) gives a formula for the shear viscosity
\be
	D_{\eta} = \frac{\eta}{T s} 
	= \frac{\sqrt{-g(r_0)}}{\sqrt{\gamma_0 \gamma_r}} \int_{r_0}^\infty \frac{f(r')}{\sqrt{-g(r')}g^{rr}(r')}dr',
\label{shearsoln}
\ee
which was already derived using similar methods in
\cite{Starinets:2008fb}, and from the membrane paradigm in
\cite{Kapusta:2008ng, Kovtun:2003wp}.  Agreement with these results in
a reassuring check on our methodology.  Explicit evaluation of this
formula for any gravity dual which satisfies the equations
(\ref{scalarsheararray1} - \ref{scalarsheararray3}) results in
$\eta/s = 1/4 \pi$, as should be expected given the general proofs of
universality \cite{Buchel:2003tz, Iqbal:2008by}.

It is worth reemphasizing that this formula has only been shown to be
valid for gravity duals which satisfy (\ref{scalarsheararray1}) -
(\ref{scalarsheararray3}), as well as the relevant background
equations.  There has been some confusion in the literature regarding
this point.  For example, one could try to use this formula to get the
shear viscosity for a theory with a non-vanishing chemical potential
(e.g. $AdS$ Reissner-N\"{o}rdstrom).  Application of (\ref{shearsoln})
in this case results in $\eta / s \neq 1/4\pi$.  The reason for this
discrepancy is that such metrics are not generated by scalar fields
alone, and hence the equations (\ref{scalarsheararray1}) -
(\ref{scalarsheararray3}) are not satisfied.  Of course, a more
careful treatment of the perturbations in the case where a chemical
potential is present gives the expected result, $\eta / s = 1/4\pi$
\cite{Mas:2006dy, Son:2006em, Saremi:2006ep, Maeda:2006by}.

A second example where confusion may occur is application of the
formula (\ref{shearsoln}) to phenomenological models such as the
soft-wall model.  In the Einstein frame, these metrics are usually
a warped version of anti-de Sitter space
(cf. (\ref{softwallmetric}),(\ref{softwallEinstein})).  Naively
plugging such a metric into the formula (\ref{shearsoln}) again
results in $\eta / s \neq 1/4\pi$ \cite{Kapusta:2008ng, Gao:2007zx}.
The reason for this strange result is exactly what was discussed in
Sec. \ref{SoftWall_section}; these metrics cannot be generated by
scalar fields alone, and hence violate the assumptions inherent in the
derivation of (\ref{shearsoln}).

In summary, if application of (\ref{shearsoln}) to a background based
on Einstein gravity results in a value different than $1/4\pi$, this
result should \emph{not} be viewed as a violation of the viscosity
bound.  Instead, it should be viewed as an inconsistency.  All classical
Einstein gravity duals saturate the conjectured viscosity bound as
shown in \cite{Iqbal:2008by}.

\section{Solution for $\tau_{\rm shear}$}
We now proceed to the next order in $q$.  By expanding the gauge invariant
equation (\ref{sheargaugeequation}) to order $q^4$, one finds an equation for $Y_4$,
\ba
	&\,&\partial_r 
	\left[ 
	  \frac{\rootg g^{rr}}{f} 
	  	\left(Y_4' + w_2^2 \frac{Y_2'}{f} - \xi \partial_r \left(\log[f]Y_2 \right) \right. \right.  \nonumber \\
		&+& \left. \left. Y_0' \left( \frac{w_2^4}{f^2} + \frac{2 w_2 w_4}{f} \right) 
		- \xi \left(\frac{w_2^2}{f} + \frac{w_4}{w_2} \right)\partial_r \left(\log[f]Y_0 \right)
		+ \frac{\xi^2}{2} \partial_r \left(\log[f]^2 Y_0 \right) \right) \right] \nonumber \\ 
	  &+& \rootg g^{tt}\left(Y_2 - \xi \log[f]Y_0 \right) = 0.
\ea
Again, the requirement of regularity of $Y_4$ is best implemented by
requiring that there are no power divergences in $Y_4'$ near $r_0$.
Integrating once, we have
\ba
	Y_4' &+& w_2^2 \frac{Y_2'}{f} - \xi \partial_r \left(\log[f]Y_2 \right) + Y_0' \left( \frac{w_2^4}{f^2} + \frac{2 w_2 w_4}{f} \right) \nonumber \\ 
	&-& \xi \left(\frac{w_2^2}{f} + \frac{w_4}{w_2} \right)\partial_r \left(\log[f]Y_0 \right)
	+ \frac{\xi^2}{2} \partial_r \left(\log[f]^2 Y_0 \right) \nonumber \\
	&=& \frac{f}{\rootg g^{rr}} \left[a_3 - \int_\infty^r \rootg g^{tt} \left(Y_2 - \xi \log[f] Y_0 \right) dr' \right].
\label{Y4primeeqn}
\ea
We should now expand this equation near the horizon in powers of
$(r-r_0)$.  Before proceeding, it is useful to know the near horizon
behavior of $Y_2$ and $Y_0$.  By construction, $Y_0(r_0)$ and
$Y_2(r_0)$ are finite.  It is easy to verify from (\ref{Y0shearsoln})
that $Y_0'(r_0)$ is also finite.  However, from (\ref{Y2shearNHeqn}),
it is clear that $Y_2'(r_0)$ could potentially be logarithmically
divergent.  Near the horizon, this equation takes the form
\be
	Y_2'(r_0)- \xi Y_0'(r_0) \log(r-r_0) + \mathcal{O}(1)
	= \frac{\gamma_r \gamma_0 g^{xx}(r_0)}{\sqrt{-g(r_0)}} \int_\infty^{r_0} \rootg g^{tt} Y_0 dr'.  
\label{Y2primeNHshear}
\ee
The integrand on the right side behaves as
\be
	\rootg g^{tt} Y_0 \approx -\frac{\sqrt{-g(r_0)}Y_0(r_0)}{\gamma_0(r'-r_0)} + \mathcal{O}(1),
\ee
and is thus potentially logarithmically divergent upon integration.  Taking this into account, (\ref{Y2primeNHshear}) becomes
\be
	Y_2'(r_0) + \log(r-r_0) \left[-\xi Y_0'(r_0) + \gamma_r g^{xx}(r_0) Y_0(r_0) \right] 
	+ \mathcal{O}(1) = 0.
\ee
However, using (\ref{w2shearsoln}), (\ref{xidef}) and the expression
for the temperature (\ref{hawkingtemp}) we see that the term in
brackets vanishes.  Thus, we have shown that $Y_2'(r_0)$ is also finite, along with
$Y_2(r_0), Y_0(r_0)$, and $Y_0'(r_0)$.

With this knowledge, we return to (\ref{Y4primeeqn}) and examine its
behavior near the horizon.  For simplicity, we use the notation
\be
	f(r \rightarrow r_0) \approx f_0 (r-r_0) + f_1 (r-r_0)^2 + \mathcal{O}(r-r_0)^3.  
\label{fexpansion}
\ee
In all, (\ref{Y4primeeqn}) becomes
\ba
	Y_4'(r_0) &+& \frac{1}{(r-r_0)^2} \left[Y_0'(r_0) \frac{w_2^4}{f_0^2}- \xi \frac{w_2^2}{f_0}Y_0(r_0) \right]
	+ \frac{\log(r-r_0)}{r-r_0} \left[ \xi^2 Y_0(r_0)-\xi \frac{w_2^2}{f_0}Y_0'(r_0) \right] \nonumber \\
	&+& \frac{1}{r-r_0} \Biggl\{ \frac{w_2^2}{f_0} Y_2'(r_0) - \xi Y_2(r_0) 
	  + \frac{w_2^4}{f_0^2} \left(Y_0''(r_0) - 2 \frac{f_1}{f_0} Y_0'(r_0)\right) \Biggr. \nonumber \\
	  &+& \Biggl. 2 \frac{w_2 w_4}{f_0} Y_0'(r_0)
	  - \xi \frac{w_2^2}{f_0}Y_0'(r_0) - \xi \frac{w_4}{w_2} Y_0(r_0) \Biggr\} +\mathcal{O}(\log(r-r_0)) = 0.  
\ea
The requirement of regularity of $Y_4$ at the horizon means that each
of these divergent terms must vanish.  Using (\ref{w2shearsoln}), one
finds that the leading and sub-leading divergent terms vanish.  The
coefficient of $(r-r_0)^{-1}$ does not vanish on its own; requiring
that this term vanish we obtain an expression for $w_4$.  Again, employing
(\ref{w2shearsoln}), we find
\ba
	-i \frac{w_4}{w_2} = \tau_{\rm shear} &=& \frac{4 \pi T}{f_0} \left[\frac{Y_2'(r_0)}{Y_0(r_0)} 
	  - \frac{Y_0'(r_0)Y_2(r_0)}{Y_0(r_0)^2} \right] \nonumber \\
	&+& \frac{1}{4 \pi T} \frac{Y_0(r_0)}{Y_0'(r_0)} 
	\left[\frac{Y_0'(r_0)}{Y_0(r_0)} + \frac{2f_1}{f_0} - \frac{Y_0''(r_0)}{Y_0'(r_0)}\right].  
\ea 
This expression can be simplified somewhat by recalling a special
property of our metric (\ref{F0Fx}).  Using the explicit expressions
for $F_t$ and $F_x$ given in the appendix, we see that our backgrounds
have the property that $\rootg g^{rr} \propto f/f'$.  Thus,
\be
	Y_0'(r) \propto f'(r).  
\ee  
Using this fact, and the expansion for $f$ (\ref{fexpansion})  we see that 
\be
	\frac{Y_0''(r_0)}{Y_0'(r_0)} = \frac{2 f_1}{f_0},
\ee
which eliminates the last two terms in the expression for $\tau_{\rm
shear}$ given above.  We can then write the expression for $\tau_{\rm
shear}$ rather succinctly as
\be
	\tau_{\rm shear} = \frac{4 \pi T}{f_0} 
	\lim_{r \to r_0}
	\partial_r \left[ \frac{Y_2(r)}{Y_0(r)} \right] 
	+ \frac{1}{4\pi T}.
\label{mastertaueqn}  
\ee  
All that remains is to get an explicit expression for $Y_2(r)$ in
order to express $\tau_{\rm shear}$ in terms of the metric only.

Returning to (\ref{Y2sheareqn}), one can solve for $Y_2/Y_0$:
\be
	\frac{Y_2}{Y_0} = \frac{a_2}{a_1} + \xi \log[f] 
	- \frac{1}{Y_0}\int_\infty^r Y_0' \left( \frac{w_2^2}{f} + \frac{1}{a_1} 
	\int_\infty^{r'} \sqrt{-g}g^{tt} Y_0 dr' \right)\, dr. 
\ee
It turns out to be useful to integrate the term involving the nested integrals by parts.  With this
simplification, we have
\ba
	\frac{Y_2}{Y_0} &=& \frac{a_2}{a_1} + \xi \log[f] -\frac{1}{a_1} \int_\infty^r \rootg g^{tt} Y_0 dr' \nonumber\\
	&+& \frac{1}{Y_0} \int_\infty^r \left( \frac{1}{a_1} \rootg g^{tt} Y_0^2 - w_2^2 \frac{Y_0'}{f} \right) dr'.	
\ea
Taking the derivative,
\be
	\partial_r \left[ \frac{Y_2}{Y_0} \right] = \xi \DL[f] - \frac{w_2^2 Y_0'}{f Y_0} 
	+ \frac{Y_0'}{Y_0^2} \int_\infty^r \left( \frac{Y_0' w_2^2}{f} - \frac{\rootg g^{tt} Y_0^2}{a_1} \right)dr', 
\ee
and using the definitions of $Y_0$ (\ref{Y0shearsoln}), and $w_2$ (\ref{w2shearsoln}) this becomes
\ba
	\partial_r \left[ \frac{Y_2}{Y_0} \right] &=& \xi \DL[f] - \frac{w_2^2 Y_0'}{f Y_0} \\
	&+& a_1 \frac{Y_0'(r)}{Y_0(r)^2} \left( \frac{ i f_0 Y_0(r_0)}{4 \pi T Y_0'(r_0)} \right)^2  
	\int_\infty^r \frac{g_{rr}}{\rootg} \left[1 + \left(\frac{Y_0 \rootg}{\sqrt{-g_{tt}g_{rr}} a_1 w_2}\right)^2 \right]dr'.\nonumber 
\ea
Defining the quantity
\be
	D(r) \equiv  -\frac{1}{a_1} \frac{\sqrt{-g(r)}}{\sqrt{-g_{tt}(r) g_{rr}(r)}} Y_0(r) 
	= \frac{\sqrt{-g(r)}}{\sqrt{-g_{tt}(r) g_{rr}(r)}} \int^\infty_r \frac{f(r')}{\sqrt{-g(r')} g^{rr}(r')} dr',  
	\label{Ddef}
\ee
and noting that
\be
	w_2 = -i D(r_0), 
\ee
we have
\ba
	\partial_r \left[ \frac{Y_2}{Y_0} \right] &=& \xi \DL[f] - \frac{w_2^2 Y_0'}{f Y_0} \nonumber \\
	&-& a_1 \frac{Y_0'(r)}{Y_0(r)^2} \left( \frac{ f_0 Y_0(r_0)}{4 \pi T Y_0'(r_0)} \right)^2  
	\int_\infty^r \frac{g_{rr}}{\rootg} \left(1 - \frac{D(r')^2}{D(r_0)^2} \right) dr'. 
\ea
The first two terms of this expression can be simplified by again invoking the special
property of our background that $F_t(r) = F_x(r)$ which means that $\rootg g^{rr} \propto f'/f$ so that  
(\ref{Y0shearsoln}) gives $Y_0(r) \propto 1-f(r)$.  Using this fact, and 
the universal fact that $w_2 = -i/4 \pi T$ for all Einstein gravity duals \cite{Iqbal:2008by, Buchel:2003tz}, we 
have
\be
	\xi \DL[f] - \frac{w_2^2 Y_0'}{f Y_0} = \left( \frac{1}{4\pi T} \right)^2 \frac{f'(r)}{f(r)-1}.  
\ee
Then, with the use of (\ref{fexpansion}), we have
\ba
	\lim_{r \to r_0} \partial_r \left[ \frac{Y_2}{Y_0} \right] &=& -\frac{f_0}{(4 \pi T)^2}
	\nonumber \\
	&+& a_1 \frac{1}{Y_0'(r_0)} \left( \frac{ f_0}{4 \pi T } \right)^2  
	\int^\infty_{r_0} \frac{g_{rr}}{\rootg} \left(1 - \frac{D(r')^2}{D(r_0)^2} \right) dr'. 
\ea
Inserting this into (\ref{mastertaueqn}) gives
\be
	\tau_{\rm shear} = a_1 \frac{1}{Y_0'(r_0)} \frac{f_0}{4 \pi T }  
	\int^\infty_{r_0} \frac{g_{rr}}{\rootg} \left(1 - \frac{D(r')^2}{D(r_0)^2} \right) dr'.
\ee
As a final step, insert the definition of $Y_0'$ (\ref{Y0shearsoln})
and substitute in for the temperature using (\ref{hawkingtemp}).
This precisely determines the quantity $\tau_{\rm shear}$ which thus completes the shear dispersion relation
up to order $q^4$,
\be
\tau_{\rm shear} = \frac{\sqrt{-g(r_0)}}{\sqrt{\gamma_0 \gamma_r}}
\int_{r_0}^\infty dr \frac{g_{rr}(r)}{\sqrt{-g(r)}}
\left[ 1- \left( \frac{D(r)}{D(r_0)} \right)^2 \right],
\label{taudef}
\ee
where $D(r)$ is defined as in (\ref{Ddef}).

The formula given in Eq. (\ref{taudef}) is the main result of this
chapter.  This is the same formula that was derived in
\cite{Kapusta:2008ng} within the context of the black hole membrane
paradigm.  Here, we have shown that the formula can also be derived
using AdS/CFT and the method of gauge invariant equations.  This
further illustrates the agreement between these two methods, at least
in the case of zero chemical potential.  For a more in depth
discussion of this agreement (which is not fully understood at
present) see \cite{ Iqbal:2008by, Kovtun:2003wp, Starinets:2008fb}.

\section{Applications}
\begin{figure}[h]
\centering
\includegraphics[width=1.0\textwidth]{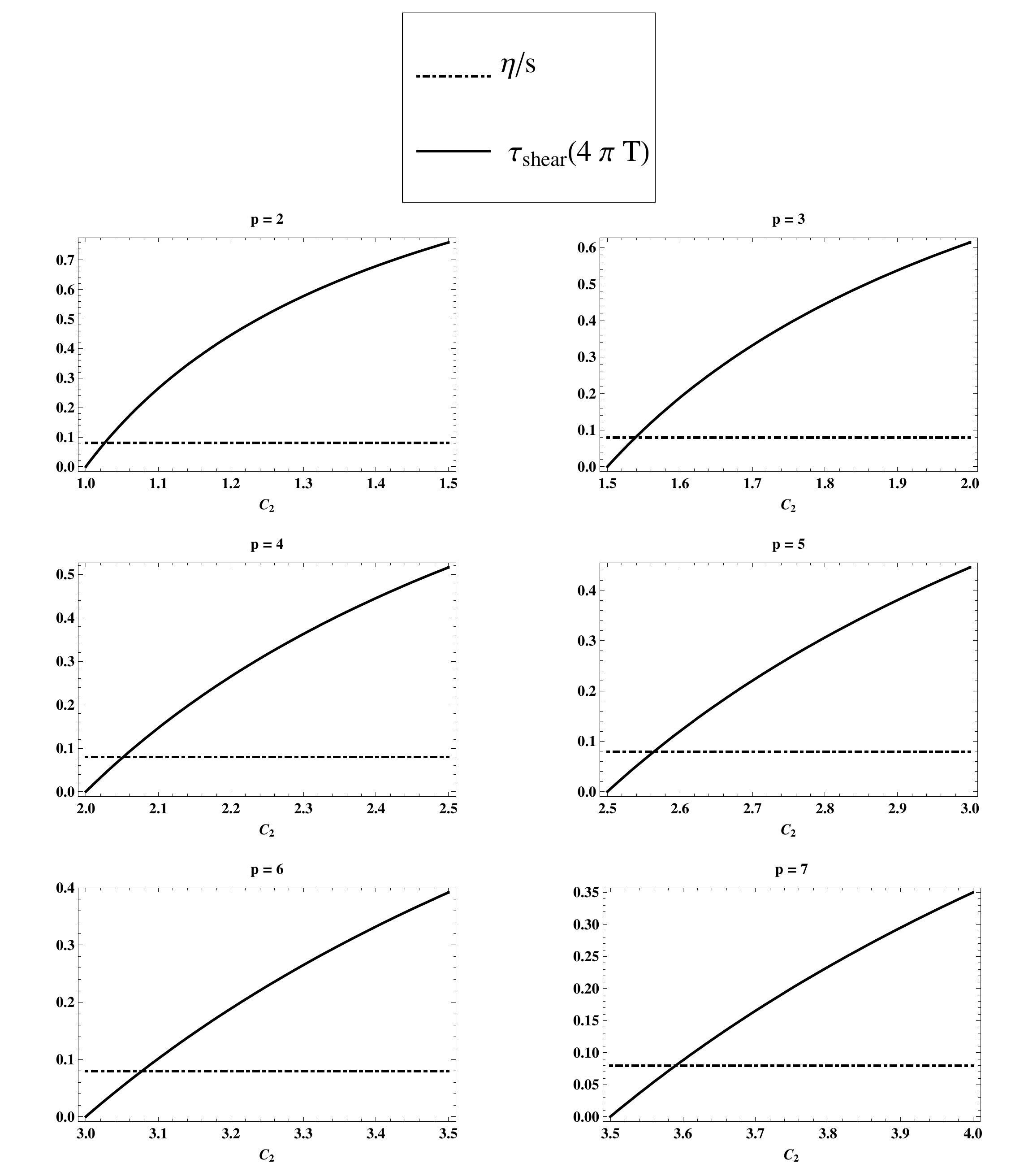}
\caption{
Plots of the dimensionless quantities: $\eta / s$
 and $ 4 \pi T \tau_{\rm shear}$, as a function of
 the free parameter $c_2$ appearing in the metric (\ref{specialf}).
 Each plot corresponds to a different number of spatial dimensions
 $p$.  This range of values of $c_2$ was chosen because outside this
 particular range, unphysical behavior is found in the sound mode.
 This will be explained in later chapters (cf.
 (\ref{vssoln}),(\ref{zetasoln})).  The metric is conformal at the upper end of the range
 of $c_2$ when $c_2 = (p+1)/2$.  Note the universality of the shear
 viscosity to entropy density ratio is evident, while the value of $\tau_{\rm shear}$
 varies with both $c_2$ and $p$.}
\label{shearfigure}
\end{figure}

Let us consider a few simple applications of this formula.  
All metrics generated by $r$ dependent scalar fields will have the 
aforementioned property that $F_t = F_x$, and thus $\rootg g^{rr} \propto f/f'$.  
In addition to this ``automatic'' constraint, let us examine metrics 
which satisfy
\be
	f(r) = 1 - \left( \frac{ g_{xx}(r_0)}{g_{xx}(r)} \right)^{c_2}.
	\label{specialf}
\ee
Here, $c_2$ is a constant.  We also assume that the metric is
asymptotically $AdS$ so that $f(r) \rightarrow 1$ as $r \rightarrow
\infty$.  We will explain the motivation behind this constraint in
subsequent chapters; for the moment, imposing the constraint simply
allows us to get analytical solutions for a special class of metrics.
Using (\ref{taudef}) for this special class of metrics we have
\be
	\tau_{\rm shear} = \frac{1}{4 \pi T} H_n \left(2 - \frac{p}{c_2}\right),
	\label{taushearspecialmetric}
\ee
where $H_n(\alpha)$ is the ``Harmonic Number'' defined as
\be
	H_n(\alpha) \equiv \int_0^1 \frac{1-x^{\alpha}}{1-x} dx.
\label{harmonicnumberdef}
\ee 
The results for $\eta / s$ and $\tau_{\rm shear}$ for this special
class of metrics is represented graphically in Fig. \ref{shearfigure}.

Two notable special metrics are the \emph{Schwarzschild anti-de
Sitter black hole}, and the \emph{Dp-brane metric}.  The first of these
is a black brane type metric which is a solution to Einstein's
equations with a negative cosmological constant.  This ($p+2$
dimensional) metric can be written
\ba
	ds^2_{SAdS} &=& \frac{r^2}{L^2} \left[ -f(r) dt^2 + dx_j dx^j \right] + \frac{L^2 dr^2}{r^2 f(r)}\\
	f(r) &=& 1 - \left(\frac{r_0}{r} \right)^{p+1}.
	\label{SAdSmetric}
\ea
Here, $L$ is the usual $AdS$ curvature radius.  Clearly, this fits the
constraint (\ref{specialf}) with $c_2 = (p+1)/2$, thus for this metric
we have
\be
	\tau_{\rm shear}^{SAdS} = \frac{1}{4 \pi T} H_n \left(\frac{2}{p+1} \right).  
	\label{taushearSAdS}
\ee
As a second example, the Dp-brane metric arises from a stack of $N$ Dp-branes in string
theory with $N \gg 1$.  The metric exists in 10 dimensions, but can be
reduced to an effective (p+2 dimensional) metric using dimensional
reduction \cite{Kovtun:2003wp, Mas:2007ng}.  The effective metric is
a solution to Einstein's equations where the metric is supported by a
single scalar field, and the scalar potential is a single exponential
\be
	U(\phi)=-\frac{(7-p)(9-p)}{2L^2}e^{-\sqrt{\frac{2(3-p)^2}{p(9-p)}} \phi}.
\ee
The metric is:
\ba
	ds^2 &=& \left(\frac{r}{L}\right)^\frac{9-p}{p}\left[-f(r) dt^2+dx_jdx^j\right] 
	+ \left( \frac{r}{L} \right)^{\frac{p^2 - 8p +9}{p}}\frac{dr^2}{f(r)}, \\
	f(r) &=& 1-\left(\frac{r_0}{r}\right)^{7-p}.  
	\label{DPmetric}
\ea
This fits our special class of metrics with $c_2 = (7p-p^2) /(9-p)$.  (Note that
$p \leq 7$, or else the metric is not asymptotically anti-de Sitter.)
 Applying the
formula (\ref{taudef}) we have
\be
	\tau_{\rm shear}^{DP} = \frac{1}{4 \pi T} H_n \left(\frac{5-p}{7-p} \right).  
	\label{taushearDP}
\ee 
The results (\ref{taushearSAdS}) and (\ref{taushearDP}) were first
presented by Natsuume in \cite{Natsuume:2008gy}.  A few comments are
in order about these applications.  First, before the formula
(\ref{taudef}) was presented in \cite{Kapusta:2008ng}, the quantity
$\tau_{\rm shear}$ had been computed in several special cases, most
notably in \cite{Natsuume:2007ty, Baier:2007ix}.  The applications
above (\ref{taushearSAdS}) and (\ref{taushearDP}) reproduce all known
results in the literature, which is another check on our calculations.
In addition, the formulas above give new information.  A particularly
curious feature is that $\tau_{\rm shear}^{Dp} = 0$ for $p = 5$, and
$\tau_{\rm shear}^{Dp} \rightarrow \infty $ for $p>5$.  It has been
known that the Dp-brane metric exhibits unusual features for $p>4$ in
the sound channel.  (For example, for $p=5$, the speed of sound
vanishes, while for $p>5$ the speed of sound becomes imaginary
\cite{Mas:2007ng}).  The analysis given here shows that these features
are also present for the shear mode.

Furthermore, once we have completed the analysis of the sound mode,
the comparison with the formula (\ref{taushearspecialmetric}) gives us
some information about the third order hydrodynamic contribution to
the shear mode as mentioned in Sec. \ref{secondorderhydro_section}.
This will be explained further in the next chapter.

Finally, perhaps the most obvious feature of these applications is
that the quantity $\tau_{\rm shear}$ is not universal in the same way
as the shear viscosity; application of the formula to different
metrics yields different results.

\section{Summary}

In this chapter, we have applied the prescription which was outlined in
Sec. \ref{prescription_section} to shear perturbations of a black
brane background supported by scalar fields.  We have thus been able
to determine the shear mode dispersion relation $w(q)$ up to order
$q^4$.  This dispersion relation is parametrized in
(\ref{sheardispersion}) with two unknown quantities which we denote $D_{\eta}$
and $\tau_{\rm shear}$.  The formulas (\ref{Ddef}) and (\ref{taudef})
are the main results of this chapter, and they allow one to compute
these unknown quantities if the background metric is given.

The formula for $D_{\eta}$, which was first derived in
\cite{Kovtun:2003wp}, allows one to compute the shear viscosity, since
it is the only hydrodynamic transport coefficient which appears to
lowest order in the shear dispersion relation.  Application of the
formula for $D_{\eta}$ to any Einstein gravity dual (in the limit of
infinite coupling) yields $\eta / s = 1/4\pi$, and is one of the main
inspirations for the conjectured shear viscosity bound.  The quantity
$\tau_{\rm shear}$ does not display the same universal behavior.

It is important to make a distinction between the dispersion relation
and the hydrodynamic transport coefficients which would appear in the
derivative expansion of the energy momentum tensor.  To lowest order,
the shear dispersion relation contains only the shear viscosity, but at the
next order ($q^4$), the quantity which we call $\tau_{\rm shear}$
contains more than one transport coefficient.  (In fact, it contains
contributions from third order hydrodynamics as explained in section
\ref{secondorderhydro_section}).  At this point, all we can state is
that the combination of transport coefficients which comprise
$\tau_{\rm shear}$ is not universal in the same way as the shear
viscosity.  If and when a consistent theory of third order hydrodynamics
is formulated, one could revisit these results to see whether additional
information about these new transport coefficients can be gained from 
the formula (\ref{taudef}).  

\chapter{Sound Mode}
\label{SoundMode_chapter}
\section{Introduction}

In this chapter, we will perform an analysis similar to that of the
previous chapter within the context of a different
hydrodynamic mode, the sound mode.  In general, the prescription
detailed in Sec. \ref{prescription_section} will be applied in
exactly the same way, but analysis is more complicated here
due to the fact that the matter perturbations are coupled with the
metric perturbations.

In \cite{Kovtun:2003wp}, a formula for the shear viscosity to entropy
density ratio $\eta/s$ was derived which is applicable to a wide
variety of gravity duals.  Application of this formula to known
Einstein gravity duals always resulted in $\eta/s =1/4 \pi$ (in the
large N/infinite t'Hooft coupling limit), and eventually led to the
celebrated viscosity bound conjecture that $\eta/s \geq 1/4 \pi$ for
all physical substances.  Clearly, a similar general formula for the
bulk viscosity would also be desirable in the hopes that other
universal behavior might be discovered.  Such a formula would also be
useful when comparing the predictions of a phenomenological gravity
dual to the lattice data \cite{Kharzeev:2007wb, Karsch:2007jc}.

The work detailed in this chapter is partially motivated by the goal
of obtaining such a formula, though such a formula has not yet been
developed.  This work should be viewed as a first step
towards this goal.  Here, we consider sound mode fluctuations of a
dual gravity theory supported by a single scalar field, and then
generalize to the case of multiple scalar fields.  We attempt to be as
general as possible by not specifying the background profiles of the
scalar field(s) except in certain special cases.

Recent works \cite{Gubser:2008yx,Gubser:2008sz} also explore the
sound mode in phenomenologically motivated single scalar models.  The
results presented here are complementary to these papers, though we allow
for the possibility of multiple scalar fields, and also do not
restrict ourselves to five dimensions.

The main results of this chapter are twofold.  First, we derive a
set of sound mode gauge invariant equations which are applicable to any Einstein gravity
dual which is supported by $r$ dependent scalar fields.  These
equations are, in general, complicated but could be solved
numerically to determine the sound mode dispersion relation for
certain phenomenological models of the quark-gluon plasma.  Second, we
solve these equations analytically for the sound mode dispersion
relation (up to order $q^3$) for a certain special class of metrics.

\section{Background field and potential}
We again work with the same set up as in the previous chapter.  We
will examine sound mode perturbations of a general black brane metric
(\ref{blackbranemetric}) with presumed near horizon behavior
(\ref{ttNH} - \ref{xxNH}).  We assume that the matter supporting the
metric is scalar in nature, so that the action, background equations,
and energy momentum tensor are given respectively by
(\ref{scalaraction}),(\ref{scalarbackgroundequations}), and
(\ref{scalarTuv}).

Because the sound mode perturbations transform like scalars under
rotations (see Sec. \ref{hydromodes_section}), the perturbations of
the scalar fields will be coupled to the metric perturbations.  For
this reason, the sound mode requires more information about the matter
perturbations than the shear mode.  In particular, we will need 
information about the scalar potential. 

The background equations of motion give a relationship 
between the scalar field profile and the potential.
\be 
	\label{Uprime}
	\frac{\partial U(\phi_i(r))}{\partial \phi_i} = \Box^{(0)} \phi_{i}(r) = \frac{1}{\rootg} \partial_r \left[\rootg g^{rr} \phi_i'(r) \right].
\ee
As usual, the prime denotes derivative with respect to $r$, and $\Box^{(0)}
\equiv \nabla^{(0)}_\mu \nabla_{(0)}^\mu$.  The covariant derivatives
are defined in Appendix \ref{app_EinsteinEqns}.

\section{Perturbations and linearized equations}

We now introduce fluctuations of the fields on this background
$g_{\mu \nu} \rightarrow g_{\mu \nu} + h_{\mu \nu}$ and $\phi
\rightarrow \phi + \delta \phi$, and assume the usual time dependence
\ba
	h_{\mu \nu}(t,z,r) &=& e^{i(qz - wt)} h_{\mu \nu}(r), \\
	\delta \phi(t,z,r) &=& e^{i(qz - wt)} \delta \phi(r). 
\ea

The linearized Einstein equations for the sound mode and an arbitrary
set of matter fields were derived in
Sec. \ref{soundmodeeqns_subsection}.  In order to apply these
equations to the case at hand, we must evaluate the perturbed energy
momentum tensor for the case of the scalar fields.  The energy
momentum tensor was expanded to first order in the perturbation in
(\ref{perturbedTuv}).  For the sound mode, we require the following
components of the energy momentum tensor: $T^{tt},T^{aa}, T^{zz},
T^{tz}, T^{rz}, T^{tr}, T^{rr}$.  As in previous chapters, we use the
subscripts (0) and (1) to distinguish background quantities from those
that are $\mathcal{O}(h_{\mu \nu})$ or $\mathcal{O}(\delta \phi)$.  A
summation over the repeated index $k$, which labels the particular
scalar field, is implied.

First, let us examine $T_{(1)}^{mn}$ with $m, n \neq r$.  Using (\ref{perturbedTuv}), 
we have
\be
	8 \pi G_{p+2} T^{m n}_{(1)} = \frac{1}{2} \left( h^{m n} \mathcal{L}_{\phi} - g^{m n} \delta \mathcal{L}_{\phi}\right).\ee
Noting that
\be
	16 \pi G_{p+2} T^{mn}_{(0)} = -g^{m n} \mathcal{L}_{\phi},
\ee
and recalling the definitions of $A$,$B$,$C$, and $D$ (\ref{Adef})-(\ref{Ddefinition}) we have
\ba
	16 \pi G_{p+2} \left(T^{tt}_{(1)} + T^{tt}_{(0)}A \right) &=& -g^{tt} \delta \mathcal{L}_{\phi}, \\
	16 \pi G_{p+2}  \left(\frac{1}{p-1}\sum_{a=1}^{p-1} T^{aa}_{(1)} + T^{zz}_{(0)} B \right) &=& -g^{xx} \delta \mathcal{L}_{\phi}, \\
	16 \pi G_{p+2} \left(T^{zz}_{(1)} + T^{zz}_{(0)}C \right) &=& -g^{xx} \delta \mathcal{L}_{\phi}, \\
	16 \pi G_{p+2} \left(T^{tz}_{(1)} + T^{zz}_{(0)}D \right) &=& 0.
\ea
The right hand side is straightforward to evaluate
\ba
	\delta \mathcal{L}_{\phi} &=& \thalf g^{\alpha \beta} \left[ \partial_{\alpha} (\delta \phi_k) \partial_\beta \phi_k +
	  \partial_{\beta} (\delta \phi_k) \partial_\alpha \phi_k \right] + \delta U \nonumber \\
	&=& \left(g^{rr} (\delta \phi_k)' \phi_k' + \frac{\partial U}{\partial \phi_k} \delta \phi_k \right).
\ea
Secondly, examine $T_{(1)}^{mr}$: 
\be
	8 \pi G_{p+2} T^{m r}_{(1)} 
	= \frac{1}{2} g^{m m} g^{rr} \partial_m (\delta \phi_k) \partial_r \phi_k, 
\ee   
This gives the relations
\ba
	16 \pi G_{p+2} T^{t r}_{(1)} &=& (-i w) g^{tt} g^{rr} (\delta \phi_k) \phi_k' ,\\ 
	16 \pi G_{p+2} T^{z r}_{(1)} &=& (i q) g^{xx} g^{rr} (\delta \phi_k) \phi_k'. 
\ea
Finally, the only remaining component to evaluate is $T^{rr}_{(1)}$, evaluating (\ref{perturbedTuv}) gives
\be
	8 \pi G_{p+2} T^{r r}_{(1)} 
	= \frac{1}{2} \left[
	2 \left(g^{r r}\right)^2 \partial_r (\delta \phi_k) \partial_r \phi_k - g^{r r} \delta \mathcal{L}_{\phi} \right],
\ee
which simplifies to 
\be
	16 \pi G_{p+2} T^{r r}_{(1)} 
	= 
	(g^{r r})^2 (\delta \phi_k)' \phi_k' 
	- g^{r r}\frac{\partial U}{\partial \phi_k} \delta \phi_k.
\ee
The insertion of these relations for the energy momentum tensor into the
general sound mode equations (\ref{g00eqn} - \ref{grzeqn}) results in
the following set of seven linearized Einstein equations.  Here we have inserted a line
above each equation that allows the reader to see which Einstein equation is the source of the 
following differential equation.
\begin{center}$G^{t\,(1)}_{\, t} = -8 \pi G_{p+2} T^{t\, (1)}_{\, t}$
\ba
	\label{g00scalareqn}
	\frac{g_{rr}}{\rootg} \sqrt{f} \partial_r \left[\frac{\rootg g^{rr}}{\sqrt{f}} ((p-1)B'+C') \right] 
	&+&g_{rr}\frac{\partial U}{\partial \phi_k}(\delta \phi_k) \\
	&+& \phi_k'(\delta \phi_k)'-(p-1) q^2 g_{rr}g^{xx} B=0, \nonumber  
\ea
$\displaystyle\sum_{i=1}^{p-1}G^{i\,(1)}_{\,i} = \displaystyle\sum_{i=1}^{p-1}\left( -8 \pi G_{p+2} T^{i\,(1)}_{\,i}\right)$
\ba	
	\label{gxxscalareqn}
	\frac{g_{rr}}{\sqrt{-g f}}\partial_r \left[\sqrt{-g f} g^{rr} A' \right] 
	&+& \frac{g_{rr}}{\rootg}\partial_r \left[\rootg g^{rr}\left((p-2)B'+C'\right) \right]
	+g_{rr}\frac{\partial U}{\partial \phi_k}(\delta \phi_k) \nonumber \\
	+ \, \phi_k'(\delta \phi_k)'&-& g_{rr}g^{xx}\left[q^2\left(A+(p-2)B\right) - \frac{w^2}{f} \left((p-2)B+C\right)+2qwD\right] \nonumber\\
	&=&0\, 
\ea
$G^{z\,(1)}_{\,z} = - 8 \pi G_{p+2} T^{z\,(1)}_{\,z} $
\ba
	\label{gzzscalareqn}
	\frac{g_{rr}}{\sqrt{-g f}}\partial_r \left[\sqrt{-g f} g^{rr} A' \right] 
	+ \frac{g_{rr}}{\rootg}\partial_r \left[\rootg g^{rr}(p-1)B' \right] 
	&+&g_{rr}\frac{\partial U}{\partial \phi_k}(\delta \phi_k) + \phi_k'(\delta \phi_k)'
	\nonumber \\
	&-& (p-1) w^2g_{rr}g^{tt}B =0, \nonumber \\
	\,
\ea
$G^{z\,(1)}_{\,t} = -8 \pi G_{p+2} T^{z\,(1)}_{\,t} $
\ba
	\label{g0zscalareqn}
	 \frac{\rootg (g^{xx})^{p+1}}{f} \partial_r \left[ \frac{(g_{xx})^{p+1}}{\rootg} \partial_r \left(f D \right)\right] 
	-2g_{rr}(F_t - F_x)D + q w (p-1) g_{rr}g^{tt}B =0, \nonumber \\
	\,  
\ea
$G^{r\,(1)}_{\,r} = -8 \pi G_{p+2} T^{r\,(1)}_{\,r} $
\ba
	\label{grrscalareqn}
	\DL \left[(g_{xx})^p\right]A' + \DL \left[g_{tt}(g_{xx})^{p-1}\right]((p-1)B'+C')
	-2\phi_k'(\delta \phi_k)'+2g_{rr}\frac{\partial U}{\partial \phi_k}(\delta \phi_k) &\,& \\
	-2g_{rr}\left(w^2g^{tt}((p-1)B+C) + q^2 g^{xx}(A+(p-1)B) +2qwg^{xx}D\right) &=&0, \nonumber \\
	\nonumber
\ea
$G^{t\,(1)}_{\,r} = -8 \pi G_{p+2} T^{t\,(1)}_{\,r}$ 
\ba
	\label{g0rscalareqn}
	w \sqrt{f} \partial_r\left[ \frac{1}{\sqrt{f}} ((p-1)B+C) \right] -q f D' + w\phi_k'(\delta \phi_k) =0,\\
	\nonumber 
\ea
$G^{z\,(1)}_{\,r} = - 8 \pi G_{p+2} T^{z\,(1)}_{\,r}$
\ba
	\label{grzscalareqn}
	\frac{q}{\sqrt{f}}\partial_r \left[\sqrt{f} A \right] + \frac{w}{f} \partial_r \left[f D\right]+ q (p-1)B'+ q \phi_k'(\delta \phi_k) =0.\\
	\nonumber 
\ea
\end{center}
In addition to these equations, one must also examine perturbations of
the background scalar equations (\ref{Uprime}).  Using the formulas
given in the Appendix, especially (\ref{deltabox}) the equation
for each scalar field (in the radial gauge, as usual) is 
\begin{center}
$\delta \left(\Box^{(0)} \phi_i \right) = \delta \left(\frac{\partial U}{\partial \phi_i} \right)$
\ba
	\label{scalareqn}
	\frac{g_{rr}}{\rootg} \partial_r \left[\rootg g^{rr} (\delta \phi_i)' \right] + \thalf \phi_i'H'
	-g_{rr} \left(w^2 g^{tt} + q^2 g^{xx}
	\right)(\delta \phi_i) 
	=   g_{rr} \frac{\partial^2 U}{\partial \phi_i \partial \phi_k} (\delta \phi_k).
\ea
\end{center}
Again, a summation over $k$ is implied and we have defined
\be
	H(r)  \equiv   A(r) + (p-1)B(r) + C(r). 
\ee

\section{Gauge Invariant Equations}
There is still residual gauge freedom in these equations under the infinitesimal diffeomorphism 
(\ref{gaugetransformation}).  The scalar field perturbations also transform under a gauge transformation.
Such a gauge transformation is essentially the freedom to choose ones coordinates.  Imagine a small coordinate shift
\be
	x^{\mu} \rightarrow x^{\mu} - \xi^{\mu}.
\ee
Then, the background scalar field will be altered
\be
	\phi(x^{\mu}) \rightarrow \phi(x^{\mu}) - \xi^{\mu} \partial_\mu \phi(x^{\mu}).  
\ee
The latter term appears to be a perturbation, but it is not a physical perturbation; it only appears
due to a change in coordinates.  Thus, under such a transformation, the scalar field perturbations
behave as
\be
	\delta \phi_i \rightarrow \delta \phi_i -  \xi^{\mu} \partial_\mu \phi_i = \delta \phi_i -  \xi^{r}  \phi_i'.  
\label{phitransformation}
\ee
Let us now examine how the different sound mode perturbations transform under this gauge transformation.
Recall  
\be
	 \nabla^{(0)}_\mu \xi_\nu + \nabla^{(0)}_\nu \xi_\mu = \partial_\mu \xi_\nu + \partial_\nu \xi_\mu 
	 - 2\Gamma^{\lambda}_{\mu \nu} \xi_\lambda,  
\ee
the Christoffel symbols given in Appendix \ref{app_EinsteinEqns}, and
the presumed form of the vector $\xi$ (\ref{xiansatz}).  Then, it is
straightforward to evaluate
\ba
	h_{tt} &\rightarrow& h_{tt} - 2 \partial_t \xi_t +2 \Gamma_{tt}^{\lambda} \xi_{\lambda} = h_{tt} + 2 i w \xi_t - g^{rr} g_{tt}' \xi_{r},  	\\
	A &\rightarrow& A + 2 i w g^{tt} \xi_t - g^{rr} \DL[g_{tt}] \xi_r.
\label{Atransformation}
\ea
In a similar manner, 
\ba
	B &\rightarrow& B - g^{rr} \DL[g_{xx}] \xi_r, \\
	C &\rightarrow& C - 2 i q g^{xx} \xi_z - g^{rr} \DL[g_{xx}] \xi_r, \\
	D &\rightarrow& D + i w g^{tt} \xi_z - i q g^{tt} \xi_t.
\label{BCDtransformation}
\ea

Given the transformation rules (\ref{phitransformation}),
(\ref{Atransformation}), and (\ref{BCDtransformation}), it is easy to
verify that the following gauge invariant combinations first presented in
\cite{Mas:2007ng} transform only into themselves under such a
diffeomorphism  (i.e. $Z_0 \rightarrow Z_0$ and $Z_\phi \rightarrow
Z_\phi$).  The choice of these gauge invariant variables is not unique.  
\be
	Z_0(r) = -f(r) \left[q^2A(r) + 2qw D(r)\right] + w^2 C(r) 
	- \left[ q^2 \frac{g_{tt}'(r)}{g_{xx}'(r)}+w^2 \right]B(r) 
\ee
\be
	Z_{\phi i}(r) = \delta \phi_i(r) - \frac{\phi_i'(r)}{\DL\left[g_{xx}(r)\right]}B(r)
\ee
By taking an appropriate combination of the equations
(\ref{g00scalareqn}-\ref{scalareqn}), one can arrive at coupled second
order equations for the gauge invariant variables.  Because of the
complexity of the sound mode equations and the gauge invariant
variables, reducing the linearized perturbation equations down to a
set of gauge invariant equations is a very difficult task.  The equations
presented below were arrived at by ``brute force'', by eliminating the 
variables $A,C,D, \delta\phi$ one at a time, and then by showing that the remaining
coefficient of $B$ vanishes by the background equations of motion.  

\subsection{Equation for $Z_0$}

First, define the quantity
\be
	\alpha(r) \equiv q^2 \left((p-1)+ \frac{\DL [g_{tt}(r)]}{\DL[g_{xx}(r)]}\right) - \frac{p w^2}{f(r)}.
\label{alphadef}
\ee
Consider the following combination of the Einstein equations
(\ref{g00scalareqn} - \ref{scalareqn}):
\ba
	&f& \left\{ \frac{\alpha}{p} \left[(\ref{g00scalareqn}) - (\ref{gxxscalareqn}) + (\ref{gzzscalareqn}) + \thalf (\ref{grrscalareqn}) \right] 
	  + \left[(\ref{g00scalareqn}) + \thalf (\ref{grrscalareqn})\right] \left(\frac{w^2}{f}\right) - 2 q w (\ref{g0zscalareqn}) 
	  \right. \nonumber \\
	  &-& q^2 \Biggl. \left[\thalf (\ref{grrscalareqn})+(\ref{gzzscalareqn}) \right] \Biggr\}
	 + 2 \left \{ q f \DL[\alpha] (\ref{grzscalareqn}) - w \DL[\alpha f ] (\ref{g0rscalareqn}) \right\}. \nonumber \\
	 \,
\ea
\emph{For simplicity, for the moment
we assume there is only one scalar field.}
After a long calculation, one can show that this reduces to
\be
	\mathcal{Z}_1 - 2 g_{rr}(F_t - F_x)\left(w^2(B-C)+Z_0\right) - B q^2 f \Delta_1 = 0,
\ee
where 
\ba
	\mathcal{Z}_1 \equiv \frac{g_{rr}}{\rootg} \alpha^2 f^2 \partial_r \left[\frac{\rootg g^{rr}}{\alpha^2 f^2}Z_0'\right] 
	+Z_0 \left( \DL[f]\DL[f\alpha] - g_{rr}\left(w^2 g^{tt} + q^2 g^{xx}\right)\right) &+& \nonumber \\
	2Z_{\phi} \phi' f \left( \alpha \partial_r \left[ \frac{1}{\alpha} \left(\frac{w^2}{f} -q^2 \right) \right] 
	+ \frac{q^2 \DL[f]}{p \DL[g_{xx}]} \DL \left[ \rootg g^{rr} \phi' \right] \right),
\ea
and 
\ba
	\Delta_1 &\equiv& 2 \DL[g_{xx}]\alpha^2 \partial_r \left[ \frac{g_{rr}}{\alpha^2 \DL[g_{xx}]^2}(F_t - F_x) \right] 
	- \frac{4 g_{rr}}{p \DL[g_{xx}]^2}(F_t - F_x) (\phi')^2 \nonumber \\  
	&+& \alpha^2 \DL[f] \partial_r \left[\frac{1}{p \alpha^2 \DL[g_{xx}]^2} \left(2 g_{rr}(F_t - F_r) - (\phi')^2\right) \right].
\ea
Here, we have been able to eliminate the potential using
(\ref{Uprime}), since it only appears in these equations via
$\partial U / \partial \phi$.  Clearly, $\Delta_1$ vanishes by the
background equations of motion (\ref{F0Fx}) and (\ref{phikeqn}),
leaving a differential equation involving only the gauge invariant
variables, namely $\mathcal{Z}_1 = 0$,
\ba
	\label{z0eqn}
	\frac{g_{rr}}{\rootg} \alpha^2 f^2 \partial_r \left[\frac{\rootg g^{rr}}{\alpha^2 f^2}Z_0'\right] 
	+Z_0 \left( \DL[f]\DL[f\alpha] - g_{rr}\left(w^2 g^{tt} + q^2 g^{xx}\right)\right) &+& \\
	2Z_{\phi} \phi' f \left( \alpha \partial_r \left[ \frac{1}{\alpha} \left(\frac{w^2}{f} -q^2 \right) \right] 
	+ \frac{q^2 \DL[f]}{p \DL[g_{xx}]} \DL \left[ \rootg g^{rr} \phi' \right] \right)
	&=&0. \nonumber
\ea
In this special case of one scalar field, one can write these equations only in terms of the metric
components because of the relation (\ref{phikeqn}).  

The generalization to multiple scalar fields is straightforward.
Since all of the equations used to derive (\ref{z0eqn}) depend only on
the sum of all scalar fields, it is clear that all we need to do is
include a summation on the last term.  An explicit calculation (using
the same combination of background Einstein equations) reveals this is
correct.  The generalization is
\ba
	\label{multiscalarz0}
	\frac{g_{rr}}{\rootg} \alpha^2 f^2\partial_r \left[\frac{\rootg g^{rr}}{\alpha^2 f^2}Z_0'\right] 
	+ Z_0 \left( \DL[f]\DL[f\alpha] - g_{rr}\left(w^2g^{tt} + q^2g^{xx}\right)\right) &+& \nonumber \\
	\sum_{k=1}^n 
	\left\{ 
		2Z_{\phi k} \phi_k' f \left( \alpha \partial_r \left[ \frac{1}{\alpha} \left(\frac{w^2}{f} -q^2 \right) \right]
		+ \frac {q^2 \DL[f]}{p \DL[g_{xx}]} \DL \left[ \rootg g^{rr} \phi_k' \right] \right)
	\right\} &=&0
\ea

\subsection{Equations for $Z_{\phi i}$}
We now proceed to derive the other gauge invariant equations.  There will
be an additional gauge invariant equation for each background scalar field present.   
The relevant combination of equations is:
\ba
	(\ref{scalareqn})
	&+& \frac{\phi_i'}{p \DL[g_{xx}]}
	\left\{
		(\ref{gxxscalareqn}) - (\ref{gzzscalareqn}) - (\ref{g00scalareqn}) - \frac{1}{2} (\ref{grrscalareqn}) 
	\right\} \nonumber \\
	&+& \frac{2}{\alpha} \partial_r 
	\left[ 
	  \frac{\phi_i'}{\DL[g_{xx}]} \right] \left\{ \frac{w}{f}(\ref{g0rscalareqn}) - q (\ref{grzscalareqn}) 
	\right\}.  
\label{scalarcombo}
\ea
Another lengthy calculation reduces this combination to
\be
	\mathcal{Z}_2 - B \Delta_2 = 0,
\ee
where
\ba
	\mathcal{Z}_2 &\equiv& \frac{g_{rr}}{\rootg}  \partial_r \left[\rootg g^{rr}Z_{\phi i}'\right] 
	-Z_{\phi i} g_{rr}\left(w^2 g^{tt} + q^2 g^{xx}\right)\nonumber -g_{rr}\sum_{k=1}^n Z_{\phi k} \frac{\partial^2 U}{\partial \phi_i \partial \phi_k}\\
	&-& \frac{2\phi_i'}{p \DL [g_{xx}] \alpha} \left\{ \sum_{k=1}^n \left[ Z_{\phi k} \phi_k' \left(\alpha \DL \left[\rootg g^{rr} \phi_k' \right]
	  + p \DL \left[\frac{\phi_i'}{\DL [g_{xx}]}\right]\left(q^2-\frac{w^2}{f}\right) \right) \right] \right\} \nonumber \\
	&+& \frac{2}{\alpha \sqrt{f}} \partial_r \left[ \frac{\phi_i'}{\DL[g_{xx}]} \right] \partial_r \left[ \frac{Z_0}{\sqrt{f}} \right],
\ea
and 
\ba
	\Delta_2 
	&\equiv& \frac{g_{rr}}{\DL[g_{xx}]} \left\{ \sum_k \frac{\partial^2 U}{\partial \phi_i \partial \phi_k}\phi_k' 
	- \partial_r \left[\frac{1}{\rootg} \partial_r \left[\rootg g^{rr} \phi_i' \right] \right] \right \} 
	\nonumber \\
	&+& \frac{\phi_i'}{p \DL[g_{xx}]^2} \left(\frac{g_{rr}}{\rootg} \right)^2 
	\partial_r \left\{\left(\rootg g^{rr} \right)^2\left[\sum_k (\phi_k')^2 - 2g_{rr}(F_t - F_r)\right] \right\} \nonumber \\
	&+& \frac{2}{\alpha \DL[g_{xx}]} 
	\partial_r \left[ \frac{\phi_i'}{\DL[g_{xx}]}\right] \left(\sum_k (\phi_k')^2 - 2g_{rr}(F_t-F_r)\right)\left(q^2 - \frac{w^2}{f} \right) \nonumber \\
	&+& \frac{2 g_{rr}}{\DL[g_{xx}]}\left( \frac{2 q^2}{\alpha} \partial_r \left[ \frac{\phi_i'}{\DL[g_{xx}]}\right]+\phi_i' \right)(F_t - F_x). 
\ea
Again, $\Delta_2$ vanishes by the background equations of motion (\ref{phikeqn}), (\ref{F0Fx}), and (\ref{Uprime}).  Thus,
we are left with an equation only containing the gauge invariant variables, $\mathcal{Z}_2 = 0$:
\ba
	&\,& \frac{g_{rr}}{\rootg}  \partial_r \left[\rootg g^{rr}Z_{\phi i}'\right] 
	-Z_{\phi i} g_{rr}\left(w^2 g^{tt} + q^2 g^{xx}\right)\nonumber -g_{rr}\sum_{k=1}^n Z_{\phi k} \frac{\partial^2 U}{\partial \phi_i \partial \phi_k}\\
	&-& \frac{2\phi_i'}{p \DL [g_{xx}] \alpha} \left\{ \sum_{k=1}^n \left[ Z_{\phi k} \phi_k' \left(\alpha \DL \left[\rootg g^{rr} \phi_k' \right]
	   + \DL \left[\frac{\phi_i'}{\DL [g_{xx}]}\right]\left(q^2- \frac{w^2}{f} \right) \right) \right] \right\} \nonumber \\
	&+& \frac{2}{\alpha \sqrt{f}} \partial_r \left[ \frac{\phi_i'}{\DL[g_{xx}]} \right] \partial_r \left[ \frac{Z_0}{\sqrt{f}} \right] = 0.
\label{multiscalarzphi}
\ea

When only one scalar field is present, (\ref{multiscalarzphi}) simplifies.  
This equation contains the combination $\partial^2 U / \partial \phi_i \partial \phi_k $.  
In the case of one scalar field, one can use
\be
	\label{Udoubleprime}
	\frac{d^2 U(\phi(r))}{d\phi^2}= \frac{1}{\phi'(r)} \partial_{r} \left[\frac{dU(\phi(r))}{d\phi} \right],
\ee  
which, along with the expression for $U'(\phi)$ (\ref{Uprime}), 
allows us to eliminate the potential completely in favor of the field
$\phi$. 
\ba
	\label{zphieqn}
	\frac{g_{rr}}{\rootg}  \partial_r \left[\rootg g^{rr}Z_\phi'\right] 
	+\frac{2}{\alpha} \partial_r \left[ \frac{\phi'}{\DL [g_{xx}]}\right] \left \{\frac{1}{\sqrt{f}} \partial_r \left[\frac{Z_0}{\sqrt{f}} \right] -
	\left(q^2-\frac{w^2}{f} \right)\phi' Z_\phi \right\} &\,& \\
	-Z_\phi \left\{g_{rr}\left(q^2 g^{xx} + w^2 g^{tt}\right) 
	+ \frac{\left(\DL [g_{xx}]\right)^2}{f \phi'} \partial_r\left[\frac{f \phi'}{\left(\DL [g_{xx}]\right)^2}\DL \left[ \rootg g^{rr} \phi' \right] \right] 
	\right\} &=&0 \nonumber
\ea

The equations (\ref{multiscalarz0}) and (\ref{multiscalarzphi}) are
the most general gauge invariant equations for any black brane
background supported by $r$ dependent scalar fields.  If only one scalar field is
present, one can use (\ref{z0eqn}) and (\ref{zphieqn}).    
To determine the dispersion relation, one needs to solve the above
equations perturbatively in $w,q$ as explained in previous chapters.
Some terms can be neglected when considering these equations to
$\mathcal{O}(w,q)$, but one should keep in mind that $Z_0$ is
$\mathcal{O}(q^2)$, $Z_\phi$ is $\mathcal{O}(1)$, and $\alpha$ is
$\mathcal{O}(q^2)$.

\section{Special case solution}

\subsection{Solution for $Z_{\phi}$}
Even for the case of a single scalar field, the gauge invariant
equations are quite complicated and we have been unable to find
a general analytic solution.  However, analytic solutions
do exist for certain special backgrounds, as we now show.  

Consider the special case of one scalar field with profile $\phi(r) = \kappa \log
\left[g_{xx}(r)\right]+\phi_0$ where $\kappa$  and $\phi_0$ are constants.  In this special
case, the equation (\ref{zphieqn}) reduces to
\be
	\frac{g_{rr}}{\rootg}  \partial_r \left[\rootg g^{rr}Z_\phi'\right] 
	-Z_\phi g_{rr}\left(q^2 g^{xx} + w^2 g^{tt}\right) = 0.
	\label{specialzphi}
\ee  
This can be more easily seen by re-writing
\ba
	&\,& \partial_r\left[\frac{f \phi'}{\left(\DL [g_{xx}]\right)^2}\DL \left[ \rootg g^{rr} \phi' \right] \right] \\
	&=&  \kappa \, \partial_r\left[\frac{f}{\DL[g_{xx}]} \DL \left[ \rootg g^{rr} \DL[g_{xx}] \right] \right] \\
	&=&  \kappa \left\{ \frac{2g_{rr}f}{\DL[g_{xx}]} (F_t - F_x) - 
		f \partial_r \left[ \frac{2 g_{rr}(F_t -F_r)}{p \left(\DL[g_{xx}]\right)^2} \right] \right\}\\
	&=&  \kappa \left\{ \frac{2g_{rr}f}{\DL[g_{xx}]} (F_t - F_x) - 
		f \partial_r \left[ \frac{\phi'(r)^2}{p \left(\DL[g_{xx}]\right)^2} \right] \right\}\\
	&=& 0,  
\ea
where we have used (\ref{phikeqn}) and (\ref{F0Fx}) in the last two steps.  

We now proceed according to the 
prescription outlined in Chapter \ref{TCReview_chapter} by applying the incoming wave ansatz
\ba
	Z_0(r) &=& f(r)^{-\frac{i w}{4 \pi T}} \left(Y_0(r)+ q Y_1(r)+ ... \right) \\
	Z_{\phi}(r) &=& f(r)^{-\frac{i w}{4 \pi T}} \left(Y_{\phi 0}(r)+ q Y_{\phi 1}(r) + ... \right)\\
	w(q) &=& w_1 q + w_2 q^2 + w_3 q^3 + ...
	\label{ansatz}
\ea
with the condition that all $Y$ functions are regular at the horizon.
Inserting this ansatz into (\ref{specialzphi}), and expanding the result in
powers of q, we have the following equation which is good up to $\mathcal{O}(q^2)$:
\be
	\partial_r \left[ \rootg g^{rr} Y_{\phi 0}' \right] 
	+ q\, \partial_r \left[ \rootg g^{rr} \left( Y_{\phi 1}' - \frac{i w_1}{2 \pi T} \DL(f) Y_{\phi 0} \right) \right]. 
\ee
It should be noted that we have omitted some terms which are
proportional to ($R^0_0 - R^x_x$) since they vanish by the background
equations of motion.  
Solving this order by order in $q$ is now quite simple.  The solution for $Y_{\phi 0}$ can be written in terms of an integral
\be 
	Y_{\phi 0}(r) = c_0 + c_1 \int_{r}^{\infty} \frac{g_{rr}(r')}{\sqrt{-g(r')}} dr',
\ee
but this integral is logarithmically divergent at the horizon by (\ref{rrNH}).  Thus,
the assumption of regularity on the $Y$ functions leads to $Y_{\phi 0}
= c_0$.  Finally, this constant must be set to zero by the Dirichlet
boundary condition at infinity.  Plugging $Y_{\phi 0} = 0$ into
the next order equation, one also finds $Y_{\phi 1} = 0$.  Proceeding now to higher
orders in $q$, one finds that the equations always reduce to the same
as that for $Y_0$, and as a result $Z_{\phi} = 0$ to all orders in
$q$.  With this simplification, the gauge invariant equation for $Z_0$ 
that must be solved is
\ba
	\label{specialz0eqn}
	\frac{g_{rr}}{\rootg} \alpha^2 f^2 \partial_r \left[\frac{\rootg g^{rr}}{\alpha^2 f^2}Z_0'\right] 
	+Z_0 \left\{ \DL[f]\DL[f\alpha] - g_{rr}\left(w^2 g^{tt} + q^2 g^{xx}\right)\right\}
	&=&0 \nonumber \\
	\, 
\ea
\subsection{Constraints on the metric and scalar potential}
\label{Potential_section}
Next, one must solve the equation for $Z_0$ using these boundary
conditions with the knowledge that $Z_\phi = 0$.  Before doing so,
it is useful to pause and ask what type of metrics these results will
be applicable to.  There are three unknown metric functions, $g_{tt},
g_{xx}$, and $g_{rr}$, but we have two constraints on them, namely
(\ref{phikeqn}) and (\ref{F0Fx}).  These two equations allow one
to solve for both $g_{tt}$ and $g_{rr}$ in terms of $g_{xx}$.  
Solving (\ref{F0Fx}) gives $g_{rr}$ in terms of the 
other metric components and one integration constant $c_1$,  
\be
	g_{rr}(r) = c_1 g_{xx}(r)^{p+1} f(r) \DL[f(r)]^2.
	\label{specialmetric1}
\ee
For convenience, we define the constant of proportionality between $\phi(r)$ and 
$\log[g_{xx}]$ as
\ba
	\phi(r) &=& \kappa \log [g_{xx}(r)] +\phi_0
	\label{specialphi}\\
	\kappa &\equiv& \pm \sqrt{\frac{p}{2}\left(p+1-2c_2 \right)}.
	\label{kappadef}
\ea
where $\phi_0$ is an integration constant. 

Taking this solution, plugging into (\ref{phikeqn}) and solving for $g_{tt}$ gives $f(r)$
in terms of $g_{xx}$ and three constants $c_2$, $c_3$ and $c_4$.  These
constants could in principle depend on $r_0$, but are independent of $r$,  
\be
	f(r) = c_3 \left[1 -c_4 \left(\frac{1}{g_{xx}(r)}\right)^{c_2}\right].	
\ee
Imposing the requirement that the function $f$ must vanish at the horizon determines $c_4$.  Furthermore
let us assume that the metric is asymptotically $AdS$, so that for $r \to \infty$,
$g_{xx}(r) \propto r^n$.  Where $n$ is some positive exponent.  
Then we must have $c_3 = 1$, and $c_2>0$ for the correct asymptotic 
behavior of $f$.  This determines $f$ up to a single constant,
\be
	f(r) = 1 - \left( \frac{ g_{xx}(r_0)}{g_{xx}(r)} \right)^{c_2}.
\label{specialmetric2}
\ee
which is exactly the same metric constraint we considered previously (\ref{specialf}).  
In the previous chapter, we simply imposed this constraint by hand.  Here we have
shown that the constraint arises naturally by first imposing a particular
condition on the scalar field which simplifies the gauge invariant equations.
The Hawking temperature for this metric can be found from (\ref{hawkingtemp}) as
\be
\label{specialT}
\frac{1}{(4 \pi T)^2} = c_1 g_{xx}(r_0)^p.
\ee 

It is also interesting to ask what sort of scalar potential can generate
such a metric.  One can determine the form of the potential by considering the following combination of
background Einstein equations 
\be
	g^{tt}G^{(0)}_{tt} +  g^{rr}G^{(0)}_{rr} = -8 \pi G_{p+2} \left(g^{tt} T^{(0)}_{tt} + g^{rr} T^{(0)}_{rr} \right) = U(\phi). 
\ee
An explicit computation of the left hand side for our special metric gives
\be
	U(\phi(r)) =  -\frac{p}{2 c_1\,c_2} g_{xx}(r_0)^{-2c_2} g_{xx}(r)^{-\frac{2}{p}\kappa^2}. 
\ee
Using (\ref{specialphi}), we find
\be
	U(\phi) =  -\frac{p}{2 c_1 c_2} g_{xx}(r_0)^{-2c_2} \exp \left(-\frac{2}{p}\kappa(\phi - \phi_0) \right).
\ee
Thus, the sort of metrics we are considering are those generated by a
potential which contains a single exponential, similar to the
Chamblin-Reall backgrounds \cite{Chamblin:1999ya} examined in
\cite{Gubser:2008ny,Gubser:2008sz}.  It is amusing that we have
arrived at the same metric considered in these references, as our 
only guiding principle has been to search for a class of 
metrics which simplify the gauge invariant equations (\ref{z0eqn}),(\ref{zphieqn}).

Finally, note that since the potential must be independent of temperature, it
is required that the constant $c_2$, and the combination $c_1
g_{xx}(r_0)^{2c_2}$ must themselves be independent of $r_0$.

\subsection{First order hydrodynamics}
One must now go back to (\ref{z0eqn}), insert $Z_{\phi} = 0$ and
the incoming wave ansatz (\ref{ansatz}), and expand the resulting equation in powers of $q$.  
In doing so, one should take into account the constraints (\ref{specialmetric1}) and (\ref{specialmetric2}).  
The first of these constraints allows one to eliminate $\rootg g^{rr}$ in favor of $\DL[f]$.  
Once this is done, the lowest order equation for $Y_0$ can be written
\be
	\frac{\left(C_0 - f(p-c_2)\right)^2}{(p-c_2) f} \partial_r \left[ \frac{1}{\DL[f]\left(C_0 - f (p-c_2) \right)^2}Y_0' \right]
	- \frac{\DL[f]}{C_0 - f (p-c_2)} Y_0 = 0,
\ee  
where 
\be
	C_0 \equiv p w_1^2 - c_2. 
\ee
The general solution to that equation contains two arbitrary constants $k_1$ and $k_2$, and can be written
\be
	Y_0(r) = [p w_1^2- c_2 + f(r)(p-c_2)]\left[k_0 
	+ k_1 \int_{r}^{\infty}\, \left( \frac{p w_1^2 - c_2 - (p-c_2) f(r')^2}{p w_1^2 - c_2 + (p-c_2) f(r')^2}\right)^2 \frac{f'(r')}{f(r')} \, dr' \right],
\label{Y0soln}
\ee
as can be found by first making the ansatz $Y_0(r) = b_0 + b_1
f(r)$, and then using the technique of reduction of order once this
solution is found (cf. Appendix \ref{app_DiffEQ}).
The integral in (\ref{Y0soln}) is logarithmically divergent near the horizon, and thus the requirement of 
regularity leads to $k_1 = 0$.  Finally, applying the Dirichlet boundary condition at $r \rightarrow \infty$, 
leads to 
\be
	w_1^2 = \frac{2 c_2}{p}-1, 
\ee  
where we have assumed that $f(r\rightarrow \infty) = 1$.  

Proceeding now to the next order in $q$, and substituting the solutions for $w_1$ and $Y_0$, one finds the
following differential equation for $Y_1$:
\be
	\partial_r \left[ \frac{ Y_1'}{\DL[f](1+f)^2} \right] 
	+ \frac{f'}{(1+f)^3}\left[Y_1 - w_1 k_0 \left( \frac{i (p-c_2)}{\pi T} + 2 p w_2 \right) \right] = 0.
\label{y1eqn}
\ee
The solution to the homogeneous part can be found using the same techniques as for the solution for $Y_0$ listed
above.  A particular solution to the inhomogeneous equation is obviously $Y_1(r)=\, $Constant.  This leads to the
general solution,
\be
	Y_1(r) = (f(r) - 1)\left(k_2 + k_3 \int_{r}^{\infty} \frac{(f(r')+1)^2}{(f(r')-1)^2} \frac{f'(r')}{f(r')} \, dr' \right)
		+ w_1 k_0 \left( \frac{i (p-c_2)}{\pi T} + 2 p w_2 \right).  
\ee
The integral above is again logarithmically divergent, leading to the requirement that $k_3=0$.  Applying now the
Dirichlet boundary condition at $r \rightarrow \infty$ leads to
\be
	w_1 k_0 \left( \frac{i (p-c_2)}{\pi T} + 2 p w_2 \right) = 0, 
\ee
\be
	w_2 = \frac{i(c_2-p)}{2 \pi T p}.
\ee
Note that we require $c_2 < p$ to avoid exponential growth of the 
perturbations.  Let us summarize the central results of this section:
\ba
	Y_0(r) &=& y_0\left(f(r)-1 \right),
	\label{y0soln}\\
	Y_1(r) &=& y_1\left(f(r)-1 \right),
	\label{y1soln}\\
	w_1 &=& \pm \sqrt{\frac{2c_2}{p}-1} ,
	\label{w1soln}\\
	w_2 &=& -\frac{i(p-c_2)}{2 \pi T p}.
	\label{w2soln}
\ea
Here $y_0 = k_0(c_2-p)$ is a constant, though its value is unimportant as it does not enter into any
formulas for physical quantities, and we have relabeled $k_2$ as $y_1$ for aesthetic reasons.  

Comparing these results to the dispersion relation expected from first order hydrodynamics (\ref{firstorderhydrosound})
\ba
	w_1 &=& \pm v_s, \\
	w_2 &=& - i \frac{\eta}{\epsilon + P} \left( \frac{p-1}{p} + \frac{\zeta}{2\eta} \right),
\ea
one gains knowledge of $v_s$ (speed of sound), $\eta$ (shear viscosity), and $\zeta$ (bulk viscosity):
\ba
	\eta/s &=& 1/4\pi \label{etasoln},\\
	v_s &=& \sqrt{\frac{2c_2}{p}-1} \label{vssoln},\\
	\zeta / \eta &=& 2 \left(\frac{1}{p} - v_s^2 \right) \label{zetasoln}.
\ea
Here we remind the reader that $s = (\epsilon + P)/T$ is the entropy density,  $\epsilon$ is the
equilibrium energy density, and $P$ is the equilibrium pressure.  Note
that the conjectured bulk viscosity bound of Buchel \cite{Buchel:2007mf} is
saturated for metrics of this type.  

\subsection{Digression: comparison with black hole thermodynamics}
It is interesting to note that the speed of sound can be derived
solely from the metric by using relations from black hole
thermodynamics.  Such a formula was derived in Chapter
\ref{GGDReview_chapter}, (cf. \ref{vsBHthermo}).  Applying it to 
the special metric and using (\ref{hawkingtemp}) and (\ref{specialT}) results in 
\be
 v_s^2 = -\frac{1}{p} \left[p+\frac{\frac{d}{dr_0} Log[c_1]}{\frac{d}{dr_0} Log[g_{xx}(r_0)]}\right].
\label{specialvsBHthermo}
\ee
Recall that $c_1$ is independent of $r$, but could in principle depend
on $r_0$.
At first sight this seems quite puzzling due to the presence of the
constant $c_1$ which is not present in the formula (\ref{vssoln}).
These two calculations of the speed of sound must agree.

The resolution of this apparent contradiction lies in the observation made at the end of
Sec. (\ref{Potential_section}).  In order to have a sensible potential
which is temperature independent, both $c_2$ and $c_1
g_{xx}(r_0)^{2c_2}$ must be independent of temperature (and thus also
independent of $r_0$).  As a result, we have
\be
	c_1 \propto g_{xx}(r_0)^{-2c_2},
\ee
where the constant of proportionality is independent of $r$ and $r_0$.  Substituting this into
(\ref{specialvsBHthermo}) provides
\be
	v_s^2 = \frac{2 c_2}{p}-1, 
\ee
which is exactly the same as that computed from the gravitational
perturbations.  It is quite reassuring that these two calculations
agree.  It is interesting that this latter method is much more
efficient for computing the speed of sound; unfortunately, it does not
give access to the bulk viscosity and any higher order transport
coefficients.

\subsection{Second order hydrodynamics}
\label{sec3}
It is the purpose of this section to extend the above calculation to the next hydrodynamic order, 
and thus to determine the next coefficient in the dispersion relation $w_3$. 
\subsubsection{Equation for $Y_2$}

We have already shown that $Z_{\phi}$ vanishes to all orders in $q$,
so then it remains to return to (\ref{specialz0eqn}), insert the
incoming wave ansatz, expand in powers of $q$, and insert the
solutions (\ref{y0soln} - \ref{w2soln}). Completing these steps,
one finds the following differential equation which must be solved for
$Y_2$:
\ba
	&\,& \partial_r \left[ \frac{ Y_2'}{\DL[f](1+f)^2} \right] + \\ 
	&\,& \frac{f'}{(1+f)^3}\left\{Y_2 +\frac{y_0}{(4 \pi T)^2} \left[x_0 
	+ \frac{1-f^2}{f} \left(w_1^2 + \frac{g_{xx}(r)^p}{g_{xx}(r_0)^p} \left(f-w_1^2 \right) \right)\right]\right\} = 0 \nonumber,
\label{y2eqn}
\ea
where
\be
	x_0 \equiv 2 \left( 5w_1^2-1 - \frac{2 w_1 w_3 (4 \pi T)^2}{1-w_1^2} \right).
	\label{x0definition}
\ee
In writing the above expression, we have replaced all occurrences of
the constant $c_2$ which appears in the metric with $w_1$ due to the
relation (\ref{w1soln}), and have removed the constant $c_1$ in favor of $T$ using (\ref{specialT}).  

\subsubsection{Solution for $Y_2$}  
The associated homogeneous equation for (\ref{y2eqn}) is the same as
the homogeneous part of the equation for $Y_1$, (\ref{y1eqn}).  The solution can easily be found by first
making the ansatz $Y_2(r) = y_{2a} (1-f(r))$, and then using the technique of
reduction of order once this solution is found.  The general
solution to the homogeneous part of the above equation is
\be
	Y_{2h}(r) = y_{2a} (1 - f(r)) + y_{2b} \Bigl[ (1-f(r))\log[f(r)] + 4 \Bigr].
\ee
Here the subscript $h$ stands for homogeneous, and $y_{2a}$, $y_{2b}$ are arbitrary constants.  

In order to find the general solution for $Y_2$, we must now find a
particular solution to the inhomogeneous equation.  Since the
homogeneous solution is known, one can construct a particular solution
using the method of variation of parameters.  For completeness, this
method is outlined in Appendix \ref{app_VarOfParams}. 
Using the notation from Appendix
\ref{app_VarOfParams}, (and changing independent variables from $x$ to
$r$), we have
\ba
	y_1(r) &=& 1- f(r),\\
	h(r) &=& \log[f(r)] + \frac{4}{1-f(r)},\\
	g(r) &=& -\frac{(f')^2}{f(1+f)} \frac{y_0}{(4 \pi T)^2} \left[x_0 
	+ \frac{1-f^2}{f} \left(w_1^2 + \frac{g_{xx}(r)^p}{g_{xx}(r_0)^p} \left(f-w_1^2 \right) \right)\right].
\ea
After some simplification, one can write the particular solution as,
\ba
	Y_{2p}(r) &=& -\frac{y_0}{(4 \pi T)^2} (1-f(r)) \left[ \int \left(\frac{1+f(r)}{1-f(r)}\right)^2 \frac{f'(r)}{f(r)} Q(r) \, dr \right] \\
	Q(r) &=& \int^r \frac{(1-f(z))}{(1+f(z))^3} f'(z)\left[x_0 
	+ \frac{1-f(z)^2}{f(z)} \left(w_1^2 + \frac{g_{xx}(z)^p}{g_{xx}(r_0)^p} \left(f(z)-w_1^2 \right) \right)\right] \,dz. \nonumber \\
\,   
\ea
The term involving $x_0$ can be integrated directly, by changing
variables from $z$ to $f$.  This simplifies the solution to
\be
	Y_{2p}(r) = - \frac{x_0 y_0}{(4 \pi T)^2} 
	+ \frac{y_0}{(4 \pi T)^2} ( f(r) - 1 ) 
	\left[ \int \left(\frac{1+f(r)}{1-f(r)}\right)^2 \frac{f'(r)}{f(r)} \hat{Q}(r) \, dr \right],
\label{yp}
\ee
\be
	\hat{Q}(r) = \int^r \left(\frac{1-f(\tilde{r})}{1+f(\tilde{r})}\right)^2 \frac{f'(\tilde{r})}{f(\tilde{r})}\left[w_1^2 + \frac{g_{xx}(\tilde{r})^p}{g_{xx}(r_0)^p} \left(f(\tilde{r})-w_1^2 \right)\right] \,d\tilde{r},
\ee
and so, the general solution for $Y_2$ is given by
\be
	Y_2(r) = Y_{2h}(r) + Y_{2p}(r).  
\ee
To proceed, it is convenient to change coordinates.  We define the coordinate $u$ by
\be
	u^2 \equiv \left(\frac{g_{xx}(r_0)}{g_{xx}(r)}\right)^{c_2} = \left(\frac{g_{xx}(r_0)}{g_{xx}(r)}\right)^{\frac{p}{2}(w_1^2+1)} ,
\ee
so that
\be
	f(u) = 1 - u^2,
\ee
and the horizon is now located at $u = 1$, and the boundary is located
at $u=0$.\footnote{Here the assumption is that $c_2 > 0$, and that
$g_{xx}(r\rightarrow \infty) \sim
r^n$ with $n>0$  These assumptions hold for the Schwarzschild $AdS$
metric for any positive $p$, and for the Dp-Brane metric provided
$p<7$.}In terms of the new coordinates, $Y_2$ is given by
\be
	Y_{2h}(u) = y_{2a} u^2 + y_{2b} \Bigl[ u^2\log[1-u^2] + 4 \Bigr],
\ee
\be
Y_{2p}(u) = -\frac{y_0}{(4 \pi T)^2} \Biggl\{x_0 
	+ 4 u^2 \int \frac{(2-u^2)^2}{u^3(1-u^2)} 
		\int^u \frac{v^5\left[w_1^2+v^{-4/(1+w_1^2)}\left(1-v^2-w_1^2\right) \right]}{(2-v^2)^2(1-v^2)}\,dv\,du \Biggr\}  
\ee
The inner integral in this latter equation can be written in terms of hypergeometric
functions, but the result is not particularly enlightening, so we do
not reproduce it here.  The full analytical form of this function is
not needed to determine the dispersion relation.

\subsubsection{Boundary conditions}
The first boundary condition which must be applied on $Y_2$ is the
condition of regularity at the horizon.  In order to do so, one must
extract the coefficient of the logarithmic divergence in the
particular solution $Y_{2p}$.  To do so, we first expand the integrand
in powers of $(u-1)$, and look for the coefficient of the $(u-1)^{-1}$
term.  After integration, this term will lead to the logarithmic
divergence.

With the aid of Mathematica, we find the nested integral can be expanded near
the horizon as
\begin{displaymath}
	\int^u \frac{v^5\left[w_1^2+v^{-4/(1+w_1^2)}\left(1-v^2-w_1^2\right) \right]}{(2-v^2)^2(1-v^2)}\,dv \approx
\end{displaymath}
\be
	\frac{w_1^2}{2}\left\{ \frac{1-3w_1^2}{w_1^2(1-w_1^2)}  -  \left[2 + i \pi - H_n\left(\frac{-2}{1+w_1^2}\right) \right]\right\} + \mathcal{O}(u-1)
\ee
where $H_n(\alpha)$ is the \emph{Harmonic Number} defined in
(\ref{harmonicnumberdef}).  Using this expansion in $Y_2$, we find
that near the horizon,
\begin{displaymath}
	Y_{2p}(u \rightarrow 1) \approx
\end{displaymath}
\be
	-\frac{y_0}{(4 \pi T)^2} \left\{x_0 + 4 \int 
	\frac{du}{4(1-u)}\left[\frac{1-3w_1^2}{1-w_1^2} - w_1^2\left(2 + i \pi - H_n\left(\frac{-2}{1+w_1^2}\right) \right)\right] \right\}
	+ \mathcal{O}(1).
\ee
Going back to the general solution for $Y_2$, one thus finds
\begin{displaymath}
	Y_{2}(u \rightarrow 1) \approx
\end{displaymath}
\be
\log [1-u] \left\{ y_{2b} +\frac{y_0}{(4 \pi T)^2}   
	\left[\frac{1-3w_1^2}{1-w_1^2} - w_1^2\left(2 + i \pi - H_n\left(\frac{-2}{1+w_1^2}\right) \right)\right] \right\} + \mathcal{O}(1).
\ee
The requirement of regularity at $u = 1$ thus gives
\be
        y_{2b} = \frac{y_0}{(4 \pi T)^2}   
	\left[ w_1^2\left(2 + i \pi - H_n\left(\frac{-2}{1+w_1^2}\right) \right)- \frac{1-3w_1^2}{1-w_1^2} \right].
\ee
For future convenience, we make use of the identity
\be
	H_n(\alpha)=H_n(\alpha+1) - \frac{1}{1+\alpha} 
\label{Hnidentity}
\ee
to write
\be
	H_n\left(\frac{-2}{1+w_1^2}\right) = H_n\left(\frac{2w_1^2}{1+w_1^2}\right) - \frac{1-2w_1^2-3w_1^4}{2w_1^2(1-w_1^2)}.
\ee
Substituting this into the equation for $y_{2b}$ we find
\be
	y_{2b} =  \frac{y_0}{(4 \pi T)^2} 
	\left[w_1^2\left(2 + i\pi -H_n\left(\frac{2w_1^2}{1+w_1^2}\right)\right) -\frac{1}{2} (1-3 w_1^2) \right].  
	\label{y2bsoln}
\ee
Finally, we must apply the Dirichlet boundary condition at $u = 0$.  We proceed as above by expanding the integrand of $Y_{2p}$ near 
$u = 0$.  Mathematica gives
\be
\int^u \frac{v^5\left[w_1^2+v^{-4/(1+w_1^2)}\left(1-v^2-w_1^2\right) \right]}{(2-v^2)^2(1-v^2)}\,dv \approx -\frac{w_1^2}{2}\left(2+i \pi\right) + \mathcal{O}(u),
\ee
so that 
\be
	Y_{2}(u \rightarrow 0) \approx \Bigl( 4 y_{2b} + \mathcal{O}(u)\Bigr) - \frac{y_0}{(4 \pi T)^2} \left[ x_0 - 8 u^2 w_1^2(2+i\pi) \int \left[\frac{1}{u^3} + \mathcal{O}(u^{-2})\right]du \right].
\ee
Doing the integral, one finally has
\be
	 Y_{2}(u \rightarrow 0) \approx 4 y_{2b} - \frac{y_0}{(4 \pi T)^2} \left[ x_0 + 4 w_1^2 \left(2+i \pi\right)\right] + \mathcal{O}(u),
\ee
and applying $Y_2(u \rightarrow 0) = 0$ gives
\be
	4 y_{2b} = \frac{y_0}{(4 \pi T)^2} \left[ x_0 + 4 w_1^2 \left(2+i \pi\right) \right].
\ee
Using (\ref{y2bsoln}),
\be
	x_0 = -4 \left[w_1^2 H_n\left(\frac{2w_1^2}{1+w_1^2}\right) + \frac{1}{2}(1-3w_1^2)\right]
\ee
and (\ref{x0definition}) allows us to solve for $w_3$.  The final result is
\be
	w_3 = \frac{w_1(1-w_1^2)}{(4 \pi T)^2} \left[1+  H_n\left(\frac{2w_1^2}{1+w_1^2}\right) \right]. 
	\label{w3soln}
\ee
This is the  main result of this section.  To summarize, we have computed the
coefficient of the $q^3$ term in the sound mode hydrodynamic
dispersion relation for a specific class of metrics (cf.
\ref{specialmetric2}).  This class of metrics contains two constants
which we denote $c_1$ and $c_2$ .  Our expression for $w_3$ should
necessarily depend on these constants, but we have eliminated $c_1$ in
favor of $T$ using (\ref{specialT}) and $c_2$ in favor of $w_1^2 =
v_s^2$ due to the relation (\ref{vssoln}).

\begin{figure}[t]
\centering
\includegraphics[width=1\textwidth]{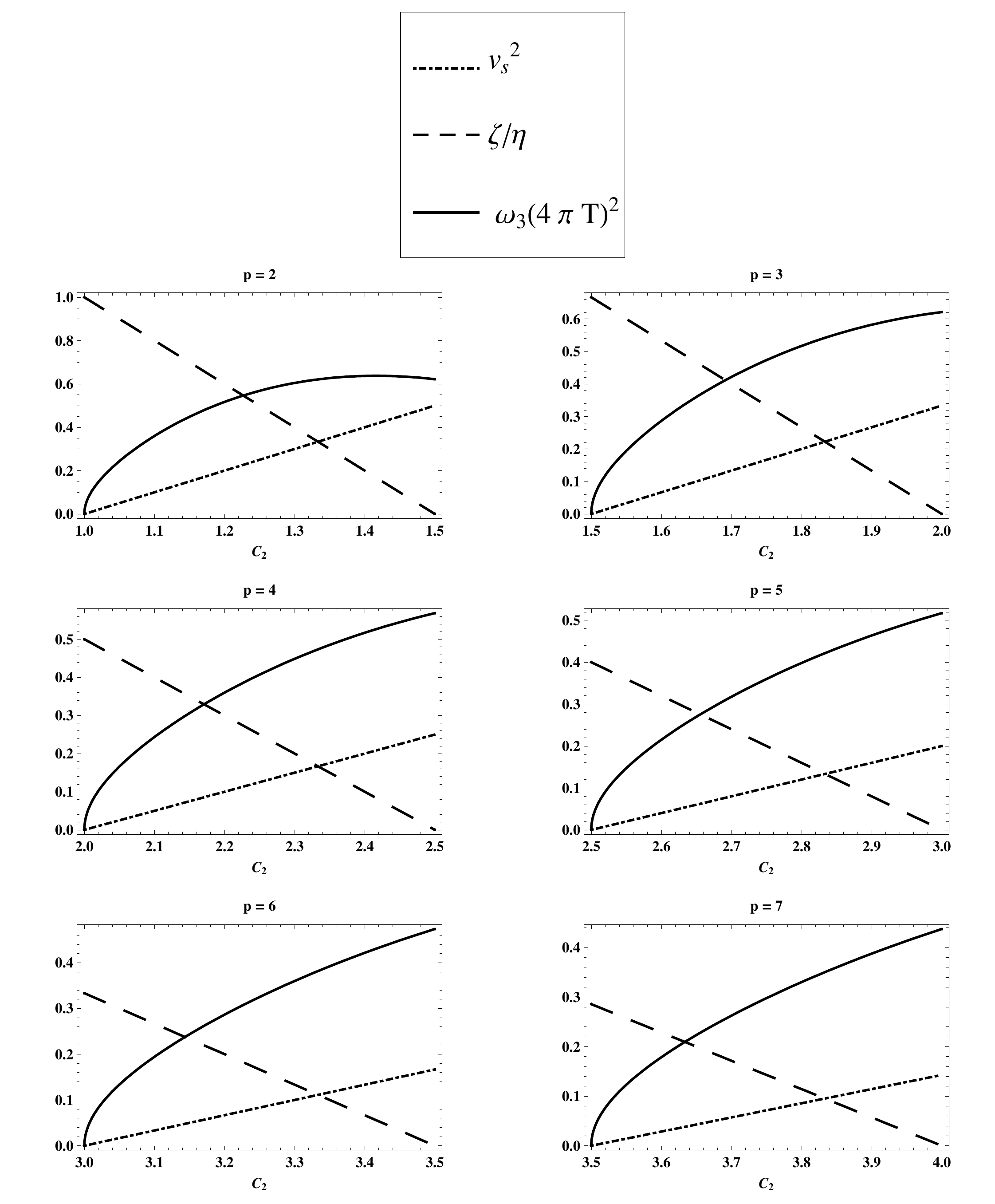}
\caption{Plots of the dimensionless quantities: the speed of sound,
 ratio of bulk to shear viscosities, and $w_3 (4 \pi T)^2$, as a
 function of the free parameter $c_2$ appearing in the metric.  Each
 plot corresponds to a different number of spatial dimensions $p$.
 Only a physical region where both $v_s^2$ and $\zeta$ are positive is
 shown.  Note that all of the specified quantities vary with $c_2$ and
 $p$, and are thus not universal in the same sense as $\eta / s$.}
\label{soundfigure}
\end{figure}

We are now able to plot the sound mode transport coefficients $v_s$,
$\zeta$, as well as this new quantity $w_3$ as a function of the
parameter $c_2$ to examine how they all behave as one deviates from
the conformal limit.  The conformal limit is the case of $c_2 = (p+1)/2$, 
giving the Schwarzschild $AdS$ metric.  Note that if $c_2 > (p+1)/2$, 
the bulk viscosity becomes negative.  Also note that if $c_2 < p/2$, 
the speed of sound becomes imaginary.  We only plot
the physical region $p/2 < c_2 < (p+1)/2$ in Fig. \ref{soundfigure}. 

\section{Transport coefficients in Israel-Stewart theory}
\label{sec4}
Now that we have computed the dispersion relation to
$\mathcal{O}(q^3)$ in the sound mode, we are in a position to examine
the implications for both first and second order hydrodynamic
transport coefficients.  As mentioned in
Sec. (\ref{secondorderhydro_section}), one of the dominant
formulations of second order hydrodynamics is the Israel-Stewart
formalism \cite{Israel:1976tn,Israel:1979wp}.  Recently, the
formulation of second order hydrodynamics presented in
\cite{Baier:2007ix} has gained popularity, though at present it is
only applicable to conformal theories.  (It should be noted that
some progress has been made in generalizing the work of
\cite{Baier:2007ix} to non-conformal theories
\cite{Kanitscheider:2009as}).  The metrics we consider are not
necessarily conformal, and thus we will use the Israel-Stewart
formulation.

The sound mode dispersion relation in Israel-Stewart theory (in the case of metrics with
no conserved charge) was given previously in (\ref{w3transport}), and included two new
transport coefficients: $\tau_{\pi}$ and $\tau_\Pi$.  
Comparing (\ref{w3transport}) to our main result (\ref{w3soln}), and
eliminating $\eta$ and $\zeta$ from the relations
(\ref{etasoln}), (\ref{zetasoln}) gives the relation
\be
	\tau_{\pi}+ \frac{\left(1-p v_s^2 \right)}{(p-1)} \tau_{\Pi} - \frac{\left(1-v_s^2\right)}{(8 \pi  T)v_s^2}\frac{p}{(p-1)}\left\{1+v_s^2\left[1+ 2 H_n \left(\frac{2 v_s^2}{1+v_s^2} \right)\right]\right\} = 0.	
\label{taupirelation}
\ee
As expected, the coefficients $\tau_{\pi}$ and $\tau_{\Pi}$ cannot in
general be determined separately using this method.  Still, if one of
these coefficients is known, the above relation allows us to determine
the other.

Let us now explicitly check that our results agree with other
calculations for specific backgrounds.  The Schwarzschild
AdS black hole metric in the near horizon limit was given in (\ref{SAdSmetric}). 
For this
metric, the parameter $c_2 = (p+1)/2$.  In this
case, (\ref{vssoln}), (\ref{etasoln}), (\ref{zetasoln}), (\ref{taupirelation}) give
\ba
	v_s^{SAdS} &=& 1/ \sqrt{p}, \\
	\eta / s &=& 1/4 \pi, \\
	\zeta^{SAdS} &=&0,\\
	\tau_{\pi}^{SAdS}&=& \frac{1}{4 \pi T} \left[ \frac{p+1}{2} + H_n\left(\frac{2}{p+1} \right) \right].  	
\label{taupiconformal}
\ea  
Because this metric is dual to a conformal field theory at finite
temperature, the results for $v_s$ and the vanishing bulk viscosity
are exactly in agreement with our expectations.  The result for
$\tau_{\pi}$ is in agreement with \cite{Haack:2008cp},
\cite{Bhattacharyya:2008mz}.\footnote{In comparing with the results of
\cite{Bhattacharyya:2008mz}, one needs to employ the identity
(\ref{Hnidentity}) to see the agreement.}This is a non-trivial check
on our calculation; the cited results were arrived at by completely
different methods than those we employ here.  Furthermore, it should
be noted that (\ref{taupiconformal}) confirms a conjecture made by
Natsuume \cite{Natsuume:2008gy}.\footnote{To be precise, the
conjecture is confirmed for the case of $p \geq 2$; the case of $p=1$
should probably be checked separately as the derivation of the gauge
invariant equations in \cite{Springer:2008js}, \cite{Mas:2007ng} rely
on at least 2 spatial dimensions.  See \cite{David:2009np}, where
first order hydrodynamics is examined for $p=1$.}

Finally, we can also consider the case of the Dp-Branes.  In the
Einstein frame, the metric was given in (\ref{DPmetric}).  For this
metric, the parameter $c_2 = (7p-p^2)/(9-p)$.  Again, using
(\ref{vssoln}), (\ref{etasoln}), (\ref{zetasoln}), and
(\ref{taupirelation}), the results for the transport coefficients are:
\ba
	v_s^{DP} &=& \sqrt{\frac{5-p}{9-p}}, \\
	\eta /s &=& 1 / 4 \pi, \\
	\zeta^{DP} / \eta &=& \frac{2(3-p)^2}{9-p},\\
	(9-p)(1-p) \tau_{\pi}^{Dp} &=& (3-p)^2 \tau_{\Pi}^{Dp} -\frac{p}{ \pi  T}\frac{(7-p)}{(5-p)}\left[1+\frac{5-p}{7-p} H_n\left(\frac{5-p}{7-p}\right)\right]  
\ea
The results for $v_s^{DP}$ and $\zeta^{DP}/ \eta$ agree with the
calculation of \cite{Mas:2007ng}.  Furthermore, the relation between
$\tau_{\pi}$ and $\tau_{\Pi}$ agrees with previous computations for
$p=1$ and $p=4$ in \cite{Natsuume:2007ty}.

This completes our analysis of transport coefficients for this special
class of metrics.  In the process, we have generated formulas for the
the speed of sound (\ref{vssoln}), bulk viscosity (\ref{zetasoln}), and a relation
between two second order transport coefficients (\ref{taupirelation}).
These formulas are applicable to metrics which obey (\ref{F0Fx}), and
(\ref{specialmetric2}), and are generalizations of formulas given in
\cite{Mas:2007ng}.

\section{Comparison with the shear mode}
\label{sec5}
Though we have exclusively dealt with the sound mode in this chapter, we
now compare our results to similar ones from the shear mode.  Recall the
shear dispersion relation (\ref{sheardispersion}),  
\be
	w(q)_{\rm shear} = -i D_{\eta} q^2 - i D_{\eta}^2 \tau_{\rm shear} q^4 + \mathcal{O}(q^6)  
\ee
where 
\be
	D_{\eta} = \frac{\eta}{T s} = \frac{1}{4 \pi T}.
\ee
A formula for $\tau_{\rm shear}$ was given in the previous
chapter (\ref{taudef}) and in the paper \cite{Kapusta:2008ng}.  This
formula is applicable to a wide variety of metrics, including the
special metrics we have considered in this chapter.  Using the special
metric (\ref{specialmetric1})-(\ref{specialmetric2}), and the
relationship (\ref{specialT}) in this formula yields the result
(\ref{taushearspecialmetric}):
\be
	\tau_{\rm shear} = \frac{1}{4 \pi T} H_n \left(2 - \frac{p}{c_2}\right) = \frac{1}{4 \pi T} H_n \left(\frac{2w_1^2}{1+w_1^2}\right).
	\label{taushearspecialmetric2}
\ee 
Before the work of \cite{Baier:2007ix}, it was thought that $\tau_{\rm shear} = \tau_{\pi}$, but
as explained in section (\ref{secondorderhydro_section}), the authors
of the aforementioned paper showed that the coefficient of the $q^4$ term
contains not only $\tau_\pi$, but also contributions from (currently
unformulated) third order hydrodynamics.  We can now determine these
unknown contributions for the special class of metrics we are considering.

As in Sec. \ref{secondorderhydro_section}, let us parametrize 
\be
	\tau_{\rm shear} = \tau_{\pi} + \Delta
\ee
where $\Delta$ denotes the unknown contributions from third order
hydrodynamics.  Combining (\ref{taushearspecialmetric2}) and (\ref{taupirelation})
allows one to solve for $\Delta$.  Evidently,
\be
	\Delta = \left(\frac{p v_s^2-1}{p\left(1-v_s^2\right)}\right)(\tau_{\pi}-\tau_{\Pi}) -\frac{1+v_s^2}{8 \pi  T v_s^2}. 
\label{deltasoln}
\ee
The correction $\Delta$ is not universal in the sense
that the first term is not present in the case of a conformal theory.
In the conformal case, $\Delta$ is still not universal, because it
depends on the number of dimensions of the theory (because $v_s$
depends on $p$).  In the future, when the particular transport
coefficients which comprise $\Delta$ are known, it will be interesting
to see whether there is any universal relationship between these
unknown coefficients, $\tau_{\pi}$, and $\tau_{\Pi}$.

We can easily check that the formula (\ref{deltasoln}) reproduces the
results in the well known $SAdS_5$ metric, which is dual to
$\mathcal{N}=4$ supersymmetric Yang-Mills theory at finite
temperature.  In this case, $p=3$ and $v_s^2 = 1/3$.  Immediately, we
have
\be
	\Delta^{SADS_5} = - \frac{1}{2 \pi T}.  
\ee
In \cite{Natsuume:2007ty,Baier:2007ix}, it was found
that\footnote{Even though the formalism of Baier \emph{et
al}. \cite{Natsuume:2007ty} is different than Israel-Stewart, one can
check that the sound mode dispersion relations coincide in the
limit of conformal theories ($\zeta \rightarrow 0$).  In an
unfortunate clash of notations, the relaxation time introduced by Baier
\emph{et al}. is denoted by $\tau_\Pi$.  Its Israel-Stewart counterpart is
$\tau_\pi$.}
\ba
	\tau^{SADS_5}_{\rm shear} &=& \frac{1-\log (2) }{2 \pi T} \\
	\tau^{SADS_5}_{\pi} &=& \frac{2- \log(2) }{2 \pi T}.
\ea
It is clear that the relaxation time computed from the shear mode and
the sound mode differ by the amount predicted by the formula
(\ref{deltasoln}).  

\section{Summary}
In this chapter, we first presented a set of gauge invariant equations for
sound mode perturbations on a generic black brane type background.
The equations (\ref{multiscalarz0}) and (\ref{multiscalarzphi}) are the
main results which pertain to this topic.  These equations can be used to determine
the speed of sound and bulk viscosity for any metric which can be
generated by a set of minimally coupled scalar fields.
In order to determine the dispersion relation, one must solve these
equations perturbatively in $q$, applying the incoming wave boundary
condition at the horizon, and Dirichlet boundary condition at $r =
\infty$.  

The gauge invariant equations are quite complicated, and so far a
general analytic solution has eluded us, though we did present a
solution for a particular class of metrics.  These metrics are not
necessarily conformal, and contain an arbitrary number of spatial
dimensions $p$ with $p>2$.  Metrics which obey (\ref{specialmetric1})
and (\ref{specialmetric2}) have speed of sound and bulk viscosity given
by (\ref{vssoln}) and (\ref{zetasoln}).  These results are a
generalization of the results of \cite{Mas:2007ng}, and include both
Dp-brane, and Schwarzschild AdS black hole metrics.  In addition, we
have computed the coefficient of the $q^3$ term in the sound mode
dispersion relation (\ref{w3soln}) for such a class of metrics.

Information about second order transport coefficients was presented
within the context of the Israel-Stewart theory.  In general, a
relationship (\ref{taupirelation}) between two transport coefficients
$\tau_{\pi}$ and $\tau_{\Pi}$ can be determined.
In the conformal case of the Schwarzschild $AdS$ metric, the relation
mentioned above allows the determination of the quantity $\tau_{\pi}$,
since the coefficient of $\tau_{\Pi}$ vanishes in this case.  We have
verified that our results agree with those calculated from different
methods.

Finally, by comparing the sound mode dispersion relation to the shear
mode discussed in Chapter \ref{ShearMode_chapter},  we were able to determine
the contribution of third order hydrodynamics to the shear mode
(it was pointed out that such contributions would be present in
\cite{Baier:2007ix}).

\chapter{Conclusion and Discussion}
\label{conclusion_chapter}
In this thesis, we started by providing a brief review of the
literature on gauge/gravity duality and its phenomenological
applications to strongly coupled gauge theories.  The central focus
throughout has been hydrodynamic dispersion relations for
perturbations of a strongly coupled plasma.  In Chapter
\ref{TCReview_chapter}, we gave a pedagogical overview of the theory
of hydrodynamics and also explained in detail the gauge/gravity
duality methods for computing such hydrodynamic dispersion relations.
In Chapters \ref{ShearMode_chapter} and \ref{SoundMode_chapter}, we
applied these methods to a general gravitational dual which is
supported by one or more scalar fields.

In Chapter \ref{ShearMode_chapter}, the shear mode dispersion relation
was computed up to order $q^4$ (sub-leading hydrodynamic order).  The
equation for $\tau_{\rm shear}$ (\ref{taudef}) is the main result of
this chapter.  This result is in agreement with the original
publication \cite{Kapusta:2008ng}, though in this thesis completely
different methods were used to derive it.  This is another instance
where the results from and AdS/CFT calculation agree with a
calculation which uses the black hole membrane paradigm.

In Chapter \ref{SoundMode_chapter}, the sound mode perturbations were
analyzed in the same type of gravitational dual.  First, we derived
the relevant gauge invariant equations (\ref{multiscalarz0}),
(\ref{multiscalarzphi}) for the perturbations which must be solved in
order to compute the dispersion relation.  These equations are
general, and can be applied to any dual metric which is supported by
($r$ dependent) scalar fields.  Secondly, we solved these equations
analytically for a special class of metrics, which allowed us to
compute the sound mode dispersion relation up to order $q^3$ for such
metrics (\ref{w1soln}),(\ref{w2soln}),(\ref{w3soln}).  Examination of
these results within the context of Israel-Stewart theory of
hydrodynamics leads to formulas for the speed of sound (\ref{vssoln}),
bulk viscosity (\ref{zetasoln}), and a relationship between two
relaxation times (\ref{taupirelation}).  Finally, by comparing the
shear mode results of Chapter \ref{ShearMode_chapter} to the sound
mode results of Chapter \ref{SoundMode_chapter}, we were able to
gain information about the third order hydrodynamic contributions to
the shear mode dispersion relation (\ref{deltasoln}).  All of these
formulas agree with previously known results in the literature, but
also offer new information about non-conformal theories.

As mentioned in the introduction, it is desirable to find universal
relations among transport coefficients, as such relations have
phenomenological implications for the quark-gluon plasma.  Some main
results of this thesis such as the shear/sound mode dispersion
relation results (\ref{taudef}), (\ref{vssoln}), (\ref{zetasoln}),
(\ref{w3soln}), (\ref{taupirelation}) and the third order hydrodynamic
contributions to the shear mode (\ref{deltasoln}) are applicable to
certain non-conformal gravity duals in an arbitrary number of
dimensions.  However, despite the fact that these relations are
applicable to many theories, they are not universal like $\eta/s$ and
the relation presented in \cite{Haack:2008xx}.  For example, the
number of spatial dimensions $p$ enters explicitly into our formulas,
whereas a universal relation should not depend on this quantity.

Furthermore, we stress that the class of theories examined herein is
limited; it seems likely that any generalization will explicitly
contain the bulk viscosity $\zeta$.  We were able to eliminate this
quantity because our analytical results apply only to theories which
satisfy the algebraic relationship (\ref{zetasoln}) between $v_s$ and
$\zeta$. 

There are many possible extensions of the work presented in this
thesis.  First, the sound mode gauge invariant equations equations can be
solved numerically for a specified metric and set of scalar field
profiles.  These equations may be useful for phenomenologically based
models of the quark-gluon plasma such as those of
\cite{Gursoy:2008za,Batell:2008zm,DePaula:2008fp}.  One chief
application of these gauge invariant equations is the determination of
the bulk viscosity of the dual field theory.  Other methods for
determining the bulk viscosity do exist, in particular the work of
\cite{Gubser:2008sz} is popular in this regard.  However, this method
is currently only applicable to gravity duals supported by a
\emph{single} scalar field, while the methods documented in this
thesis can be applied to models with \emph{multiple} scalar fields.  For
example, consider the model of \cite{Batell:2008zm} which uses two
scalar fields to dynamically generate a soft-wall setup.
Once a finite temperature generalization of this setup is created, one
could use the methods detailed in this thesis to compute the bulk
viscosity of this phenomenological model.

In this thesis, we focused exclusively on the hydrodynamic
approximation of the dispersion relation $w(q)$, which is an expansion
in powers of $q/T \ll 1$.  However, it is also possible to compute the
full dispersion relation $w(q)$ numerically.  This was done for a
conformal theory in \cite{Kovtun:2005ev}, but the analysis could be
extended to the special non-conformal theories considered in this
thesis.  Numerical computation of the dispersion relation allows one
to examine the perturbation's group velocity, (for example, see
\cite{Romatschke:2009im,Hod:2009ig}).  One should check that the
theory obeys causality for all values of the deformation parameter
$c_2$.

It would also be desirable to increase the generality of the gauge
invariant equations by including other types of matter such as gauge
fields, as this would increase their utility.  It is a simple matter to 
add additional matter fields to the linearized Einstein equations 
presented in Chapter \ref{TCReview_chapter}, since we left the
energy-momentum tensor arbitrary at this stage and only later specified
to the case of scalar fields.  

Another issue which should be investigated is whether there are other
special cases or regimes in which the gauge invariant equations can be
solved analytically.  Recently, in \cite{Cherman:2009kf},
the authors find that in the high temperature regime of a single
scalar gravity dual, their numerical results for the bulk viscosity
agree with a simple analytic formula to a remarkable degree of
precision.  This observation may be an indication that a general
analytical formula for the bulk viscosity exists for single scalar
gravity duals.  If this is the case, one should be able to determine
it from the gauge invariant equations presented in this thesis.  The 
establishment of such a formula (which was, indeed, a primary motivation
for the investigation of the work in this thesis) is still highly desirable.  

In this thesis we examined the dispersion relations within the context
of Israel-Stewart hydrodynamics to gain information about the
transport coefficients in this theory.  It would be beneficial to
repeat this analysis for the formulation of second order hydrodynamics
by Baier \emph{et al}.  \cite{Baier:2007ix}.  We have checked that for
conformal theories, the sound mode dispersion relations in both
formulations of hydrodynamics are identical, but it is possible that
differences may exist for non-conformal theories.  At present, the
formulation of second order hydrodynamics presented in
\cite{Baier:2007ix} have not been generalized to non-conformal
theories, though a first step was taken by Kanitscheider \emph{et al}.
\cite{Kanitscheider:2009as} (see also \cite{Romatschke:2009im}); it
would be interesting to combine these results with those presented
here to see if any new information emerges regarding second order
transport coefficients.  These are all issues on which we hope to
report in the future.




\clearpage
\phantomsection
\addcontentsline{toc}{chapter}{References}
\bibliography{thesis}


\appendix
\chapter{General Relativity Reference}
\label{app_EinsteinEqns}
This appendix contains all information necessary to construct 
the linearized Einstein equations (to first order in the metric perturbation)
for a general energy momentum tensor.  The general relativistic notations
used herein are those of Weinberg \cite{Weinberg:1972}.

\section{Metric Perturbations}
This appendix is devoted to the expressions for the relevant general relativistic 
tensors when the background metric $g_{\mu \nu}$ is perturbed
\begin{equation}
  \hat{g}_{\mu \nu} = g_{\mu \nu} + h_{\mu \nu}.  
\end{equation}
Before proceeding one must examine what happens to the inverse metric
\begin{equation}
  \hat{g}^{\mu \nu} \rightarrow g^{\mu \nu} + \delta g^{\mu \nu}.   
\end{equation}
What is $\delta g^{\mu \nu}$?  By definition, the inverse metric is defined such that
\begin{eqnarray}
  \hat{g}_{\mu \nu} \hat{g}^{\mu \rho} &=& \delta^{\rho}_{\nu}, \\ 
  \left(g_{\mu \nu} + h_{\mu \nu} \right)\left( g^{\mu \rho} + \delta g^{\mu \rho} \right) &=& \delta^{\rho}_{\nu}. 
\end{eqnarray}
Expanding to linear order in $h$, we have
\begin{eqnarray}
  g_{\mu \nu}g^{\mu \rho} + h_{\mu \nu} g^{\mu \rho} + g_{\mu \nu} \left(\delta g^{\mu \rho}\right) &=& \delta^{\rho}_{\nu}, \\
  h_{\mu \nu} g^{\mu \rho} + g_{\mu \nu} \left(\delta g^{\mu \rho}\right) &=& 0.  
\end{eqnarray}
Contracting both sides by $g^{\sigma \nu}$ yields the relation
\begin{equation}
  \delta g^{\sigma \rho} = - g^{\sigma \nu} g^{\mu \rho} h_{\mu \nu} \equiv -h^{\sigma \rho}.  
\end{equation}
Thus 
\begin{equation}
  \hat{g}^{\mu \nu} = g^{\mu \nu} - h^{\mu \nu}.  
\end{equation}

\section{Christoffel Symbols}
\label{app_Christoffel}
The Christoffel connection is defined as
\begin{equation}
  \Gamma^{\lambda}_{\mu \nu} = \frac{1}{2} g^{\lambda \sigma} \left[\partial_\mu g_{\sigma \nu} + \partial_\nu g_{\sigma \mu} - \partial_\sigma g_{\mu \nu} \right].  
\end{equation}
These can be evaluated explicitly for the black brane metric (\ref{blackbranemetric}).  The non-vanishing results are
\ba
	\Gamma^{t}_{tr} &=& \frac{1}{2} \DL[g_{tt}(r)], \\
	\Gamma^{i}_{ir} &=& \frac{1}{2} \DL[g_{xx}(r)], \\
	\Gamma^{r}_{rr} &=& \frac{1}{2} \DL[g_{rr}(r)], \\
	\Gamma^{r}_{tt} &=& - \frac{1}{2} g^{rr}(r) g_{tt}'(r), \\
	\Gamma^{r}_{ii} &=& - \frac{1}{2} g^{rr}(r) g_{xx}'(r). 
\ea
Here $i$ denotes the spatial coordinates $x_1...x_p$, and we are using
the notation $\DL$ to denote the logarithmic derivative as defined in
the text (\ref{DLdef}).

Introducing perturbations over the background, one finds the change in the Christoffel symbols:
\begin{equation}
   \delta \Gamma^{\lambda}_{\mu \nu} = \frac{1}{2} g^{\lambda \sigma} \left[
     \partial_\mu h_{\sigma \nu} + \partial_\nu h_{\sigma \mu} - \partial_\sigma h_{\mu \nu} \right]
   -\frac{1}{2} h^{\lambda \sigma} \left[
     \partial_\mu g_{\sigma \nu} + \partial_\nu g_{\sigma \mu} - \partial_\sigma g_{\mu \nu} \right].  
\end{equation}

\section{Covariant derivative}
\label{app_CovariantDeriv}
The definition of the covariant derivative changes with the object on which it acts.  For simplicity,
we will act on a vector.  Generalization to higher rank tensors is straightforward and is found
in any text on general relativity (cf. \cite{Weinberg:1972}).  As in the text, the superscript $(k)$ denotes
quantities which are $k$th order in the perturbation.  For the background metric:
\be
	\nabla^{(0)}_\mu \xi_\nu = \partial_\mu \xi_\nu - \Gamma^{\lambda}_{\mu \nu} \xi_{\lambda}.  
\ee
Supposing now that we want to examine this quantity under perturbations we straightforwardly have
\be
	\delta(\nabla^{(0)}_\mu \xi_\nu) = \partial_\mu (\delta \xi_\nu) 
	- (\delta \Gamma^{\lambda}_{\mu \nu}) \xi_{\lambda}.  
	- \Gamma^{\lambda}_{\mu \nu} (\delta \xi_{\lambda}).  
\ee
One application of these formulas in the text is the evaluation of 
\be
	\delta(\Box^{(0)} \phi) \equiv \delta \left(g^{\mu \nu} \nabla^{(0)}_\mu \nabla^{(0)}_\nu \phi \right).
\label{boxdef}
\ee
Using the above formula with $\xi = \partial_\nu \phi$, one has
\ba
	\delta \left(\Box^{(0)} \phi \right) &=& -h^{\mu \nu} \partial_\mu \partial_\nu \phi +
	g^{\mu \nu} \partial_\mu \partial_\nu (\delta \phi) + h^{\mu \nu} \Gamma_{\mu \nu}^{\lambda} \partial_{\lambda}\phi
	\nonumber \\
	&-& g^{\mu \nu} (\delta \Gamma_{\mu \nu}^{\lambda}) \partial_{\lambda}\phi
	-g^{\mu \nu} \Gamma_{\mu \nu}^{\lambda} \partial_{\lambda}(\delta \phi).  
\label{deltabox}
\ea

\section{Ricci Tensor and Ricci Scalar}
By definition, the background Ricci tensor is given by
\begin{equation}
	R^{(0)}_{\mu \nu} = \partial_{\nu} \Gamma^{\lambda}_{\lambda \mu}
	- \partial_{\lambda} \Gamma^{\lambda}_{\mu \nu}
	+ \Gamma^{\eta}_{\mu \lambda} \Gamma^{\lambda}_{\nu \eta} 
	- \Gamma^{\eta}_{\mu \nu} \Gamma^{\lambda}_{\lambda \eta}.
\end{equation}
It's components can be explicitly evaluated for the `black brane' type
metric given in (\ref{blackbranemetric}).  The results are
\begin{eqnarray}
	F_t(r) &\equiv& g^{tt}R^{(0)}_{tt} = \frac{1}{2\rootg} \partial_r \left( \rootg g^{rr} \DL[g_{tt}] \right), 
	\label{Ftdef}\\
	F_x(r) &\equiv& g^{ii}R^{(0)}_{ii} = \frac{1}{2\rootg} \partial_r \left( \rootg g^{rr} \DL[g_{xx}] \right),
	\label{Fxdef}\\
	F_r(r) &\equiv& g^{rr}R^{(0)}_{rr} = \frac{1}{4 g_{tt}^{\,\prime}}\partial_r \left(g_{tt}g^{rr}\DL[g_{tt}]^2 \right)
	  +  \frac{p}{4 g_{xx}^{\, \prime}}\partial_r \left(g_{xx}g^{rr}\DL[g_{xx}]^2 \right) 
	  \label{Frdef}. 
\end{eqnarray}
where we have defined the shorthand notation $F_t$, $F_x$ and $F_r$ for convenience.  

Combinations which appear frequently in the text are 
\be
	F_t - F_x = \frac{1}{2 \rootg} \partial_r \left( \rootg g^{rr} \DL[f] \right),  
\ee
and 
\be
	F_t - F_r = \frac{p}{2}g^{rr} \DL[g_{xx}] \DL \left[ \frac{\sqrt{f g_{rr}}}{\DL[g_{xx}]} \right],  
\ee  
where $f$ is defined as in the text (\ref{fdef}).  

To first order in the perturbation, the Ricci tensor is expanded as expected
\begin{equation}
	R^{(1)}_{\mu \nu} = \partial_{\nu} \delta \Gamma^{\lambda}_{\lambda \mu}
	- \partial_{\lambda} \delta \Gamma^{\lambda}_{\nu \mu}
	+ \delta \Gamma^{\eta}_{\mu \lambda} \Gamma^{\lambda}_{\nu \eta} 
	+ \Gamma^{\eta}_{\mu \lambda} \delta \Gamma^{\lambda}_{\nu \eta} 
	- \delta \Gamma^{\eta}_{\mu \nu} \Gamma^{\lambda}_{\lambda \eta}
	- \Gamma^{\eta}_{\mu \nu} \delta \Gamma^{\lambda}_{\lambda \eta}.
\end{equation}
The Ricci Scalar is defined as
\begin{equation}
	R^{(0)} \equiv g^{\mu \nu} R^{(0)}_{\mu \nu},
\end{equation}
and in the black brane background takes the form
\begin{equation}
	R^{(0)} = F_t + p F_x + F_r.
\end{equation}
To first order in the perturbation, 
\begin{equation}
  R^{(1)} = g^{\mu \nu} R^{(1)}_{\mu \nu} - h^{\mu \nu} R^{(0)}_{\mu \nu}.
\label{firstorderRicciScalar}
\end{equation}
One must take great care when raising and lowering indices here.  For example, consider
the quantity $	R^{\mu \nu}_{(1)}$.  This notation means $\delta \left(R^{\mu \nu}_{(0)}\right)$, as follows  
\ba
	R^{\mu \nu}_{(1)} &=& \delta \left[g^{\mu \alpha} g^{\nu \beta} R_{\alpha \beta}^{(0)} \right] \nonumber\\
	&=& R_{\alpha \beta}^{(1)} g^{\mu \alpha} g^{\nu \beta} - R_{\alpha \beta}^{(0)} \left[h^{\mu \alpha}g^{\nu \beta} + 
	  g^{\mu \alpha}h^{\nu \beta}\right]. 
\ea
Notice that $R^{\mu \nu}_{(1)} \neq R_{\alpha \beta}^{(1)} g^{\mu
\alpha} g^{\nu \beta}$, there are extra terms that come from the
perturbation metric which raises the indices.

\section{Einstein Equations}
The background Einstein equations are
\begin{equation}
  R^{(0)}_{\mu \nu} - \frac{1}{2} R^{(0)} g_{\mu \nu} = -8 \pi G_{p+2} T^{(0)}_{\mu \nu}.
\end{equation}
Here, $G_{p+2}$ denotes gravitational constant in $p+2$ dimensions,
and $T^{(0)}_{\mu \nu}$ is the background energy momentum tensor
containing information about what matter supports the metric.
The perturbed equations are straightforwardly given as
\begin{equation}
  R^{(1)}_{\mu \nu} - \frac{1}{2} R^{(1)} g_{\mu \nu} - \frac{1}{2} R^{(0)} h_{\mu \nu} = -8 \pi G_{p+2} T^{(1)}_{\mu \nu}.
\end{equation}
To arrive at the linearized Einstein equations given in the text, examine the quantity
\be
	G^{\mu \,(1)}_{\,\nu} = - 8 \pi G_{p+2} T^{\mu \,(1)}_{\,\nu},
\ee
which means
\ba
	\delta \left(g^{\mu \lambda} R_{\nu \lambda}^{(0)} 
	- \frac{1}{2} g^{\alpha \beta}R_{\alpha \beta}^{(0)} \delta^\mu_{\,\nu} \right) 
	&=& -8 \pi G_{p+2} \delta \left( g_{\nu \lambda}T^{\mu \lambda}_{(0)} \right), 
	\\
	g^{\mu \lambda}R^{(1)}_{\nu \lambda} - h^{\mu \lambda} R_{\nu \lambda}^{(0)} 
	- \frac{1}{2}\delta^{\mu}_{\, \nu} \left[g^{\alpha \beta}R^{(1)}_{\alpha \beta} - h^{\alpha \beta} R_{\alpha \beta}^{(0)}\right] 
	&=& - 8\pi G_{p+2} \left(T^{\mu \lambda}_{(1)}g_{\nu \lambda} + T^{\mu \lambda}_{(0)}h_{\nu \lambda} \right).  
	\nonumber \label{masterlinearizedeqn}\\
	\,
\ea
In deriving this equation, we have used the fact that the Kronecker
delta is unchanged under any perturbation.  Explicit evaluation of the
components of this tensor equation for the black brane metric
(\ref{blackbranemetric}) leads to the equations given in the text
(\ref{sheararray1} - \ref{sheararray3}) and (\ref{g00eqn} -
\ref{grzeqn}).

We have left the background energy momentum tensor arbitrary at this
phase.  It needs to be computed separately for the particular set of
matter fields under consideration.  This is done for scalar field
theory in Chapter \ref{ShearMode_chapter}.

\chapter{Techniques used in solving differential equations}
\label{app_DiffEQ}

This appendix contains some details regarding the techniques
used to solve the differential equations in the text.
\section{Reduction of Order}
\label{app_RedOfOrder}

The technique of reduction of order can be used to help solve
second order linear differential equations.  This technique
is useful if one solution of a homogeneous equation is known, 
and one is searching for the second linearly independent solution.  

Consider the homogeneous equation
\be
	y''(x) + p(x) y'(x) + q(x) y(x) = 0,
\ee
and assume that we know one solution $y_1(x)$ which
satisfies this equation.  To find a second solution, make
the ansatz
\be
	y_2(x) = y_1(x)v(x).
\ee
Substituting this into the above equation results in 
\ba
	v''(x)y_1(x) &+& 2 v'(x) y_1'(x) + y_1''(x)v(x) \nonumber \\
	&+& p(x) \left[y_1'(x)v(x) + y_1(x) v'(x) \right] + q(x)y_1(x)v(x) = 0.
\ea
or
\ba
	v''(x)y_1(x) &+& v'(x) \left[2y_1'(x) + p(x)y_1(x) \right] \nonumber \\
	&+& v(x) \left[y_1''(x) + p(x)y_1'(x) +q(x)y_1(x) \right] =0.
\ea
Using the fact that $y_1(x)$ satisfies the differential equation, the last term vanishes, and we are left with the first
order differential equation for $v'(x)$:  
\be
	v''(x) + v'(x) \left[2 \frac{y_1'(x)}{y_1(x)} + p(x) \right] =0.
\ee
This can then be easily integrated to find $v'(x)$
\ba
	v'(x) &=& v_0 \exp \left\{-\int \left(2 \frac{y_1'(x)}{y_1(x)} + p(x)\right)\,dx \right\} \\
	v'(x) &=& \frac{v_0}{y_1(x)^2} \exp \left\{-\int p(x) \,dx \right\}
\ea
where $v_0$ is a arbitrary constant. $v(x)$ is then trivially found by performing another integration.  

\section{Variation of parameters}
\label{app_VarOfParams}

One fundamental technique used in obtaining general solutions to
inhomogeneous differential equations is the method of `Variation of
Parameters'.  The method can be used to find a general solution to a
second order linear differential equation if the solution to the
associated homogeneous equation is known.

The theory behind the method can be found in any textbook on
differential equations.  Here, we simply present the essential formulas.
Consider a differential equation
\be
	y''(x) + p(x) y'(x) + q(x) y(x) = g(x),
\ee
and assume that the functions $y_1(x)$ and $y_2(x)$ are linearly
independent, and satisfy the associated homogeneous equation.  That
is,
\be
	y_1''(x) + p(x)y_1'(x) + q(x)y_1'(x) = 0,
\ee
and similarly for $y_2(x)$.  It can be shown that the function
$y_p(x)$ is a solution to the inhomogeneous equation, where
\be
	y_p(x) = y_2(x) \int \frac{y_1(x)g(x)}{W(x)} \,dx -  y_1(x) \int \frac{y_2(x)g(x)}{W(x)}.
\ee
Here, $W(x)$ is the Wronskian
\be
	W(x) \equiv y_1(x)y_2'(x) - y_1'(x)y_2(x).
\ee
To arrive at the form of the function $y_p$ used in the text (\ref{yp}), let us define
\be
	h(x) \equiv \frac{y_2(x)}{y_1(x)}.  
\ee 
Then,
\be
	W(x) =  h'(x)y_1^2(x),  
\ee
and
\be
	y_p(x) = y_1(x)\left[h(x) \int \frac{g(x)}{y_1(x)h'(x)} \,dx - \int \frac{h(x)g(x)}{h'(x)y_1(x)}\right].
\ee
One can see that this is equivalent to
\be
	y_p(x) = y_1(x) \int h'(x) \int^x \frac{g(z)}{y_1(z)h'(z)} \,dz \,dx
\ee
by performing an integration by parts. 

\end{document}